\documentclass[10pt]{article}

\usepackage{arxiv}
\usepackage{etoolbox,enumitem}

\AtBeginDocument{

  \setlength{\parskip}{0pt}%
  \setlength{\parindent}{15pt}%

  \setlength{\abovedisplayskip}{6pt plus 2pt minus 2pt}%
  \setlength{\belowdisplayskip}{6pt plus 2pt minus 2pt}%

  \setlength{\textfloatsep}{8pt plus 2pt minus 2pt}%
  \setlength{\floatsep}{8pt plus 2pt minus 2pt}%
  \setlength{\intextsep}{8pt plus 2pt minus 2pt}%
  \setlength{\abovecaptionskip}{4pt}%
  \setlength{\belowcaptionskip}{0pt}%
}

\setlist{itemsep=2pt, topsep=2pt, parsep=0pt, partopsep=0pt}

\makeatletter
\def\fps@figure{!t}
\def\fps@table{!t}
\makeatother

\usepackage[utf8]{inputenc} 
\usepackage[T1]{fontenc}    
\usepackage{hyperref}       
\usepackage{url}            
\usepackage{booktabs}       
\usepackage{amsfonts}       
\usepackage{nicefrac}       
\usepackage{silence}
\usepackage{microtype}      
\microtypesetup{activate=true}    
\usepackage{lipsum}		
\usepackage{float} 
\usepackage{graphicx}
\usepackage{doi}
\usepackage{amsmath}
\usepackage{float}
\usepackage{bbding}
\usepackage{pifont}
\usepackage{wasysym}
\usepackage{amssymb}
\usepackage{titling}
\usepackage{lmodern}
\usepackage{caption}
\usepackage{subcaption}
\usepackage[table,xcdraw]{xcolor}
\usepackage{cleveref}

\usepackage[section]{placeins}
\usepackage{caption}

\hypersetup{colorlinks=true, linkcolor=blue, citecolor=blue, urlcolor=blue}

\title{Quantum Physics-Informed Neural Networks for Maxwell's Equations: Circuit Design, ``Black Hole'' Barren Plateaus Mitigation, and GPU Acceleration}
\author{
  Ziv Chen$^{\text{a,c}}$\thanks{Corresponding author: \texttt{ziv.chen@campus.technion.ac.il}} \quad
  Gal G. Shaviner$^{\text{b}}$ \quad 
  Hemanth Chandravamsi$^{\text{b}}$ \quad
  Shimon Pisnoy$^{\text{b}}$ \quad \\
  Steven H. Frankel$^{\text{b,c}}$ \quad
  Uzi Pereg$^{\text{a,c}}$
  \vspace{5pt} \\
  \small{$^{\text{a}}$The Andrew and Erna Viterbi Faculty of Electrical \& Computer Engineering,} \\ [-0.25em]
  \small{Technion - Israel Institute of Technology, Haifa, Israel.} \\
  \small{$^{\text{b}}$Faculty of Mechanical Engineering, Technion - Israel Institute of Technology, Haifa, Israel.} \\
  \small{$^{\text{c}}$Helen Diller Quantum Center, Technion - Israel Institute of Technology, Haifa, Israel.} \\
}

\date{}

\begin{document}
\maketitle

\begin{abstract}
Physics-Informed Neural Networks (PINNs) have emerged as a promising approach for solving partial differential equations (PDEs) by embedding the governing physics into the loss function associated with a deep neural network. In this work, a \textit{Quantum Physics-Informed Neural Network} (QPINN) framework is proposed to solve two-dimensional (2D) time-dependent Maxwell's equations. Our approach utilizes a parametrized quantum circuit (PQC) in conjunction with the classical neural network architecture and enforces physical laws, including a global energy conservation principle, during training. A quantum simulation library, \textit{TorQ - Tensor Operations for Research in Quantum systems,} was developed to efficiently compute circuit outputs and derivatives by leveraging GPU acceleration based on PyTorch, enabling end-to-end training of the QPINN. The method was evaluated on two 2D electromagnetic wave propagation problems: one in free space (vacuum) and the other has an added dielectric medium. Multiple quantum circuit ans\"atze, input scales, and an added loss term were compared in a thorough ablation study.
Furthermore, recent techniques to enhance PINN convergence, including random Fourier feature embeddings and adaptive time weighting, have been incorporated. Our results demonstrate that the QPINN achieves accuracy comparable to, and even greater than, the classical PINN baseline, while using a significantly smaller number of trainable parameters. This study also demonstrates that adding an energy conservation term to the loss stabilizes training and improves the physical fidelity of the solution in the lossless free-space case. This added term helps mitigate a new kind of barren plateau (BP) related phenomenon - \textit{``black hole''} (BH) loss landscape for the quantum experiments in that scenario. By optimizing the quantum-circuit ansatz and embedding energy-conservation constraints, our QPINN achieves up to a $19\%$ higher accuracy on 2D Maxwell benchmark problems compared to a classical PINN.
\end{abstract}


\section{Introduction}
Recent advances in quantum computing have drawn growing interest due to its potential to accelerate scientific machine learning. Among the promising approaches in this domain are \textit{Physics-Informed Neural Networks} (PINNs)~\cite{raissi2019physics,wang2024respecting, toscano2025pinns}, which have emerged as mesh-free, differentiable function-approximation frameworks for solving forward and inverse problems governed by ordinary differential equations (ODEs)/ PDEs~\cite{chen2020physics, yuan2022pinn}. PINNs have been successfully applied to problems in wave propagation \cite{rasht2022physics}, fluid dynamics \cite{jin2021nsfnets}, solid mechanics \cite{haghighat2021physics}, and electromagnetic simulations \cite{noakoasteen2020physics}. Building on this foundation, QPINNs extend the classical PINNs paradigm by embedding parametrized quantum circuits (PQCs) within the neural network ansatz, either partially or fully replacing classical components~\cite{panichi2026quantum,trahan2024quantum}. The integration of quantum models into PINNs is motivated by the prospect of enhanced expressivity and improved scalability, particularly for high-dimensional problems where classical neural networks often face limitations.

Recent works on QPINNs have explored hybrid quantum-classical PINN architectures on simple benchmark problems, demonstrating feasibility for ODEs and small-scale PDEs. For instance, Trahan et al.~\cite{trahan2024quantum} employed a QPINN architecture to solve a an ODE (spring-mass oscillator) and a PDE (Poisson's equation), and employed a hybrid architecture for the one-dimensional Burgers' equation. Sedykh et al.~\cite{sedykh2024hybrid} extended the framework to a three-dimensional fluid flow problem using a hybrid QPINN, reporting improved accuracy over classical PINNs in a complex geometry. These early studies indicate that QPINNs are capable of solving PDEs and can sometimes yield improved accuracy or parameter efficiency. However, despite these promising results, prior work has primarily focused on relatively \textit{simple and canonical problems}, and the performance of QPINNs on high-dimensional, unsteady, and nonlinear systems remains largely unexplored. Motivated by this, in the present work, we study challenges associated with using QPINNs to solve the set of coupled two-dimensional, time-dependent Maxwell's equations - representative of multi-dimensional unsteady field problems. Notable works related to solving Maxwell's equations using classical PINN networks include:~\cite{piao2024domain, brendel2022physics, lim2022maxwellnet, zheng2024implementation, barmada2024novel, shaviner2025pinns}.

Despite the success of PINNs and their broad applicability, training them remains notoriously hard \cite{wang2021eigenvector} due to several practical issues. These include (a) \textit{spectral bias}, wherein high-frequency components are learned significantly more slowly than low-frequency modes, (b) the complex loss landscapes associated with high-dimensional, multiscale problems, and (c) strong sensitivity to hyperparameter choices. To address these limitations, a range of architectural and training strategies have been proposed in the literature to enhance convergence, stability, and generalization~\cite{tancik2020fourier, wang2021eigenvector, wang2023expert, toscano2025pinns, anagnostopoulos2024residual, krishnapriyan2021characterizing, dong2021method}. These techniques have been demonstrated to substantially enhance PINN performance across various applications. In this work, we incorporate a subset of these architectural enhancements within the QPINN framework, focusing specifically on their application to two-dimensional time-dependent Maxwell's equations.

In this paper, we develop a QPINN for solving the 2D time-dependent Maxwell's equations and conduct a detailed analysis of the quantum network ansatz and enforcement of physical conservation laws. The study focuses on a model problem of electromagnetic pulse propagation both in free space (vacuum) and through a dielectric medium. Our contributions are as follows: 
\begin{itemize}
    \setlength{\itemsep}{0pt}
    \setlength{\parskip}{0pt}
        \item We develop the first hybrid quantum-classical PINN for Maxwell's equations, embedding a PQC as a neural network layer and leveraging an in-house GPU-accelerated PyTorch-based quantum simulator with native automatic differentiation for PDE residuals - something no prior work has combined.  
        \item We observe a new phenomenon, BH, in which after several hundred epochs of descending toward an approximate PDE solution, the solution suddenly descends to the trivial solution instead. Due to this effect, the solution has no physical meaning, and its values become approximately 0 at every point in space and time but $t=0$ as it captures only the initial condition. This collapse is distinct from known BP and laziness effects~\cite{liu2024laziness} (elaborated in section~\ref{sec:black_hole}). 
        \item We introduce a novel, physics-informed penalty based on the Poynting energy conservation law. We demonstrate that enforcing this global constraint is crucial to avoid BH collapse in the free-space case.  
        \item We perform a comprehensive ablation, including six quantum ans\"atze and five input-angle encoding schemes, demonstrating exactly how circuit depth, entanglement pattern, and encoding choice affect both convergence and final accuracy.  
        \item We carry out a head-to-head comparison of our QPINN against a leading classical PINN architecture on both vacuum and dielectric test cases, quantifying errors, convergence rates, and the impact of the energy-conservation term.  
\end{itemize}

The rest of the paper is structured in the following manner: Section~\ref{sec:Methodology} describes the problem setup and methodology (\ref{subsec:maxwells_eq}), including the Maxwell equations, the PINN and QPINN formulations (\ref{subsec:PINN}, \ref{subsec:QPINN}), and the quantum circuit ansatz designs. In Section~\ref{sec:simulator}, we detail the custom quantum simulation approach and optimization strategy. Section~\ref{sec:results} presents the experiments and results for the two test scenarios (\ref{subsec:vacuum}, \ref{subsec:dielectric}), with comparisons of different ans\"atze and analysis of the impact of the energy conservation term. This section also elaborates on the ``black hole'' phenomenon (\ref{sec:black_hole}). Section~\ref{sec:discussion_and_conclusion} provides a discussion of the implications of our findings, including advantages and current limitations of the QPINN approach, followed by concluding remarks and potential directions for future research.

\section{Methodology and problem formulation} \label{sec:Methodology}
\subsection{Maxwell's equations in two dimensions and problem setup} \label{subsec:maxwells_eq}
The microscopic form of Maxwell's equations governs the behavior of electric and magnetic fields in space and time. They are expressed as:

\addtocounter{equation}{1}
\begin{equation} \label{eq:maxwell_all}
\begin{aligned}
\nabla \!\cdot\! \mathbf{E} &= \dfrac{\rho}{\varepsilon}, \quad
\nabla \!\cdot\! \mathbf{B} = 0, \quad
\nabla \!\times\! \mathbf{E} = -\dfrac{\partial \mathbf{B}}{\partial t}, \quad
\nabla \!\times\! \mathbf{B} = \mu \left( \mathbf{J} + \varepsilon \dfrac{\partial \mathbf{E}}{\partial t} \right).
\end{aligned} \tag{\arabic{equation}\text{a--d}}
\end{equation}

Here $\mathbf{E}$ and $\mathbf{B}$ denote the electric and magnetic fields; $\rho$ is the electric charge density; $\mathbf{J}$ is the current density; $\varepsilon$ is the electric permittivity; and $\mu$ is the magnetic permeability. In regions free of charges and currents, these equations simplify to:
\addtocounter{equation}{1}
\begin{equation} \label{eq:maxwell_simplified_all}
\begin{aligned}
\nabla \!\times\! \mathbf{B} &= \mu \varepsilon \dfrac{\partial \mathbf{E}}{\partial t}, \quad
\nabla \!\times\! \mathbf{E} = -\dfrac{\partial \mathbf{B}}{\partial t}
\end{aligned} \tag{\arabic{equation}\text{a--b}}
\end{equation}

For linear, isotropic, and homogeneous materials, the constitutive relations connecting the electric field $\mathbf{E}$, magnetic field $\mathbf{B}$, electric displacement field $\mathbf{D}$, and the magnetizing field $\mathbf{H}$ are given by:
\addtocounter{equation}{1}
\begin{equation} \label{eq:BmuH}
\begin{aligned}
\mathbf{D} = \varepsilon \mathbf{E}, \quad
\mathbf{B} = \mu \mathbf{H}
\end{aligned} \tag{\arabic{equation}\text{a--b}}
\end{equation}
where $\varepsilon$ and $\mu$ denote the permittivity and permeability of the medium, respectively.

Substituting Eqs.~\ref{eq:BmuH} into Eq.~\ref{eq:maxwell_simplified_all}, the Maxwell's equations in terms of the electric field $\mathbf{E}$ and the magnetic field $\mathbf{H}$ are obtained:

\addtocounter{equation}{1}

\begin{equation} \label{eq:maxwell_eh2}
\begin{aligned}
\frac{\partial \mathbf{E}}{\partial t} &= \frac{1}{\varepsilon} \nabla \!\times\! \mathbf{H}, \quad
\frac{\partial \mathbf{H}}{\partial t} = -\frac{1}{\mu} \nabla \!\times\! \mathbf{E}
\end{aligned} \tag{\arabic{equation}\text{a--b}}
\end{equation}

For simplicity, a transverse electric (TE$_z$) polarization, where the electric field has only a $z$-component $E_z(x,y,t)$ (out of plane) and the magnetic field has in-plane components $(H_x(x,y,t), H_y(x,y,t))$ was used. The Maxwell equations for this 2D TE case can be written as: 
\addtocounter{equation}{1} 
\begin{equation} \label{eq:maxwell_2d_3}
\begin{aligned}
\frac{\partial E_z}{\partial t} &= \frac{1}{\varepsilon} \left( \frac{\partial H_y}{\partial x} - \frac{\partial H_x}{\partial y} \right), \quad
\frac{\partial H_x}{\partial t} = -\frac{1}{\mu} \frac{\partial E_z}{\partial y}, \quad
\frac{\partial H_y}{\partial t} = \frac{1}{\mu} \frac{\partial E_z}{\partial x}
\end{aligned} \tag{\arabic{equation}\text{a--c}}
\end{equation}

The permittivity and permeability are different by several orders of magnitude, so the equations are manipulated by changing the electric field:
\begin{equation}
    \mathbf{\tilde{E}} = \sqrt{\frac{\varepsilon_0}{\mu_0}} \mathbf{E} \label{eq:norm_E}
\end{equation} \label{eq:tildeE}
Hence Eqs.~\ref{eq:maxwell_2d_3} become: 
\addtocounter{equation}{1}
\begin{equation} \label{eq:maxwell2DWithTildeE_all}
\begin{aligned}
\frac{\partial \tilde{E_z}}{\partial t} &= \frac{1}{\sqrt{\varepsilon_0\mu_0}} \left( \frac{\partial H_y}{\partial x} - \frac{\partial H_x}{\partial y} \right), \quad
\frac{\partial H_x}{\partial t} = -\frac{1}{\sqrt{\varepsilon_0\mu_0}} \frac{\partial \tilde{E_z}}{\partial y}, \quad
\frac{\partial H_y}{\partial t} = \frac{1}{\sqrt{\varepsilon_0\mu_0}} \frac{\partial \tilde{E_z}}{\partial x}
\end{aligned}
\tag{\arabic{equation}\text{a--c}}
\end{equation}

After the transformation in Eq.~\ref{eq:norm_E}, the field amplitudes $\tilde{E_z}$, $H_x$, and $H_y$ become comparable in magnitude. This allows us to normalize the permittivity and permeability by setting $\varepsilon_0 = \mu_0 = 1$, which improves numerical stability. In vacuum, the material parameters are given by $\varepsilon = \varepsilon_0$ and $\mu = \mu_0$. In dielectric regions, we assume $\mu = \mu_0$ remains constant, while the permittivity becomes $\varepsilon = \varepsilon_r \varepsilon_0$. In our setup, the relative permittivity is set to $\varepsilon_r = 4$ inside the dielectric medium. For notational simplicity, we drop the tilde and henceforth denote the normalized electric field simply as $E_z$.

The domain in which both cases were tested is spatially periodic \( x,y \in [-1,1] \). The time domain in the vacuum case was \( t \in [0,1.5] \) to capture wall-reflection effects and in the dielectric case \( t \in [0,0.7] \) so the waves were well contained within the boundaries to simulate a boundaryless space.

$C_{\text{loss, per point}} \;\approx\; 1 \;+\; \sum_{\text{needed derivatives}} \big(2^{\text{order}} \times \text{\#occurrences of the relevant input dependence}\big)$
\subsection{PINN architecture and convergence enhancements} \label{subsec:PINN}
As a baseline, a classical PINN for the 2D Maxwell problem is first established, which is also the basis for our hybrid QPINN. The solution fields are represented by a neural network $f_\theta(x,y,t)$ with trainable parameters $\theta$. The network takes continuous coordinates $(x,y,t)$ as input and outputs a vector $(E_z, H_x, H_y)$ which constitutes the PINN's prediction for the fields at that point.

\noindent
Residuals for each equation are defined to calculate the PDE loss - $\mathcal{L}_{\text{phys}}$. For the first equation, in the vacuum test case, a single residual is defined: 
\addtocounter{equation}{1}
\begin{equation} \label{eq:residual_1_vac}
\begin{aligned}
res_{1,\;\text{vac}}=\frac{\partial \tilde{E_z}}{\partial t} - \left( \frac{\partial H_y}{\partial x} - \frac{\partial H_x}{\partial y} \right), 
\end{aligned}
\end{equation}
While in the dielectric test case, two residuals are defined. The first ($res_{1,\;\text{vac}}$) is the same as in the vacuum case (\ref{eq:residual_1_vac}) and consists of the points in the vacuum part of the domain, and consists of the points in the dielectric part of the domain:
\addtocounter{equation}{1}
\begin{equation} \label{eq:residual_1_diel}
\begin{aligned}
res_{1,\;diel}=\frac{\partial \tilde{E_z}}{\partial t} - \frac{1}{\varepsilon_r} \left( \frac{\partial H_y}{\partial x} - \frac{\partial H_x}{\partial y} \right), 
\end{aligned}
\end{equation}
Residuals 2 and 3 are the same for both cases:
\addtocounter{equation}{1}
\begin{equation} \label{eq:residuals_2_and_3}
\begin{aligned}
res_2=\frac{\partial H_x}{\partial t} + \frac{\partial \tilde{E_z}}{\partial y}, \quad
res_3=\frac{\partial H_y}{\partial t} - \frac{\partial \tilde{E_z}}{\partial x}
\end{aligned}
\tag{\arabic{equation}\text{a--b}}
\end{equation}
A physics-informed loss $\mathcal{L}_{\text{phys}}$ that penalizes the PDE residuals is defined as follows (one for each case):
\begin{equation}
\begin{aligned}
\mathcal{L}_{\text{phys, vac}} = MSE(res_{1,\;\text{vac}}) + MSE(res_2) + MSE(res_3)
\label{eq:phys_loss_vac}
\end{aligned} 
\end{equation}
\begin{equation}
\begin{aligned}
\mathcal{L}_{\text{phys, diel}} = MSE(res_{1,\;\text{vac}}(x,y,t)_{\in N_{vac}}) + MSE(res_{1,\;diel}(x,y,t)_{\in N_{diel}}) + MSE(res_2) + MSE(res_3)
\label{eq:phys_loss_diel}
\end{aligned} 
\end{equation}
where $N_{vac}$ is the number of collocation points in the grid in which $\varepsilon=1$, and $N_{diel}$ are the rest of the collocation points. The MSE function is defined as follows:
\begin{equation}
\begin{aligned}
MSE(F(x,y,t)_{\in N}) = \frac{1}{N}\sum_{x,y,t\in N}F^2(x,y,t)
\label{eq:mse}
\end{aligned} 
\end{equation}
In the dielectric case, this loss function weights equally the vacuum part and the dielectric part of the domain, even though there are fewer collocation points in the dielectric part than in the vacuum part. The emphasis on the dielectric part stabilizes convergence and avoids BH as explained later, in subsection~\ref{subsec:stable_diel_loss}.
The grid size used was $64^3$ (64 per coordinate), spread equally.

The initial conditions are the same for both cases, a 2-D Gaussian pulse in the electric field, while the magnetic fields are initialized as zero:
\begin{equation}
{E_{z}}_{0}(x,y,t=0) = e^{-25 (x^2 + y^2)}
\label{eq:pulse_IC_Ez}
\end{equation}
\begin{equation}
{H_{x}}_{0}(x,y,t=0) = 0
\label{eq:pulse_IC_Hx}
\end{equation}
\begin{equation}
{H_{y}}_{0}(x,y,t=0) = 0
\label{eq:pulse_IC_Hy}
\end{equation}
Therefore, the initial-condition loss $\mathcal{L}_{\text{IC}}$ is defined as follows:
\begin{equation}
\mathcal{L}_{\mathrm{IC}} = \frac{1}{N_{x,y}}\sum_{x,y}\bigg[\left( E_z(x,y,0) - {E_{z}}_{0}(x,y,t=0)\right)^2 + \left( H_x(x,y,0) \right)^2 + \left( H_y(x,y,0) \right)^2 \bigg]
\end{equation} \label{eq:IC_loss}
where $N_{x,y}$ is the amount of collocation points in the $x$-$y$ plane, i.e. $64^2$ in this case.

It is possible to add any other loss terms that fit the specific problem to aid in convergence, while accounting for the caveat that each added loss term enlarges the optimization tradeoff against other loss terms if they do not fully comply with each other. In this case, two more loss terms are added.\\ 

The first term is symmetry loss, which penalizes deviations from spatial (anti-)symmetry values.
In the vacuum case, the PDE+IC+domain are invariant under reflections $x\!\to\!-x$ and $y\!\to\!-y$. 
With a centered Gaussian initial condition in $E_z$ and uniform $(\varepsilon,\mu)$ imply the following parities:

\begin{align*}
&\text{(i) } E_z \text{ is even in } x \text{ and } y: && E_z(x,y,t)=E_z(-x,y,t)=E_z(x,-y,t),\\
&\text{(ii) } H_x \text{ is even in } x \text{ and odd in } y: && H_x(x,y,t)=H_x(-x,y,t),\ \ H_x(x,y,t)=-H_x(x,-y,t),\\
&\text{(iii) } H_y \text{ is odd in } x \text{ and even in } y: && H_y(x,y,t)=-H_y(-x,y,t),\ \ H_y(x,y,t)=H_y(x,-y,t).
\end{align*}
We encode these parities with a symmetry penalty $\mathcal{L}_{\mathrm{sym}}$ defined below, where for any field $f$ we write \\ 
$f^{\mathsf{mirr}_x}(x,y,t):=f(-x,y,t)$ and $f^{\mathsf{mirr}_y}(x,y,t):=f(x,-y,t)$.
\begin{equation}
\begin{aligned}
\mathcal{L}_{\mathrm{sym}} 
= \frac{1}{N}\sum_{(x,y,t)}
\Big[ 
&\underbrace{\big(E_z - E_z^{\mathsf{mirr}_x}\big)^2
+ \big(E_z - E_z^{\mathsf{mirr}_y}\big)^2}_{\text{(i)}} \\
+ & \underbrace{\big(H_x - H_x^{\mathsf{mirr}_x}\big)^2 + \big(H_x + H_x^{\mathsf{mirr}_y}\big)^2}_{\text{(ii)}}  \\
+ & \underbrace{\big(H_y + H_y^{\mathsf{mirr}_x}\big)^2 + \big(H_y - H_y^{\mathsf{mirr}_y}\big)^2}_{\text{(iii)}}
\Big],
\label{eq:2D_symLoss}
\end{aligned}    
\end{equation}

In the dielectric case, the terms mirrored along the $x$ axis should be omitted, as the dielectric medium addition cancels the symmetry along this axis. \\

The second term is \textit{energy conservation loss} - $\mathcal{L}_{\mathrm{energy}}$. An important physical property of Maxwell's equations (in lossless media) is the conservation of electromagnetic energy. In a closed domain with periodic boundaries and no absorption, the total energy of the fields should remain constant in time. Numerically, PINN solutions might violate this principle due to approximation errors. A global energy conservation term in the loss is introduced to push the network towards this invariance according to the Poynting theorem: 
\begin{equation}
\begin{aligned}
-\frac{\partial u}{\partial t}=\nabla \!\cdot\! \mathbf{S}+\mathbf{J} \!\cdot\! \mathbf{E} \\
\end{aligned}
\label{eq:Poynting_thm}
\end{equation}
 
where $-\frac{\partial u}{\partial t}$ is the rate of decrease of electromagnetic energy density in the volume, 
$\nabla\!\cdot\!\mathbf{S}$ is the net energy flux out of the volume - i.e., Poynting vector divergence, and 
$\mathbf{J}\!\cdot\!\mathbf{E}$ is the electrical work-rate density (W\,m$^{-3}$) done by the fields on the charges - i.e., the power per unit volume transferred from the field to matter, which is positive when energy leaves the field and enters the charges. For Maxwell's equations,
\begin{equation}
\begin{aligned}
u&\equiv \frac{1}{2}(\mathbf{E} \!\cdot\! \mathbf{D}+\mathbf{B} \!\cdot\! \mathbf{H}) \\
 &=\frac{1}{2}\left(\varepsilon E_z^2 + \mu\left(H_x^2 + H_y^2\right)\right) =\frac{1}{2}\left(\varepsilon E_z^2 + H_x^2 + H_y^2\right).
\end{aligned}
\label{eq:Poynting_value_u}
\end{equation}
Furthermore
\begin{equation} \label{eq:Poynting_value_S}
\mathbf{S} = \mathbf{E} \!\times\! \mathbf{H} \quad \Rightarrow \quad S_x = -E_z H_y, \quad S_y = E_z H_x
\end{equation}
and
\begin{equation}
\begin{aligned}
\mathbf{J} \!\cdot\! \mathbf{E}=0,
\end{aligned}
\label{eq:Poynting_value_JE}
\end{equation}
$\mu=1$, hence the last expression in Eq. (\ref{eq:Poynting_value_u}), and Eq. (\ref{eq:Poynting_value_JE}) is correct due to the fact that in the problems considered here $\mathbf{J}=0$.
Now the energy residual can be evaluated, which represents our energy conservation loss:
\begin{equation}
\begin{aligned}
\mathcal{L}_{\mathrm{energy}}
= \;\;&\left(\varepsilon E_z \frac{\partial E_z}{\partial t} + H_x \frac{\partial H_x}{\partial t} + H_y \frac{\partial H_y}{\partial t}\right) \\
- &\left(\frac{\partial E_z}{\partial x} H_y + E_z\frac{\partial H_y}{\partial x} \right)
+ \left(\frac{\partial E_z}{\partial y} H_x + E_z\frac{\partial H_x}{\partial y} \right)
\end{aligned}
\label{eq:energy_loss}
\end{equation}

In the lossless, time-independent setting considered here, the local Poynting balance $\partial_t u+\nabla\!\cdot(E\times H)=0$ is implied by the TE$_z$ Maxwell system (Eq.~\ref{eq:maxwell_2d_3}). Hence, $\mathcal{L}_{\mathrm{energy}}$ does not add new physics; it reweights a continuum identity. This reweighting was found to be beneficial in the QPINN runs of the vacuum test case (Subsec.~\ref{subsec:vacuum}) because it penalizes spurious global amplitude decay (the ``black-hole'' failure mode, Subsec.~\ref{sec:black_hole}), but it can be detrimental in heterogeneous media where the PDE residual scales with $1/\varepsilon(x)$ while the energy residual scales with $\varepsilon(x)$, leading to stiff, unbalanced gradients near interfaces, as in the dielectric test case (Subsec.~\ref{subsec:dielectric}).

Including the loss term $\mathcal{L}_{\mathrm{energy}}$ aids the training process in two ways: it provides a global feedback that couples the network's errors across the domain (e.g., if the network tries to introduce a spurious attenuation or growth in the wave amplitude, it will be penalized), and it encodes a physical constraint that might otherwise require many PDE residual points to capture. 
The derivatives in this loss term were already calculated, hence its computational overhead is negligible. In addition, it turns out that the \textit{energy conservation loss} had a significant impact on the network's convergence, and it was thus an integral part of this ablation study.

The total loss combines the contributions from the initial condition, symmetry, and the equations' residual ($\mathcal{L}_{\mathrm{phys}}$) losses:
\begin{equation}
    \mathcal{L}_\mathrm{tot}
    = \mathcal{L}_{\mathrm{phys}}\
    + 10\,\mathcal{L}_{\mathrm{IC}}
    + 10\,\mathcal{L}_{\mathrm{sym}}
    + 10\,\mathcal{L}_{\mathrm{energy}}
    \label{eq:2D_lossTot}
\end{equation}

\noindent Several recently developed techniques to improve PINN training were also incorporated:
\begin{itemize}\setlength\itemsep{0em}
    \item \textbf{Random Fourier Features (RFF):} A mapping layer that embeds $(x,y,t)$ into a higher-dimensional feature space was applied using randomized sinusoidal features~\cite{rahimi2007random, tancik2020fourier}. Specifically, the inputs were transformed as $(x,y,t) \mapsto \big(\cos(\mathbf{\Omega} [x,y,t]^T), \sin(\mathbf{\Omega} [x,y,t]^T)\big)$ where \(\mathbf{\Omega}\) is a fixed random projection tensor whose entries are sampled from a Gaussian distribution and remain constant throughout training (i.e., they are not trainable parameters). This transformation incorporates high-frequency components into the input representation, enhancing the network's capacity to capture multiscale features of the PDE solution and thereby mitigating spectral bias inherent in PINNs.
    \item \textbf{Strict spatial periodicity enforcement:} The $x$ and $y$ coordinates are each mapped through sine and cosine functions, scaled according to the domain length, to enforce periodic boundary conditions. 
    Specifically, for an input $(x,y,t)$ we replace $x$ by $x_{sin}$ and $x_{cos}$: 
    \begin{align}
        x_{sin} &= \sin(2\pi x / L_x) \\
        x_{cos} &= \cos(2\pi x / L_x)
        \label{x_sin_cos}
    \end{align} 
    and similarly for $y$, where $L_x, L_y$ are the domain lengths. This ensures 
    $f_\theta(x)\equiv f_\theta(x+L_x)$ and $f_\theta(y)\equiv f_\theta(y+L_y)$ for the domain boundaries, eliminating 
    the need for an explicit boundary loss term as suggested by Dong et al.~\cite{dong2021method}. 
    \item \textbf{Periodic time domain mapping:} the $t$ coordinate is also passed through a $\sin$ and $\cos$ mapping, but instead of dividing by the domain length, the denominator (the periodicity time length) is a learned parameter to the neural network, as the simulation length is not long enough to complete a full period.
    \item \textbf{Adaptive temporal weighting:} A time-domain curriculum strategy 
    similar to Wang et al.~\cite{wang2024respecting} is adopted, such that the network learns 
    early-time dynamics first. In practice, the collocation points are divided into $M=5$ time bins, with lower loss residual weights assigned to subsequent time bins at the beginning of training; these weights are gradually increased as the network converges on the early-time dynamics. This method 
    allows the model to focus on the early-time behavior (closer to initial conditions), thus propagating the solution forward in a causality-respecting manner. For further details, refer to \cite{shaviner2025pinns}.
\end{itemize}

\begin{figure}[!htbp]
    \centering
    \includegraphics[width=1.0\textwidth]{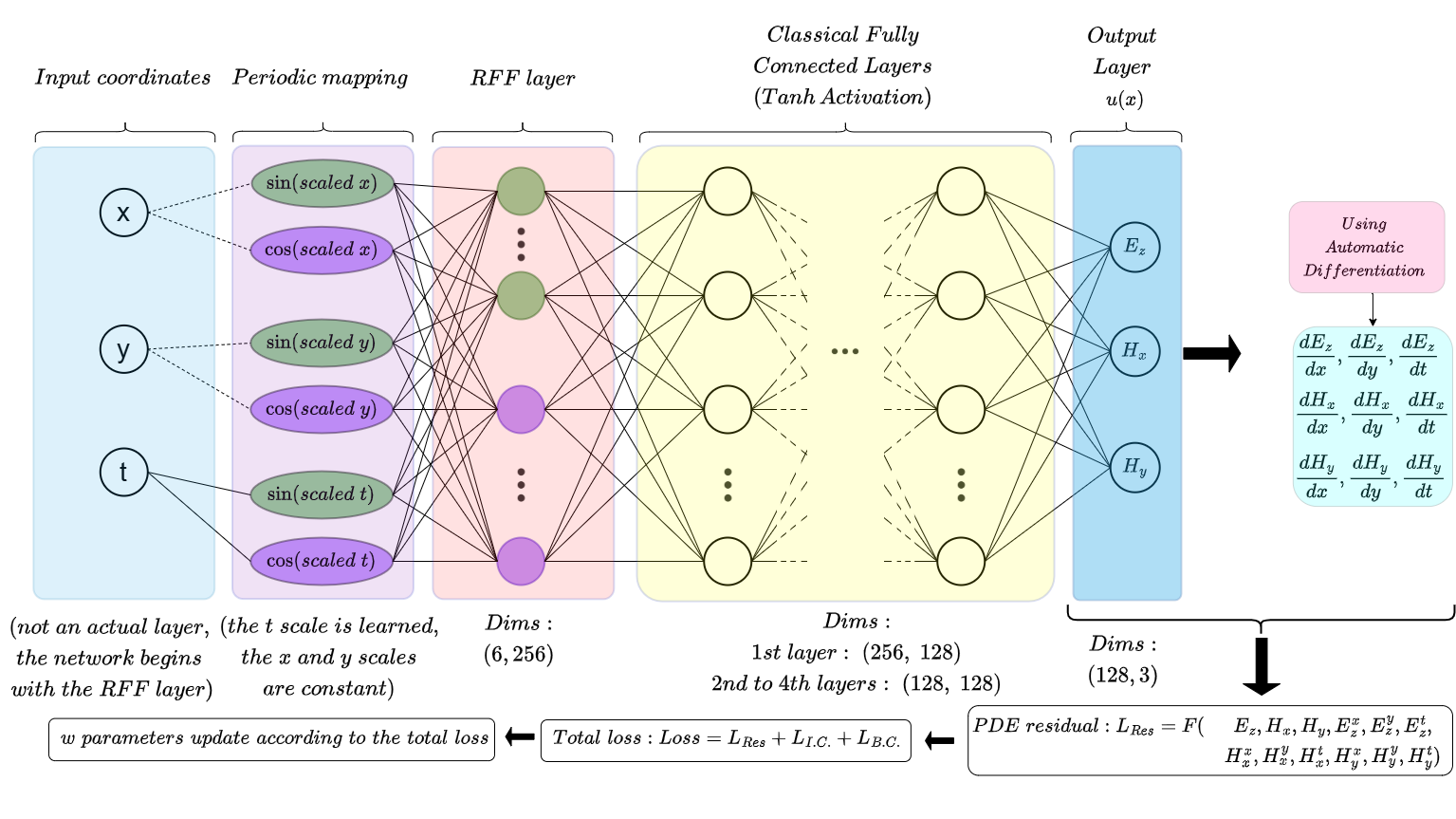}
    \caption{Schematic of a classical PINN architecture employed in the present work for solving 2D Maxwell's equations. The inputs \( (x, y, t) \) are mapped to outputs \( (E_z, H_x, H_y) \). A periodic mapping layer first enforces strict periodic boundary conditions in space and learned periodicity in time, followed by a random Fourier feature (RFF) layer that expands the inputs into a higher-dimensional sinusoidal feature space. The transformed inputs then pass through fully connected hidden layers before producing the final field outputs.}
    \label{fig:pinn_architecture}
\end{figure}

Using these enhancements, the classical PINN with a feed-forward architecture (fully connected layers) was trained. In our experiments, a network with $4$ hidden layers as used in Shaviner et al.~\cite{shaviner2025pinns}, and two additional networks for experimentation, with an extra layer and a reduced layer. Each layer consists of $H=128$ neurons, except the first one after the RFF layer that is set to $256$ neurons, because the RFF layer has $128$ $sine$ and $128$ $cosine$ outputs. Aside from the sinusoidal embedding, all models (Fig.~\ref{fig:pinn_architecture}) used hyperbolic tangent activations on hidden layers. Training was performed using the Adam optimizer~\cite{kingma2014adam}, with a learning rate of $0.001$, that decayed by a factor of $0.85$ every $2000$ epochs.

\subsection{QPINN architecture and ansatz designs} \label{subsec:QPINN}
\begin{figure}[!htbp]
    \centering
    \includegraphics[width=1.0\textwidth]{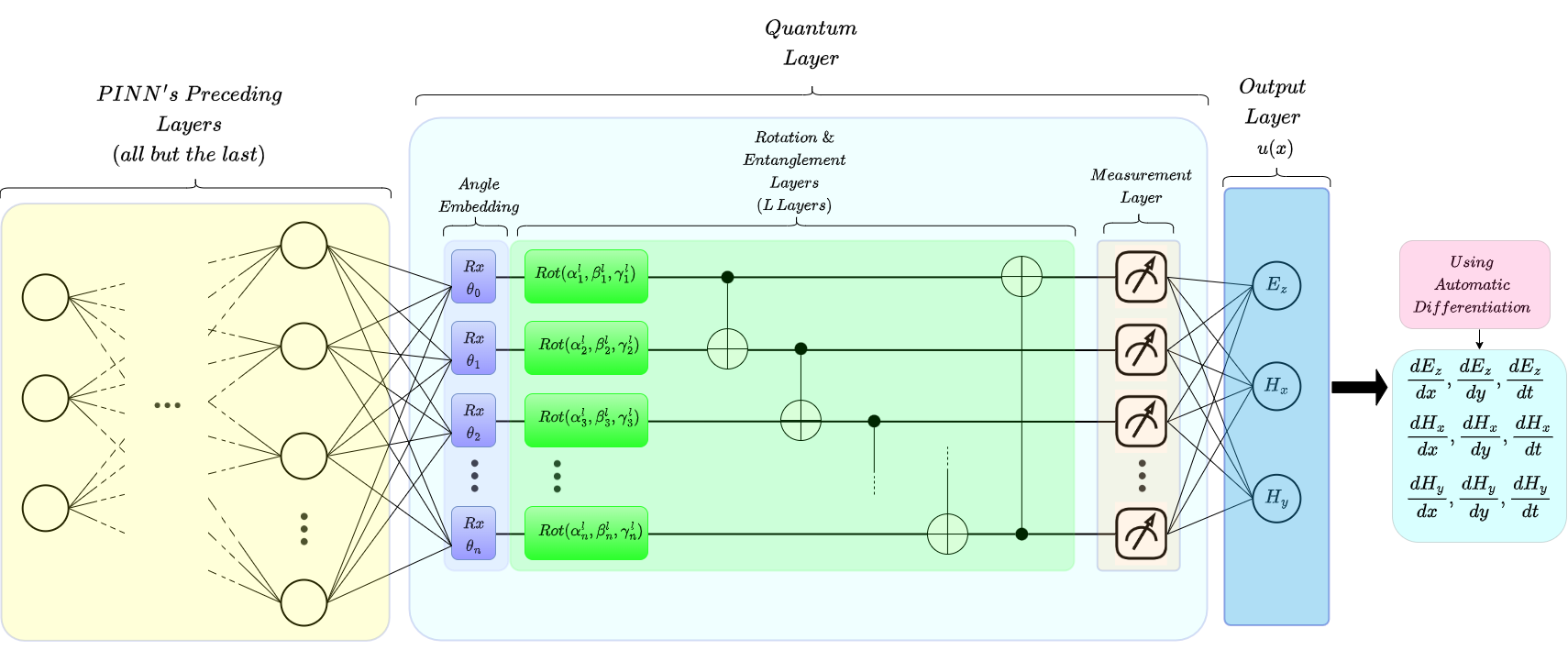}
    \caption{Schematic of the QPINN architecture. It shares the same overall structure as the PINN in Fig.~\ref{fig:pinn_architecture}, but a quantum circuit layer is inserted near the output. The preceding classical layer and subsequent output layer adjust dimensions for the quantum layer's 7 qubits (each qubit acting as a neuron).}
    \label{fig:qpinn_architecture}
\end{figure}

Our QPINN extends the classical PINN by replacing the second-to-last layer of the classical neural network with a PQC. The location was chosen such that the layer is near the end of the network, which is \textit{assumed} to bear more impact. Nevertheless, we do not use it as the last one to guarantee a generic amount of outputs. This is much easier to achieve using a classical layer rather than a quantum one. In essence,  a hybrid model $f_{\boldsymbol{\theta}}(x,y,t)$, where $\boldsymbol{\theta}$ is a parameter vector that includes both classical network parameters and parameters of the constructed quantum circuit. The PQC's architecture is structured by first setting an angle embedding layer, followed by $L$ ansatz layers, and finally a Pauli-Z measurement. This architecture is depicted in Fig.\ref{fig:qpinn_architecture}, where 7 qubits and 4 layers are used in all experiments. The amount of classical learnable parameters in all QPINN models is 66848 for all ansatz types.

The integration between the classical and quantum neural networks layer, i.e., the PQC, is done based on an angle embedding such that the output values propagate from the activation layer preceding the PQC, into the PQC itself. The angle embedding layer in the PQC rotates each qubit around the $X$ axis by an angle that corresponds to the scaled activation layer output. Five different scaling methods are tested in this experiment, as the variable $a$ takes the $tanh$ output values, ranging between $[-1,1]$: 
\addtocounter{equation}{1}
\begin{equation} \label{eq:scale}
\begin{aligned}
    scale_{none}(a) &= a \;\;\Rightarrow\; \in [-1,1], \quad 
    scale_{\pi}(a) = a\!\cdot\!\pi \;\;\Rightarrow\; \in [-\pi,\pi], \quad 
    scale_{bias}(a) = \frac{(a + 1)}{2}\!\cdot\!\pi \;\;\Rightarrow\; \in [0,\pi], \\
    scale_{asin}(a) &= \arcsin{(a)}+\frac{\pi}{2} \;\;\Rightarrow\; \in [0,\pi], \quad 
    scale_{acos}(a) = \arccos{(a)} \;\;\Rightarrow\; \in [0,\pi]
\end{aligned} \tag{\arabic{equation}\text{a--e}}
\end{equation} 

\begin{figure}[!htbp]
  \centering
  \begin{subfigure}[t]{0.49\textwidth}
    \centering
    \includegraphics[width=\linewidth]{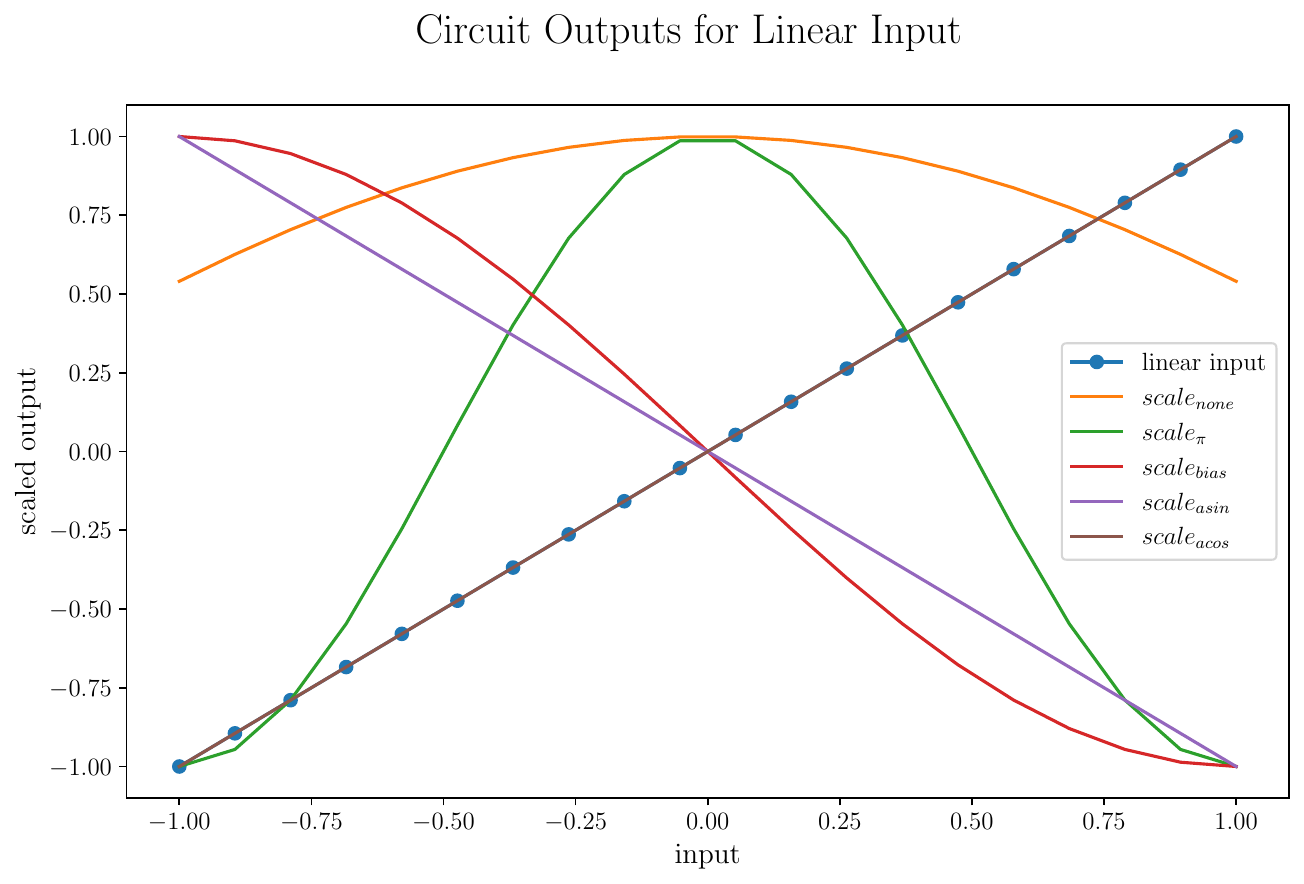}
    \caption{}
    \label{fig:scales_linear}
  \end{subfigure}\hfill
  \begin{subfigure}[t]{0.49\textwidth}
    \centering
    \includegraphics[width=\linewidth]{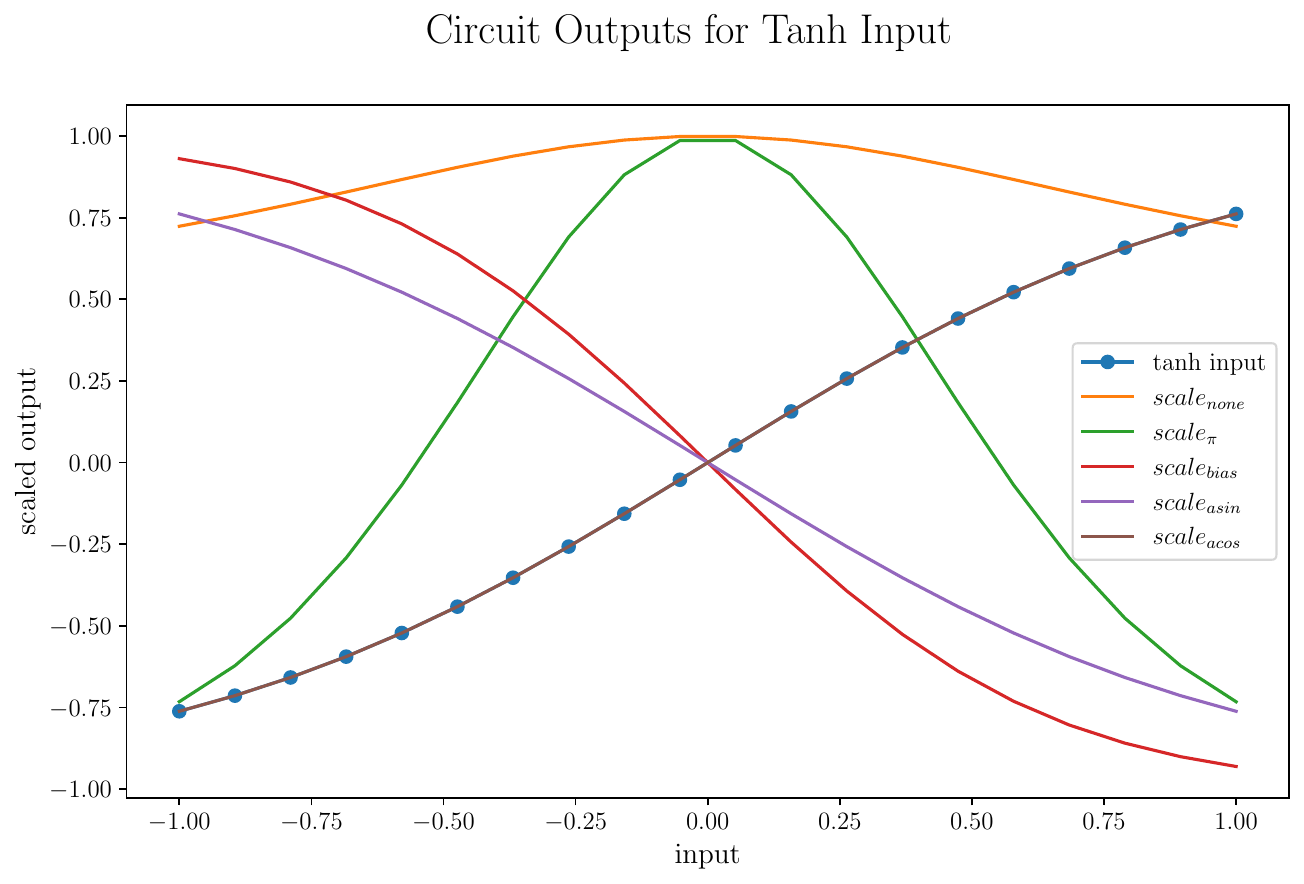}
    \caption{}
    \label{fig:scales_tanh}
  \end{subfigure}

  \medskip

  \begin{subfigure}[t]{0.85\textwidth}
    \centering
    \includegraphics[width=\linewidth]{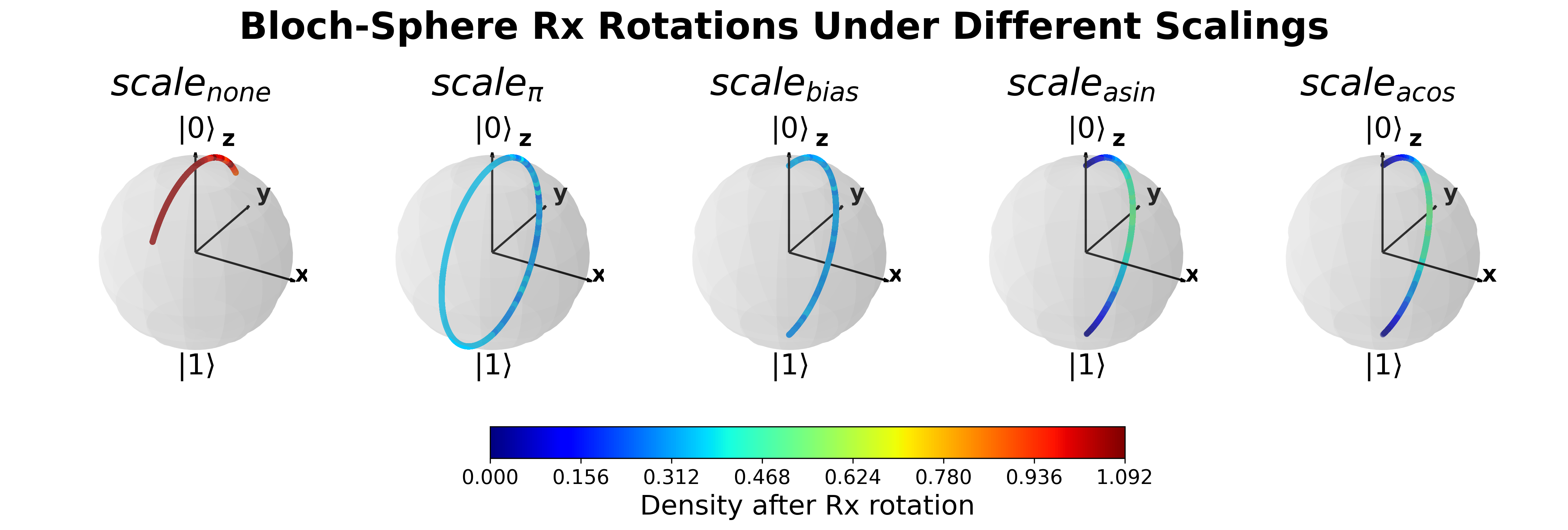}
    \caption{}
    \label{fig:bloch_spheres_spreads}
  \end{subfigure}\hfill

  \medskip
  
  \begin{subfigure}[t]{0.85\textwidth}
    \centering
    \includegraphics[width=\linewidth]{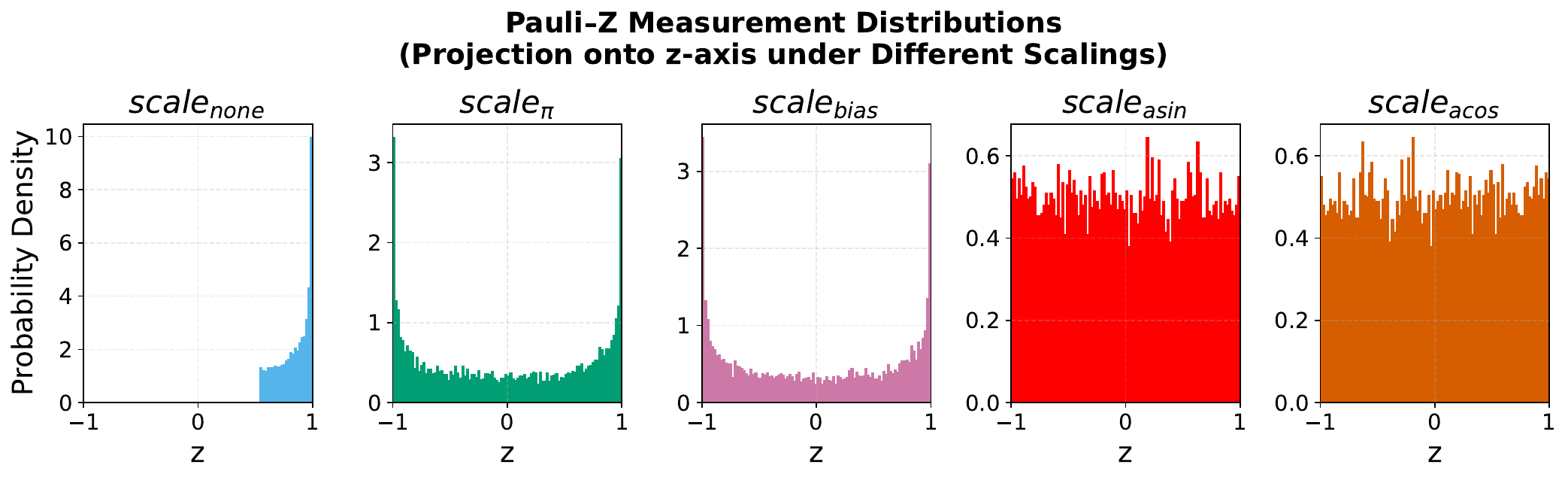}
    \caption{}
    \label{fig:bloch_pauli_z_histogram}
  \end{subfigure}

  \caption{Effect of input-angle scalings for the $\mathrm{Rx}$ embedding and $Z$ readout. 
(a) General case with linear inputs $a\in[-1,1]$: with $\langle Z\rangle=\cos\theta$, 
$scale_{acos}$ ($\theta=\arccos a$) yields $\langle Z\rangle=a$ (identity) and 
$scale_{asin}$ ($\theta=\arcsin a+\tfrac{\pi}{2}$) yields $\langle Z\rangle=-a$ (sign flip). 
(b) Same analysis for our actual setting, where the PQC receives $\tanh$-bounded activations $a\in[-1,1]$ from the preceding layer. 
(c) Distribution of the scaled angles $\theta$ induced by each mapping in Eqs.~\ref{eq:scale} for $a\sim \mathrm{Unif}[-1,1]$. 
(d) Probability distribution of the Pauli-Z measurement outcomes for the corresponding inputs in (c).}
  \label{fig:scales}
\end{figure}

This choice of scaling aims to utilize the Bloch sphere with a better distinction between the inputs, as illustrated in figure \ref{fig:bloch_spheres_spreads}. Furthermore, when using the Pauli-Z measurement, several considerations apply. The first is that only the projection along the z-axis in the Bloch sphere matters, hence symmetric angle rotations yield the same probabilities for different results, such as $-1$ and $1$, which are supposed to be the most distant from one another in the $tanh$ output. The second consideration is that using Pauli-Z measurement after angle embedding results in a sinusoidal shape of the results (as can be seen in Figs.\ref{fig:scales_linear}, \ref{fig:scales_tanh}). Thereby, $scale_{asin}$ and $scale_{acos}$ are also tested to help shape the probability density into a uniform one, as shown in Fig.\ref{fig:bloch_pauli_z_histogram}. 

\begin{figure}[!htbp]
  \centering
  \begin{subfigure}[t]{0.49\textwidth}
    \centering
    \includegraphics[width=\linewidth]{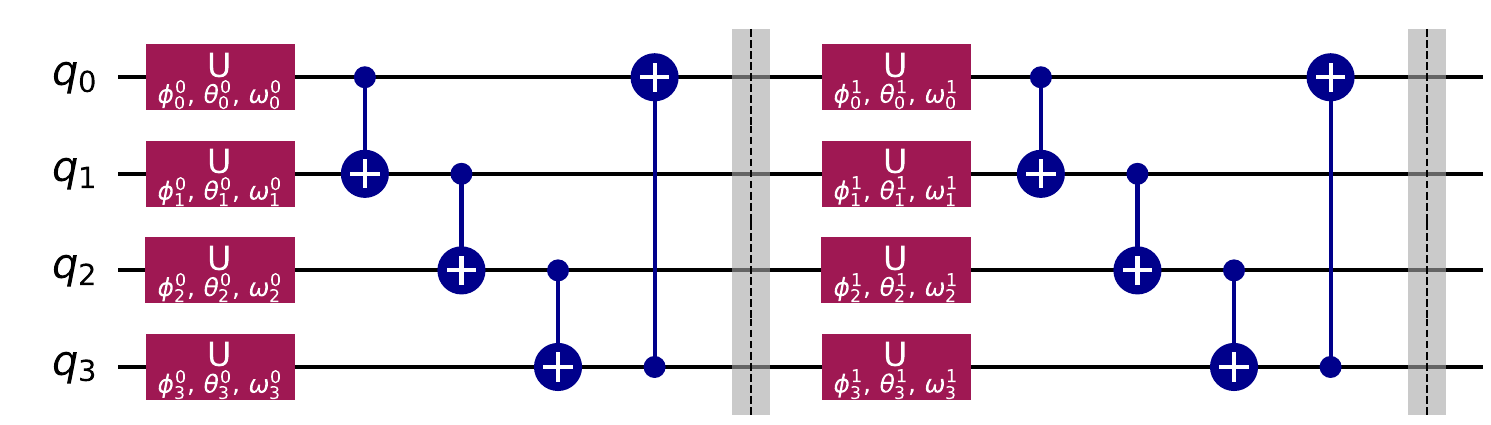}
    \caption{}
    \label{fig:basic_ent_layer}
  \end{subfigure}\hfill
  \begin{subfigure}[t]{0.49\textwidth}
    \centering
    \includegraphics[width=\linewidth]{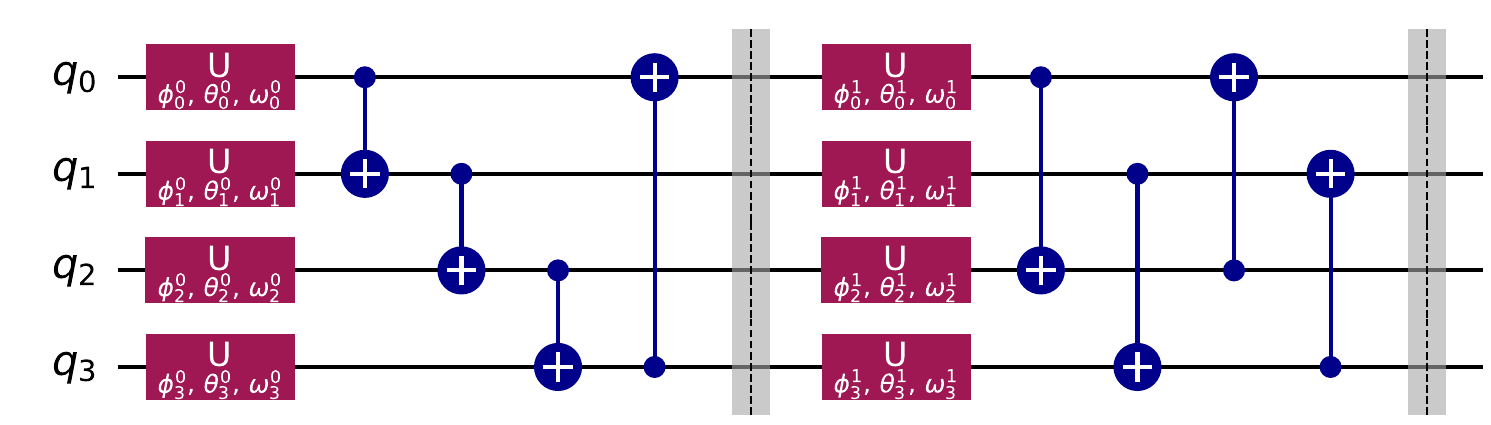}
    \caption{}
    \label{fig:strongly_ent_layer}
  \end{subfigure}

  \medskip

  \begin{subfigure}[t]{0.49\textwidth}
    \centering
    \includegraphics[width=\linewidth]{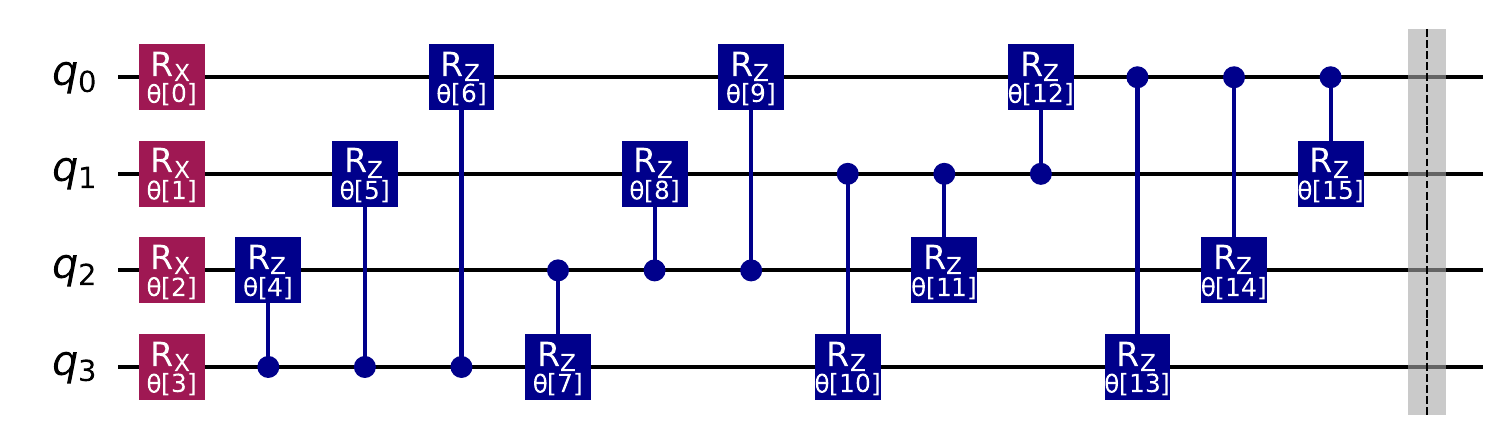}  
    \caption{}
    \label{fig:cross_mesh}
  \end{subfigure}\hfill
  \begin{subfigure}[t]{0.49\textwidth}
    \centering
    \includegraphics[width=\linewidth]{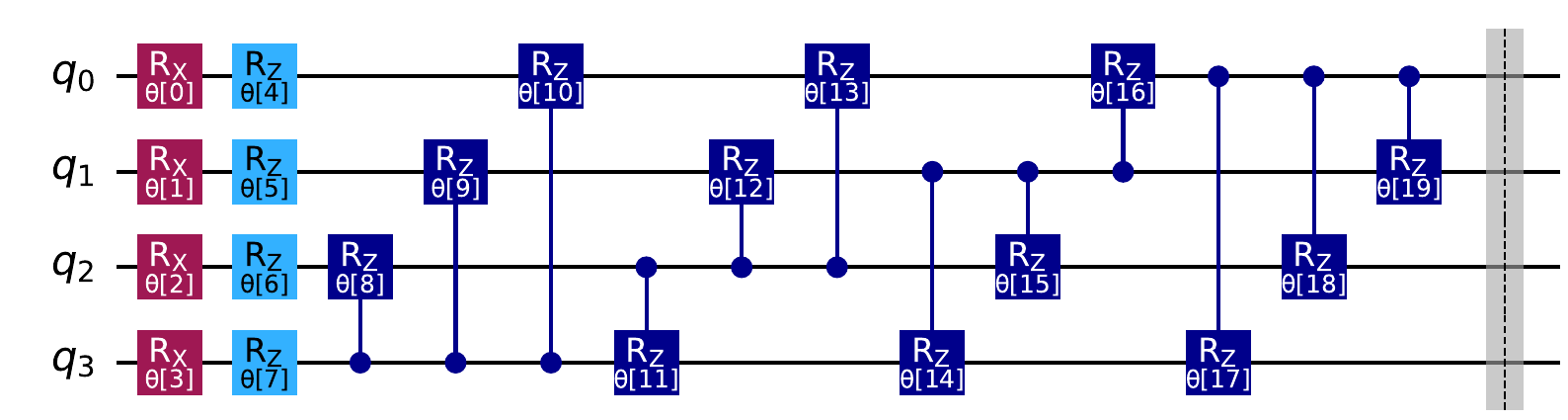}
    \caption{}
    \label{fig:cross_mesh_2rots}
  \end{subfigure}
  
  \medskip

  \begin{subfigure}[t]{0.49\textwidth}
    \centering
    \includegraphics[width=\linewidth]{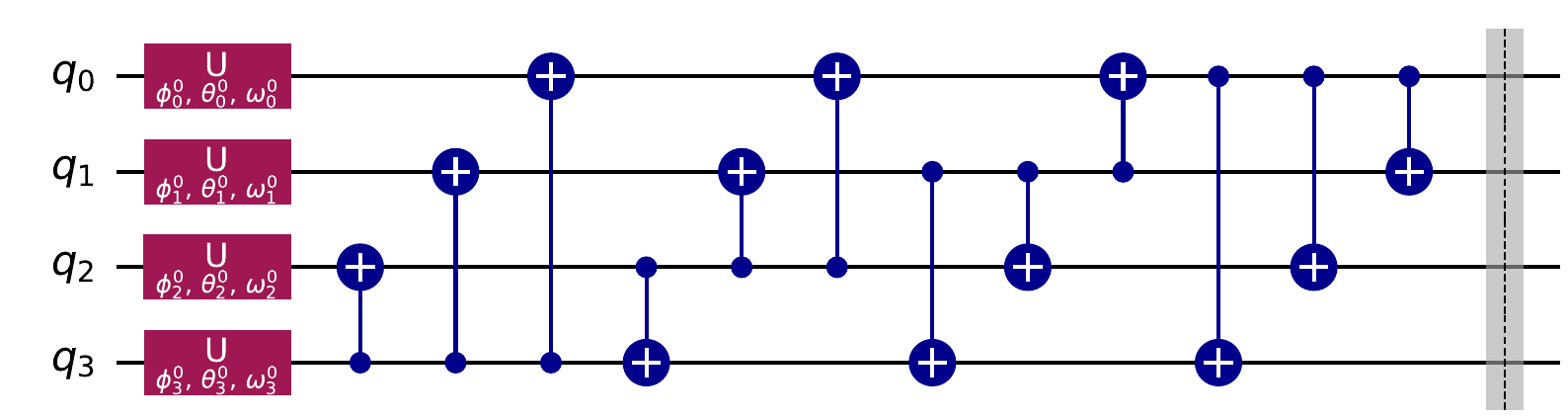}
    \caption{}
    \label{fig:cross_mesh_cx_rot}
  \end{subfigure}\hfill
  \begin{subfigure}[t]{0.49\textwidth}
    \centering
    \includegraphics[width=\linewidth]{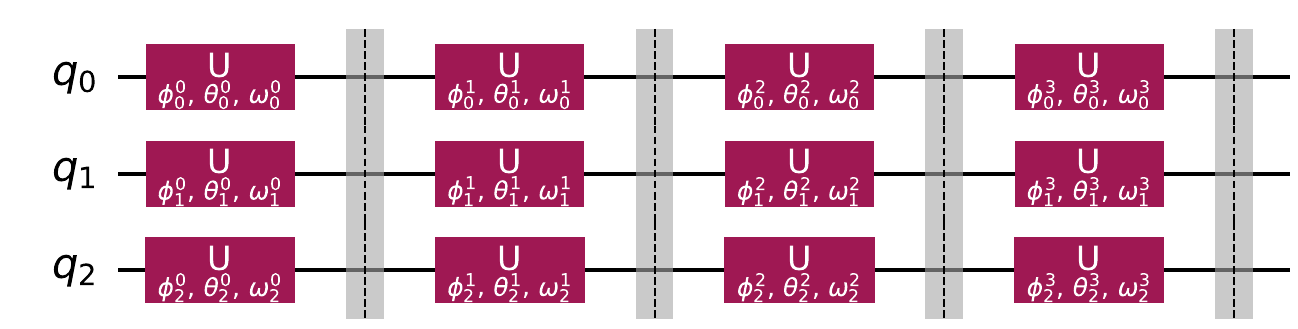}
    \caption{}
    \label{fig:no_entanglement}
  \end{subfigure}

  \caption{Schematics of all ans\"atze used in this study. (a) Basic Entangling Layers. (b) Strongly Entangling Layers. (c) Cross-Mesh. (d) Cross-Mesh-2-Rotations. (e) Cross-Mesh-CNOT. (f) No Entanglement Ansatz.}
  \label{fig:ans\"atze}
\end{figure}

The six ans\"atze picked for this ablation study are based on previous selections~\cite{sim2019expressibility, schuld2020circuit, perez2020data}. The ans\"atze are presented in Fig.\ref{fig:ans\"atze}. Each ansatz layer consists of two parts: parametrized single qubit rotation gates applied on each qubit and two-qubit gates, that can be either parametrized or not.

\noindent The ans\"atze are:
\begin{enumerate}
    \item Basic Entangling Layers (\ref{fig:basic_ent_layer}): in this ansatz, the single qubit rotations are \textit{Rot} gates: 
    \begin{equation}
        Rot(\alpha, \beta, \gamma)=RZ(\gamma)RY(\beta)RZ(\alpha)
        \label{eq:rot_gate}
    \end{equation}
    which are arbitrary rotations around the Bloch sphere, using 3 parameters per gate. The two-qubit gates are constructed by a cyclic chain of \textit{CNOT} gates connecting nearest-neighbor qubits. 
    
    \item Strongly Entangling Layers (\ref{fig:strongly_ent_layer}): this ansatz is similar to the \textit{Basic Entangling Layers} ansatz, but the gap between the target and control qubits in each layer is incremented by 1. This means that the first layer will be the same as the first layer of the \textit{Basic Entangling Layers}, but in the second one, the first gate will be \textit{CNOT(0, 2)}, then \textit{CNOT(1, 3)}, and so on. The motivation behind choosing the Basic and Strongly is that they are both widely used, especially the basic one, many times referred to as ``hardware efficient ansatz", usually when considering the architecture of superconducting qubits.
    
    \item Cross-Mesh (\ref{fig:cross_mesh}): in this ansatz, the single qubit rotations are $RX_{\alpha(i)}$ gates, which have a single parameter. The entangling part of each layer consists of $CRZ_{\beta(i,j)}$ gates, connecting each qubit to each of the other qubits:
    \begin{equation}
        \text{Entangling Part}=\prod^{N-1}_{i=0}\prod^{N-1}_{j\neq i}CRZ(i,j)_{\alpha(i,j)}
        \label{eq:crz_cross_mesh}
    \end{equation}
    There are three main reasons behind testing the \textit{Cross-Mesh} ansatz here: the first is to check if a ``fully-connected'' circuit achieves better results than less-connected circuits, analogous to the difference between fully-connected layers and convolutional layers in a classical neural network. The second reason is checking if and how the full connectivity affects BH, and the third is that there are hardware implementations able to apply all of those gates in parallel as Nemirovsky et al.~\cite{nemirovsky2025efficient}. This ansatz had $196$ quantum learnable parameters.
    
    \item Cross-Mesh-2-Rotations (\ref{fig:cross_mesh_2rots}): this ansatz is almost identical to the \textit{Cross-Mesh}, but instead of having only $RX_{\alpha(i)}$ per qubit in the single qubit rotations part, it has $RX_{\alpha(i)}$$RZ_{\beta(i)}$. This ansatz had $224$ quantum learnable parameters.
    
    \item Cross-Mesh-CNOT (\ref{fig:cross_mesh_cx_rot}): this ansatz is similar to the previous \textit{Cross-Meshes}, but it has \textit{Rot} gates in the single qubit rotations part, and instead of parametrized $CRZ_{\beta(i,j)}$ gates, it consists of \textit{CNOT} gates connected in the same manner.
    
    \item No Entanglement Ansatz (\ref{fig:no_entanglement}): this ansatz consists only of a \textit{Rot} gate per qubit in each layer, without any two-qubit gates. The motivation for using such an ansatz is to investigate the effect of entanglement on our QPINN model, as the role of entanglement in QML is not fully understood~\cite{bowles2024better}.    
\end{enumerate}
The ans\"atze besides \textit{Cross-Mesh}, \textit{Cross-Mesh-2-Rotations}, and \textit{Cross-Mesh-CNOT} had $84$ quantum learnable parameters besides the classical learnable parameters (summary of learnable parameters can be found in Table \ref{tab:parameters_amount}).

The measurement method used here was to measure the expectation value of a Pauli-Z observable measurement for each qubit, and each qubit was considered a ``neuron'' for the subsequent layer. The circuit is differentiable with respect to both the circuit parameters and the input angles, so automatic differentiation is enabled to update classical and quantum parameters and to get the derivatives for the PDE. The runs throughout this work were all made by a classical simulator, so derivatives were calculated using backpropagation. When running on real quantum hardware, this would not be possible, and other methods such as the \textit{parameter-shift rule}~\cite{mitarai2018quantum} should be used.

\begin{table}[H]
\centering
\caption{Summary of the number of learnable parameters in each network architecture.}
\begin{tabular}{|c|c|c|c|}
\hline
\rowcolor[HTML]{C0C0C0} 
\begin{tabular}[c]{@{}c@{}}Ansatz / network's \\ hidden layers\end{tabular} & \begin{tabular}[c]{@{}c@{}}\# Classical\\ parameters\end{tabular} & \begin{tabular}[c]{@{}c@{}}\# Quantum\\ parameters\end{tabular} & \begin{tabular}[c]{@{}c@{}}\# Total\\ parameters\end{tabular} \\ \hline
Classical - regular                                                        & 82820                                                             & 0                                                               & 82820                                                         \\ \hline
\begin{tabular}[c]{@{}c@{}}Classical - reduced layer\end{tabular}       & 66308                                                             & 0                                                               & 66308                                                         \\ \hline
\begin{tabular}[c]{@{}c@{}}Classical - extra layer\end{tabular}          & 99332                                                             & 0                                                               & 99332                                                         \\ \hline
Cross-Mesh                                                                 & 66848                                                             & 196                                                             & 67044                                                         \\ \hline
Cross-Mesh-2-Rotations                                                     & 66848                                                             & 224                                                             & 67072                                                         \\ \hline
Cross-Mesh-CNOT                                                            & 66848                                                             & 84                                                              & 66932                                                         \\ \hline
No Entanglement Ansatz                                                     & 66848                                                             & 84                                                              & 66932                                                         \\ \hline
Basic Entangling Layers                                                    & 66848                                                             & 84                                                              & 66932                                                         \\ \hline
Strongly Entangling Layers                                                 & 66848                                                             & 84                                                              & 66932                                                         \\ \hline
\end{tabular}
\label{tab:parameters_amount}
\end{table}

\section{Quantum simulator and performance optimization} \label{sec:simulator}
All experiments in this work were run on a classical GPU-based ideal quantum simulator. Moreover, all runs were noiseless and calculated analytically, meaning that the measurements' values were determined according to the circuit's final state-vector (no ``shots'' used). However, it should be noted that this architecture doesn't require additional measurements, but rather multiple shots per circuit and the use of a hardware-compatible differentiation method when running on quantum hardware.
Gradients were computed via backpropagation on the simulator. We developed \textit{TorQ - Tensor Operations for Research in Quantum systems}, an in-house quantum simulation library using PyTorch, which we found to be over 50$\times$ faster than using PennyLane's \texttt{default.qubit} simulator with Torch integration (see Table~\ref{tab:run_time_comparison}). Qiskit runs were slower than PennyLane's. The custom simulator 
also had significantly better memory usage: the largest grid of collocation points we could run without memory overflow was $87^3$ on our simulator, versus $43^3$ on PennyLane's \texttt{default.qubit}. All simulations were run on NVIDIA L40s and A100 GPUs. We did not use mini-batching or multi-GPU training, since, according to Hao et al.~\cite{zhongkai2024pinnacle}, PINN batch training yields worse results, and using multiple GPUs 
behaves similarly to mini-batching.

\begin{table}[h!]
\caption{Comparison between different PennyLane simulators and TorQ}
\begin{tabular}{|c|c|c|c|c|c|}
\hline
\rowcolor[HTML]{C0C0C0} 
Package name & Hardware                                                                     & Simulator        & Diff. method    & Grid size             & \begin{tabular}[c]{@{}l@{}}Average seconds \\per  epoch\end{tabular} \\ \hline
PennyLane    & \begin{tabular}[c]{@{}l@{}}CPU for the PQC, \\ GPU for the rest\end{tabular} & default.qubit    & backpropagation & $40^3$ & 7.729721                                                     \\ \hline
PennyLane    & \begin{tabular}[c]{@{}l@{}}CPU for the PQC, \\ GPU for the rest\end{tabular} & default.qubit    & backpropagation & $43^3$ & 9.580446                                                     \\ \hline
PennyLane    & GPU                                                                          & lightning.gpu    & adjoint         & $5^3$  & 113.327774                                                   \\ \hline
PennyLane    & GPU                                                                          & lightning.kokkos & adjoint         & $5^3$  & 99.751072                                                    \\ \hline
TorQ  & GPU                                                                          & -                & backpropagation & $40^3$ & 0.145136                                                     \\ \hline
TorQ  & GPU                                                                          & -                & backpropagation & $87^3$ & 0.849967                                                     \\ \hline
\end{tabular}
\label{tab:run_time_comparison}
\end{table}

\FloatBarrier
\section{Experiments and results} \label{sec:results}

In this section, the ablation study results are presented for both test cases, and the metric each experiment is evaluated according to is $L_{2}$ relative error norm (later referred to as $L_2$ error for convenience) of the electric field $E_z$ compared to a high-fidelity $4^{th}$-order Padé scheme, considered as a high-fidelity reference solution. The error is computed over a dense grid of $512\!\times\!512$ spatial points and 1500 time steps:

\begin{equation}
L_{2}\;error = \sqrt{\frac{\sum_{n=1}^{N} \left(E_z^{(n)} - E_{z,\text{Padé}}^{(n)}\right)^2}{\sum_{n=1}^{N} \left((E_{z,\text{Padé}}^{(n)}\right)^2}} \,,
\label{eq:l2_error}
\end{equation}
where $n$ indexes the grid points in space and time.

\begin{figure}[!htbp]
  \centering
  \begin{subfigure}[t]{0.32\textwidth}
    \centering
    \includegraphics[width=\linewidth]{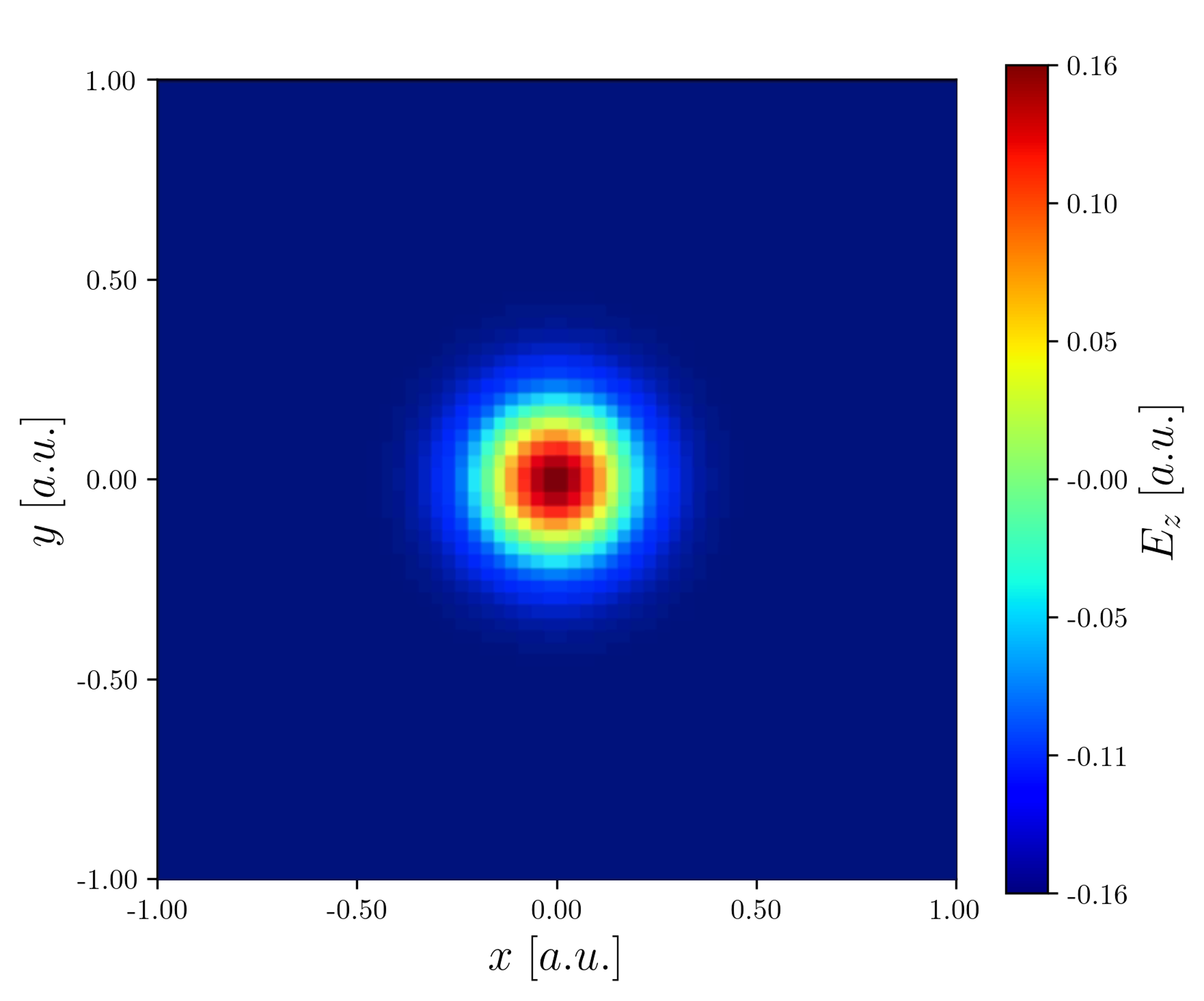}
    \caption{}
    \label{fig:init_cond}
  \end{subfigure}
  \begin{subfigure}[t]{0.32\textwidth}
    \centering
    \includegraphics[width=\linewidth]{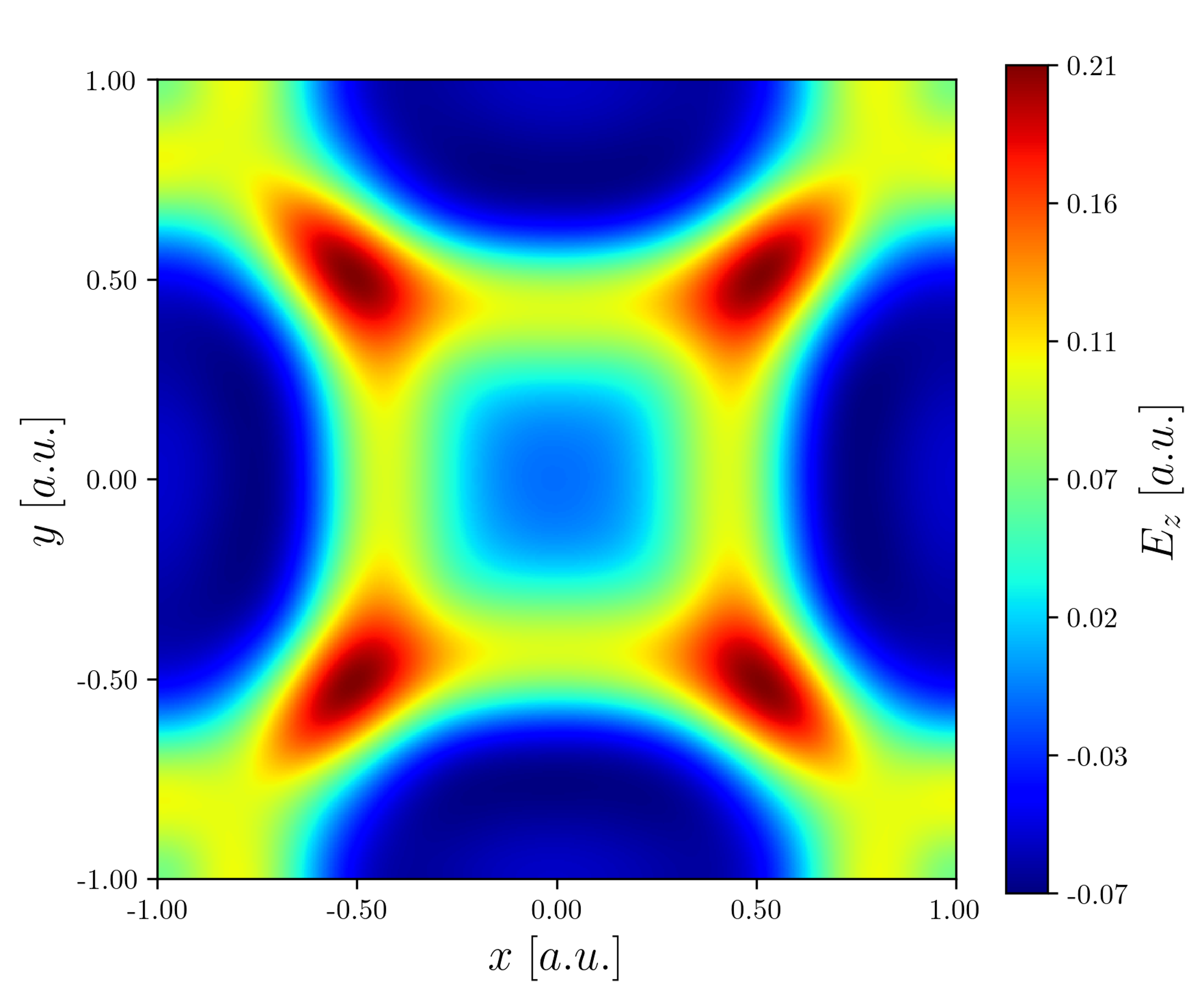}
    \caption{}
    \label{fig:vacuum_res}
  \end{subfigure}\hfill
  \begin{subfigure}[t]{0.32\textwidth}
    \centering
    \includegraphics[width=\linewidth]{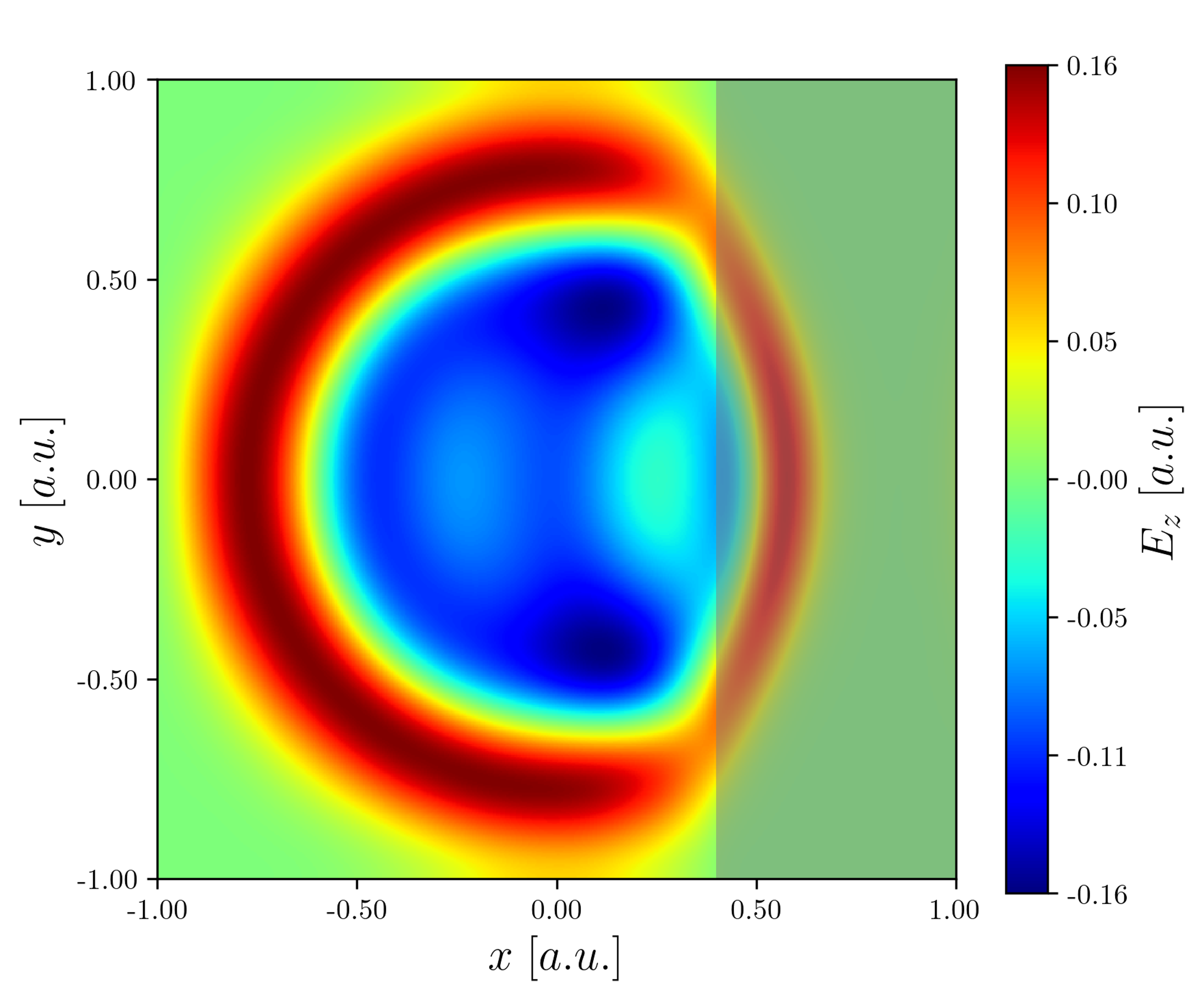}
    \caption{}
    \label{fig:dielectric_res}
  \end{subfigure}
  
  \caption{(a) Initial conditions (i.e. $t=0$) for both test cases, with the training size grid ($64^2$). Contours of the electric field $E_z$ at the final time of the wave propagation are shown for (b) vacuum at $t=1.5$ and (c) dielectric medium at $t=0.7$, as obtained from the QPINN simulations. The corresponding classical PINN and Padé scheme solutions appear visually identical; please refer to \cite{shaviner2025pinns} for the classical results. The region occupied by the dielectric medium is shaded in (c).}
  \label{fig:results}
\end{figure}

Each ablation study (vacuum and dielectric) consists of multiple QPINN runs varying: (1) the ansatz, (2) the input scaling, and (3) whether the energy conservation loss term $\mathcal{L}_{\mathrm{energy}}$ (Eq.~\ref{eq:energy_loss}) is included. For comparison, classical PINN runs with different network depths (regular, extra layer, and reduced layer) were also performed, both with and without the energy conservation loss term. Each configuration was run 5 times to compute the mean and standard deviation. Results of the final time slice for both cases are shown in Fig.~\ref{fig:results}.

\begin{figure}[!htbp]
  \centering
  \begin{subfigure}[t]{1.0\textwidth}
    \centering
    \includegraphics[width=\linewidth]{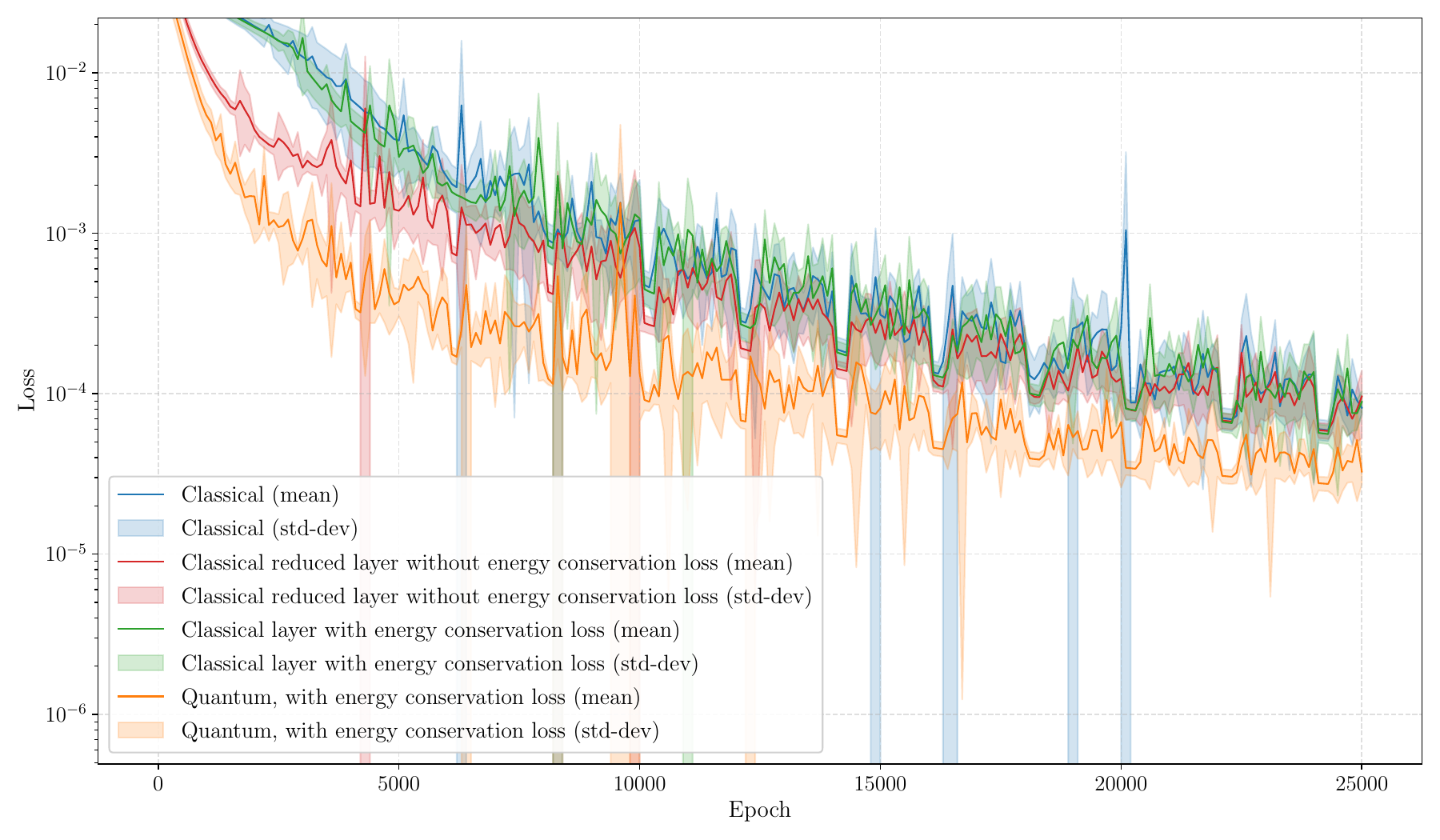}
    \caption{}
    \label{fig:vacuum_avg_loss}
  \end{subfigure}\hfill
  
  \medskip

  \begin{subfigure}[t]{1.0\textwidth}
    \centering
    \includegraphics[width=\linewidth]{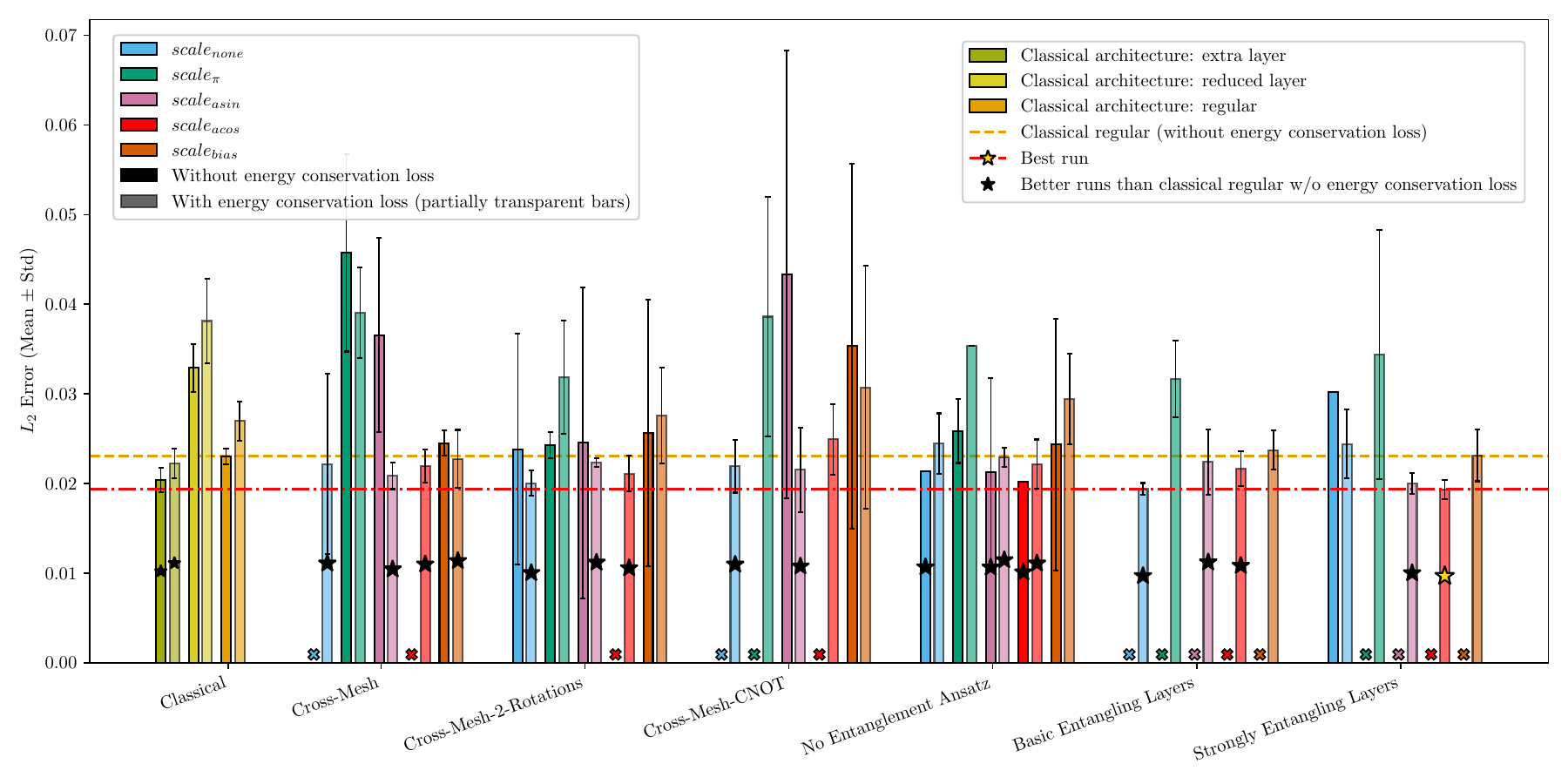}
    \caption{}
    \label{fig:vacuum_bar_chart_all}
  \end{subfigure}\hfill
  
  \caption{(a) Mean loss (log scale) for the runs of the best combination for the vacuum case: Strongly Entangling Layers ansatz with $scale_{acos}$ and energy loss included. The semi-transparent band indicates the standard deviation. (b) $L_2$ errors for all combinations in the vacuum case. The dashed yellow line shows the classical result for the regular-depth network; black stars mark runs that outperformed this baseline. The dash-dot line indicates the best run's error (highlighted by a gold star). An ``X'' denotes configurations where none of the 5 runs converged.}
  \label{fig:avg_loss_and_bar_chart_all_vacuum}
\end{figure}

\begin{figure}[!htbp]
  \centering
  \begin{subfigure}[t]{0.49\textwidth}
    \centering
    \includegraphics[width=\linewidth]{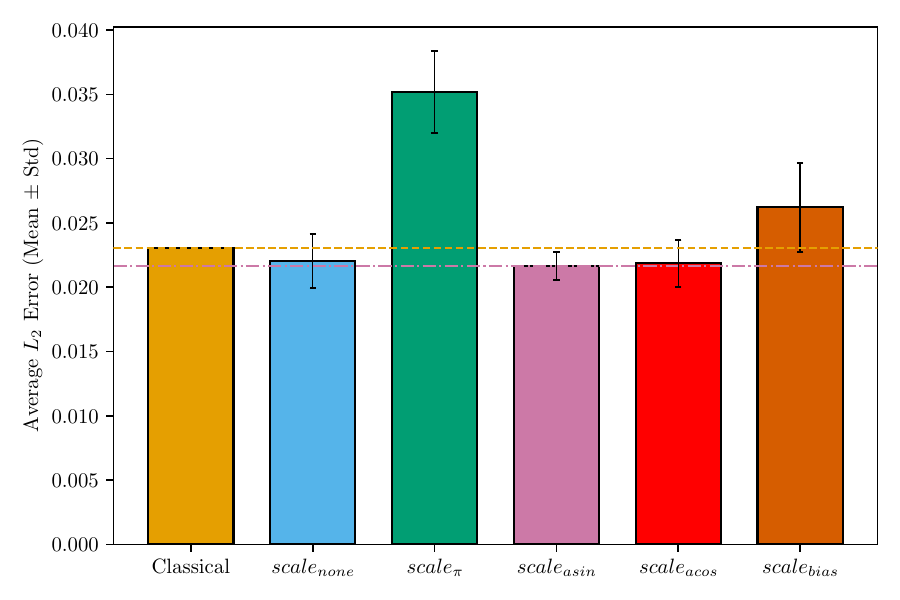}
    \caption{}
    \label{fig:bar_chart_vacuum_avg_by_scale}
  \end{subfigure}\hfill
  \begin{subfigure}[t]{0.49\textwidth}
    \centering
    \includegraphics[width=\linewidth]{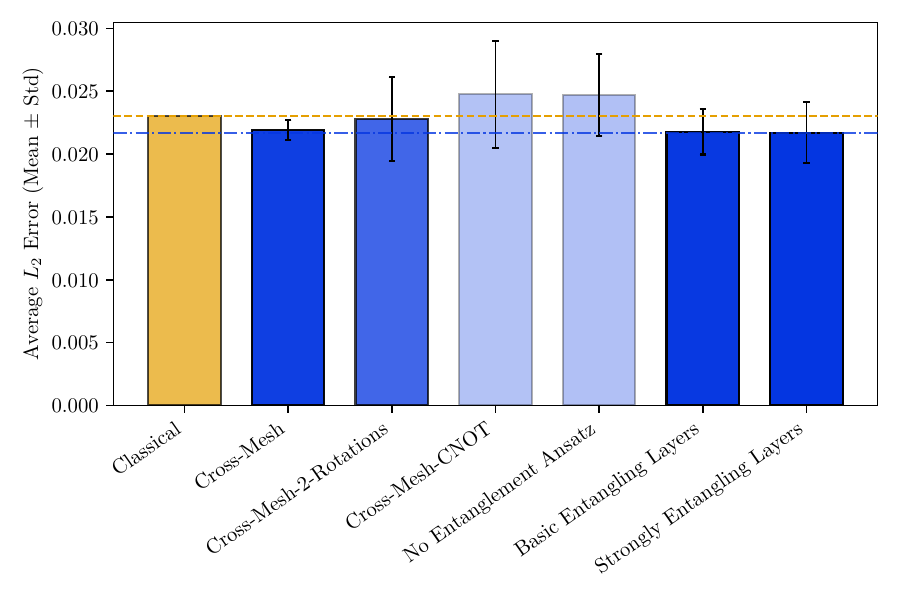}
    \caption{}
    \label{fig:bar_chart_vacuum_avg_by_ansatz}
  \end{subfigure}

  \caption{Average $L_2$ errors for the vacuum case grouped by (a) input scale and (b) ansatz, the $scale_{\pi}$ runs were omitted from the average as they performed much worse than the rest of the scales in all ans\"atze. The yellow dashed line shows the classical average error, and the dash-dot line (same color as the lowest bar) shows the lowest average error achieved. In (b), the color for each bar is set according to the $L_2$ value.}
  \label{fig:bar_charts_avg_vacuum}
\end{figure}
\subsection{Case 1: Pulse in vacuum} \label{subsec:vacuum}
According to Figs.~\ref{fig:avg_loss_and_bar_chart_all_vacuum} and \ref{fig:bar_charts_avg_vacuum}, four key observations can be made for the vacuum case:
\begin{enumerate}
    \item \textbf{Scales and ans\"atze:} The performance using different scaling methods can be divided into two groups: $scale_{\pi}$ and $scale_{bias}$ performed poorly compared to the other scaling methods; the other scalings achieved similar results. For comparison, the $scale_{\pi}$ achieved higher $L_2$ errors than the $scale_{asin}$ by more than $62\%$. From the comparison of different ans\"atze after omitting the $scale_{\pi}$ because it wouldn't be used for such case according to its performance across all ans\"atze, the Cross-Mesh-CNOT and No Entanglement Ansatz performed worse than the other ans\"atze, the Cross-Mesh-2-Rotations achieved similar accuracy to the classical and the rest were much better as the Cross-Mesh-CNOT had higher $L_2$ error than the Strongly Entangling Layers by approximately $14\%$. However, the different ans\"atze didn't vary as much as the different scales, meaning that choosing a scale that fits the most is more impactful than choosing the most fitting ansatz.
    \item \textbf{Parameter reduction:} $42.2\%$ of converged QPINN runs achieved lower $L_2$ errors than the classical regular-depth network, despite having approximately $19\%$ fewer total trainable parameters (Table~\ref{tab:parameters_amount}). To isolate the effect of parameter count, we also ran classical PINNs with a reduced layer (similar in terms of parameters to the quantum models) and an extra layer (approximately $48\%$ more parameters than the quantum models). The QPINN runs still outperformed both, indicating the quantum model's efficiency is not solely due to fewer parameters.
    \item \textbf{Faster convergence:} The QPINN runs converged faster to the PDE solution than the PINN runs ( Fig.~\ref{fig:vacuum_avg_loss}). This cannot be attributed simply to having fewer parameters, as the QPINN run also surpassed the PINN reduced-network case (that achieved a much worse final $L_2$ error). Note that the QPINN run's loss included the additional energy term, which the PINN run did not have. 
    \item \textbf{Impact of energy loss term:} Including the energy conservation term $\mathcal{L}_{\mathrm{energy}}$ (Eq.~\ref{eq:energy_loss}) is critical in the vacuum case. Otherwise, the QPINN runs would suffer from the BH phenomenon. With the energy term included, the QPINN runs not only avoid this collapse, but also significantly outperform the PINN runs in $L_2$ error. Interestingly, the PINN runs show the opposite effect: adding the energy loss term degrades their performance. In particular, this is the case for the `No Entanglement ansatz', which behaves like a classical network in many aspects. For the PINN runs, the energy term remains a redundant bilinear penalty that distorts gradient scales and slows learning; we therefore observe consistent degradation. In contrast, the QPINN runs in the vacuum case are susceptible to the BH collapse: the PQC head has an expressive entangling ansatz that admits a degenerate low-loss trajectory that fades amplitudes after the initial slice. The energy penalty acts as a structural regularizer that removes this discrete failure mode and is therefore beneficial specifically for QPINNs.
\end{enumerate}

\subsection{Case 2: Pulse-dielectric medium interaction} \label{subsec:dielectric}

\begin{figure}[!htbp]
  \centering
  \begin{subfigure}[t]{1.0\textwidth}
    \centering
    \includegraphics[width=\linewidth]{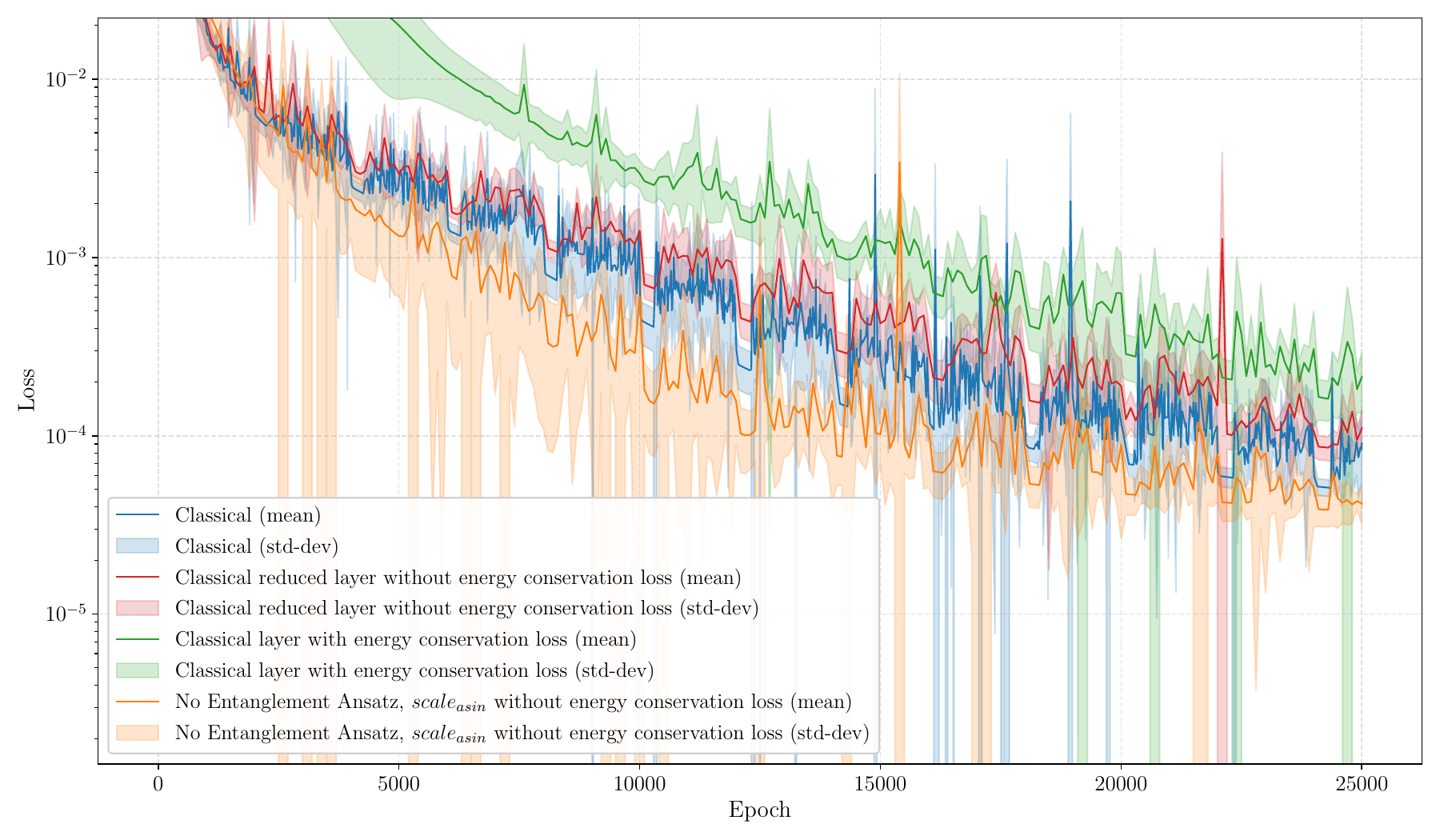}
    \caption{}
    \label{fig:dielectric_avg_loss}
  \end{subfigure}\hfill
  
  \medskip

  \begin{subfigure}[t]{1.0\textwidth}
    \centering
    \includegraphics[width=\linewidth]{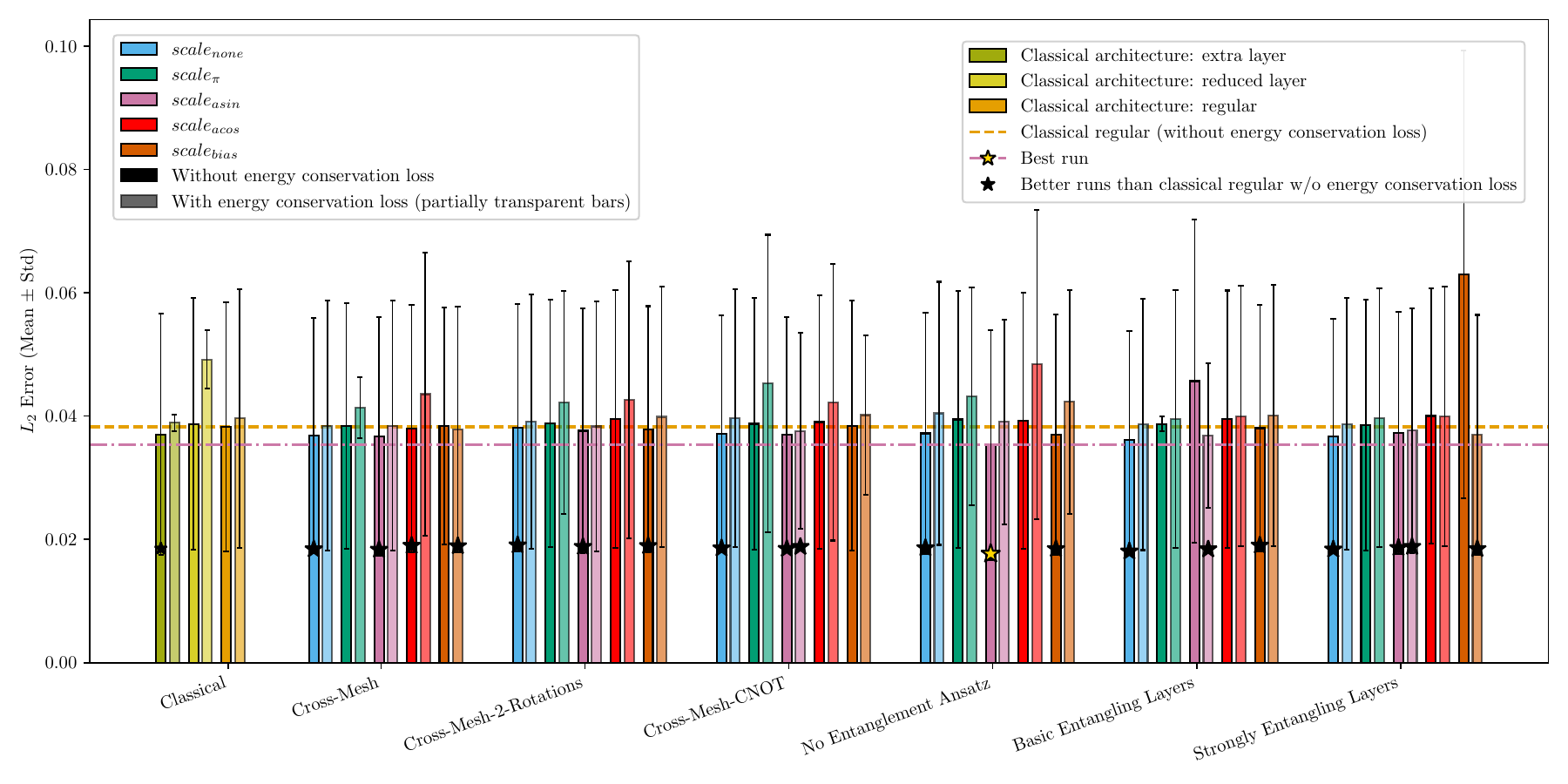}
    \caption{}
    \label{fig:dielectric_bar_chart_all}
  \end{subfigure}\hfill
  
  \caption{(a) Mean loss (log scale) for the runs of the best combination for the dielectric case: No Entanglement ansatz with $scale_{asin}$, without energy loss. The semi-transparent band indicates the standard deviation. (b) $L_2$ errors for all combinations in the dielectric case. The dashed yellow line shows the classical result for the regular-depth network; black stars mark runs that outperformed this baseline. The dash-dot line indicates the best run's error (highlighted by a gold star).}
  \label{fig:avg_loss_and_bar_chart_all_dielectric}
\end{figure}

\begin{figure}[!htbp]
  \centering
  \begin{subfigure}[t]{0.49\textwidth}
    \centering
    \includegraphics[width=\linewidth]{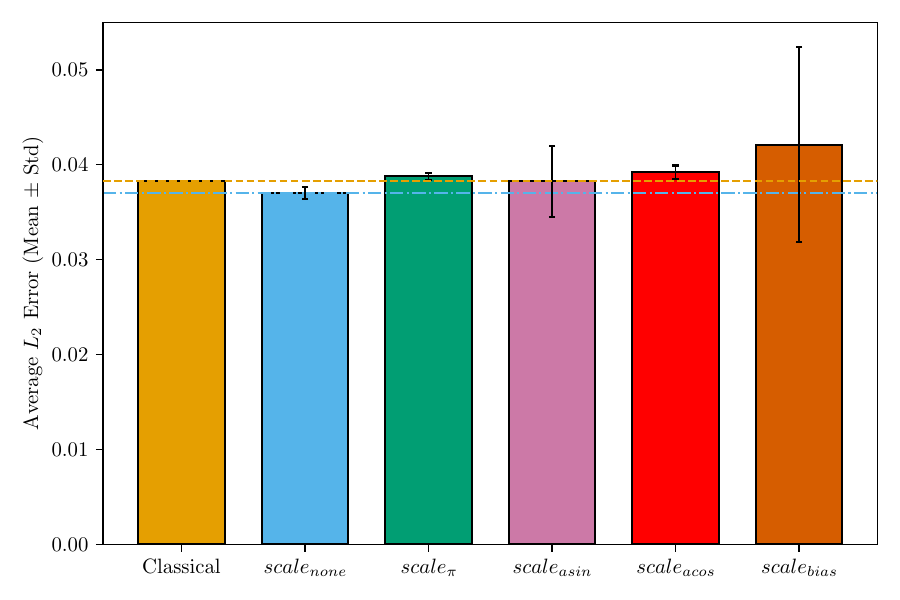}
    \caption{}
    \label{fig:bar_chart_dielectric_avg_by_scale}
  \end{subfigure}\hfill
  \begin{subfigure}[t]{0.49\textwidth}
    \centering
    \includegraphics[width=\linewidth]{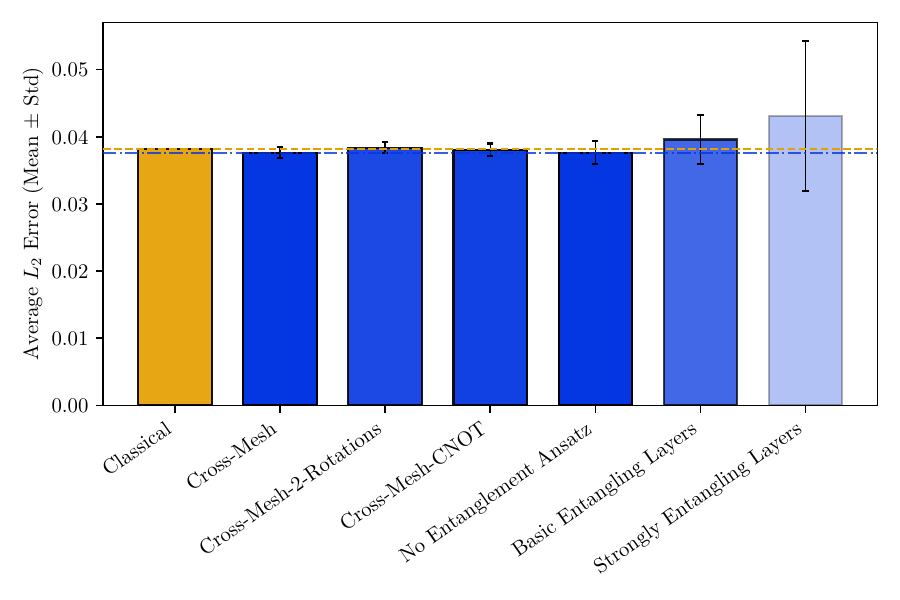}
    \caption{}
    \label{fig:bar_chart_dielectric_avg_by_ansatz}
  \end{subfigure}
  
  \caption{Average $L_2$ errors for the dielectric case grouped by (a) input scale and (b) ansatz. The yellow dashed line shows the classical average error, and the dash-dot line (same color as the lowest bar) shows the lowest average error achieved. In (b), the color for each bar is set according to the $L_2$ value.}
  \label{fig:bar_charts_avg_dielectric}
\end{figure}

Based on Figs.~\ref{fig:avg_loss_and_bar_chart_all_dielectric} and \ref{fig:bar_charts_avg_dielectric}, we make four observations for the dielectric case: 
\begin{enumerate}  
    \item \textbf{Scales and ans\"atze:} Unlike the vacuum case, in the dielectric case, the variance between different scales is much smaller, as the worst scale - $scale_{bias}$ has a higher $L_2$ error compared to the best scaling method - $scale_{none}$, by approximately $13\%$. The gaps between ans\"atze are similar to the vacuum, where the Strongly Entangling Layers have approximately $14\%$ higher $L_2$ error than the No Entanglement Ansatz and the Cross-Mesh (for which the performance is almost the same on average).
    \item \textbf{Parameter reduction:} Similar to the vacuum case, there is a $19\%$ reduction in the total number of parameters achieved compared to the classical regular network (Table~\ref{tab:parameters_amount}). Nonetheless, the QPINN still performs competitively, and in many cases even outperforms the classical baseline.
    \item \textbf{Stability:} The dielectric runs are significantly more stable than the vacuum runs in terms of convergence and variance. Nearly all dielectric runs converge (there are no severe BH issues), and the spread in final errors is smaller. 
    \item \textbf{Effect of energy loss term:} Including the energy loss term $\mathcal{L}_{\mathrm{energy}}$ has a negative impact on the dielectric runs. Runs with the energy term generally yield higher errors than those without, for both classical and quantum models (with the possible exception of the Basic and Strongly Entangling ans\"atze, though those improvements could be statistical anomalies). Since BH does not manifest in the dielectric case, as almost all runs converge regardless, the added energy term 
    remains a redundant bilinear penalty that distorts gradient scales and slows learning by creating a stiff constraint without added information.
\end{enumerate}

\section{``Black hole'' loss landscape} \label{sec:black_hole}

\begin{figure}[!htbp]
  \centering
  \begin{subfigure}[t]{0.49\textwidth}
    \centering
    \includegraphics[width=\linewidth]{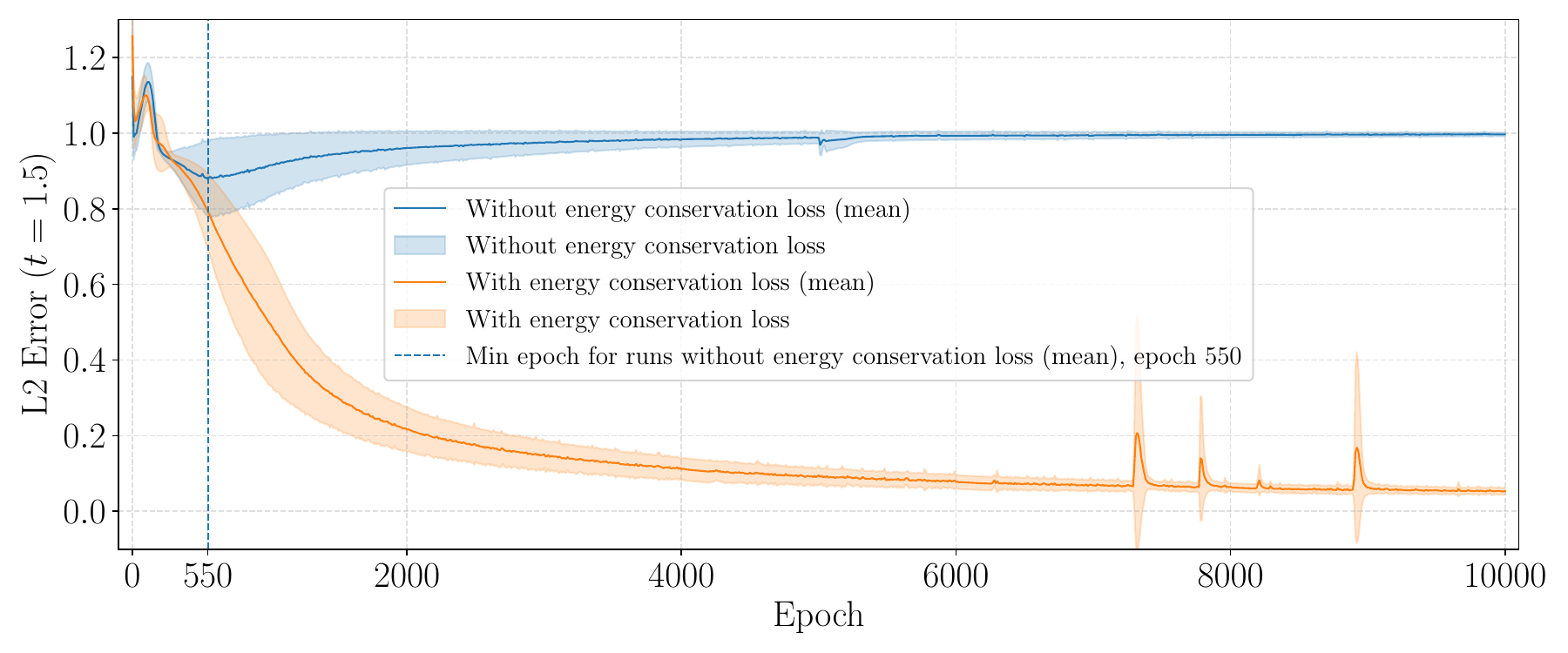}
    \caption{}
    \label{fig:bh_L2}
  \end{subfigure}\hfill
  \begin{subfigure}[t]{0.49\textwidth}
    \centering
    \includegraphics[width=\linewidth]{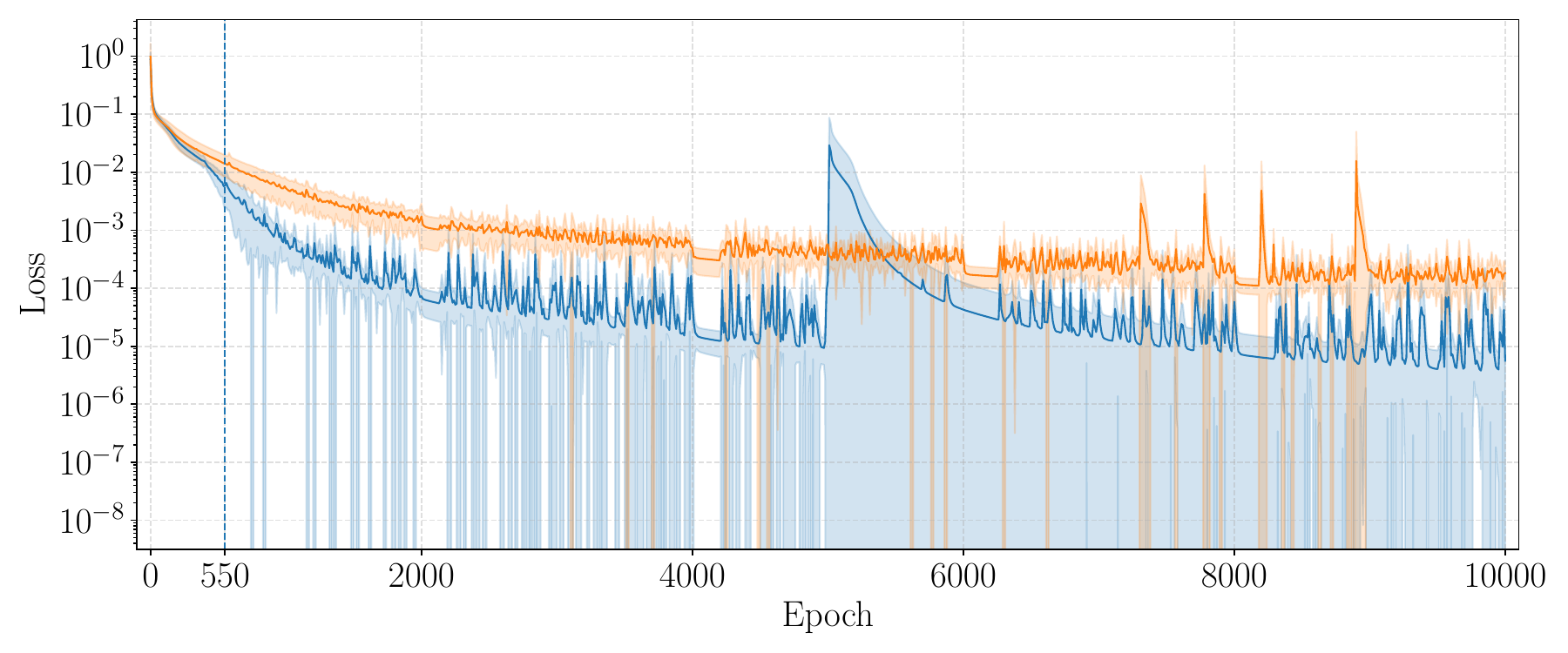}
    \caption{}
    \label{fig:bh_average_loss}
  \end{subfigure}

  \medskip

  \begin{subfigure}[t]{0.33\textwidth}
    \centering
    \includegraphics[width=\linewidth]{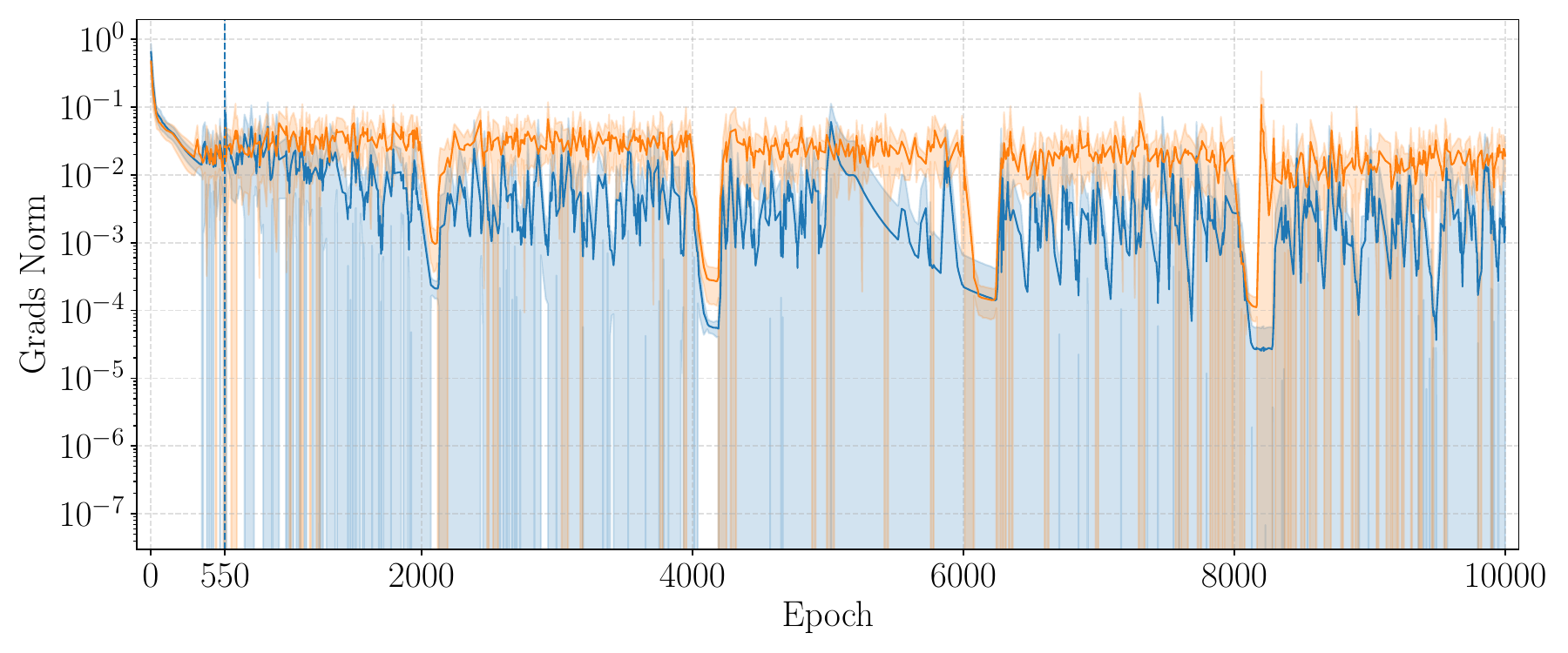}  
    \caption{}
    \label{fig:bh_grad_norm}
  \end{subfigure}\hfill
  \begin{subfigure}[t]{0.33\textwidth}
    \centering
    \includegraphics[width=\linewidth]{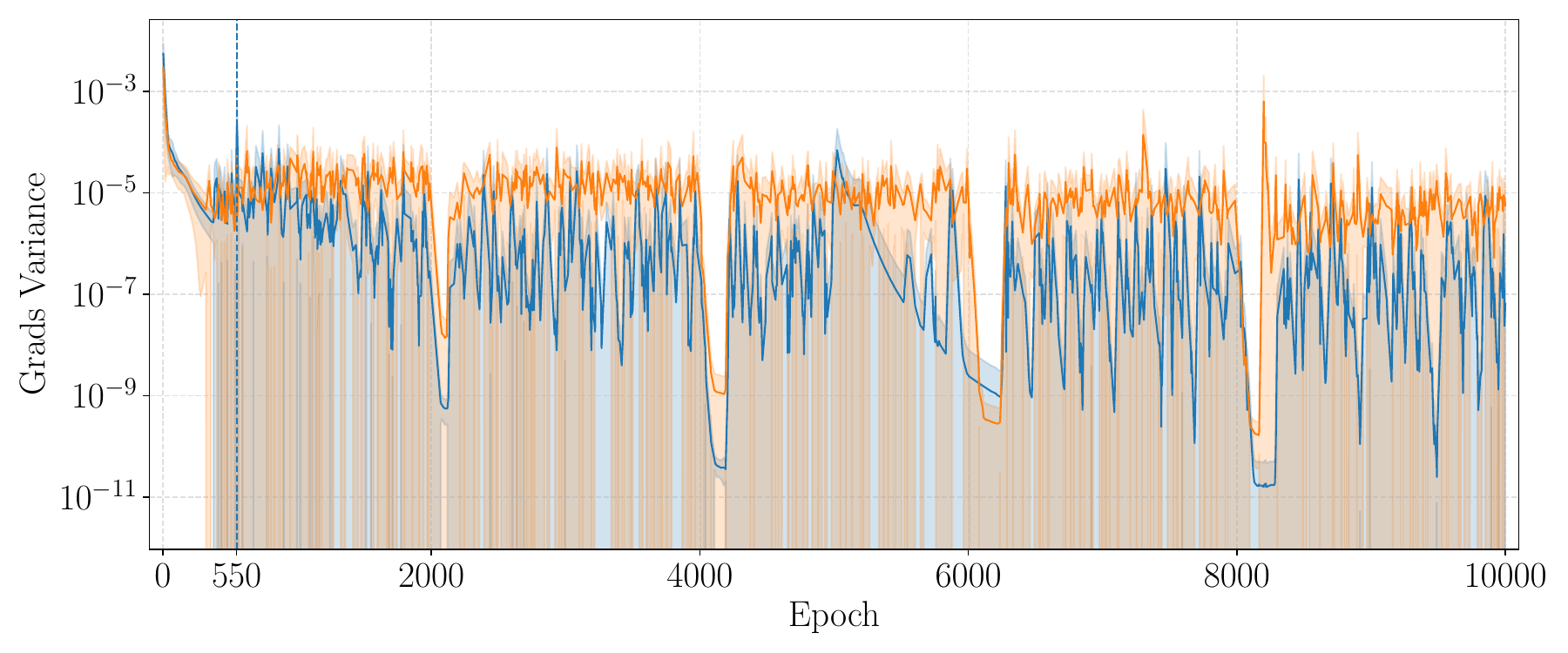}  
    \caption{}
    \label{fig:bh_grad_var}
  \end{subfigure}
  \begin{subfigure}[t]{0.33\textwidth}
    \centering
    \includegraphics[width=\linewidth]{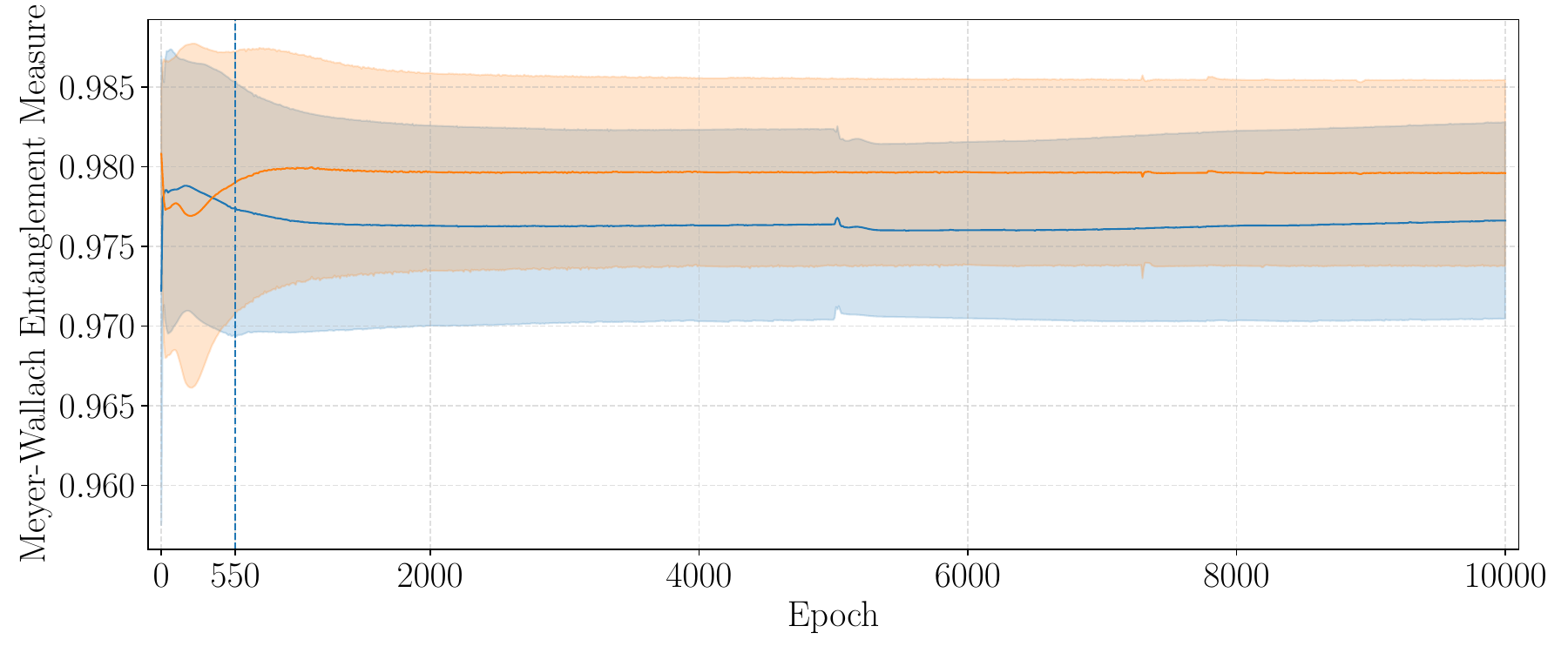}  
    \caption{}
    \label{fig:bh_lin}
  \end{subfigure}\hfill

  \caption{(a) $L_2$ error, (b) loss, (c) gradient norm, (d) gradient variance, and (e) global Meyer-Wallach entanglement entropy, which is correlated to the amount of entanglement during optimization. The graphs compare the results with and without the energy conservation loss term for the vacuum test case, averaged over 5 runs. The legend to graphs (a) - (e) is the same, and appears in (a). Graphs (b) - (d) are presented in log scale.}
  \label{fig:bh_figs}
\end{figure}

\begin{figure}[!htbp]
  \centering
  \begin{subfigure}[t]{0.33\textwidth}
    \centering
    \includegraphics[width=\linewidth]{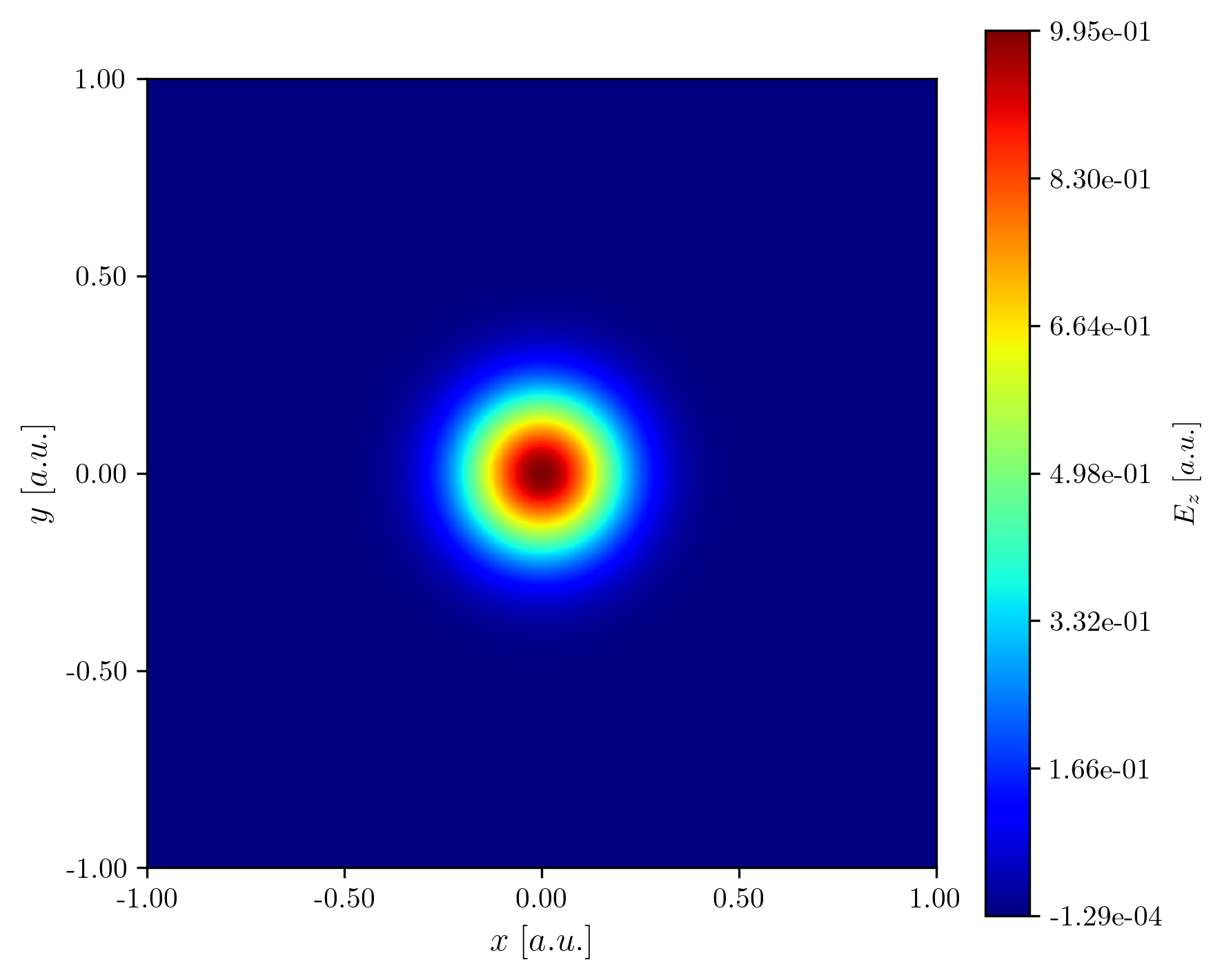}  
    \caption{}
    \label{fig:bh_ez_t0}
  \end{subfigure}\hfill
  \begin{subfigure}[t]{0.33\textwidth}
    \centering
    \includegraphics[width=\linewidth]{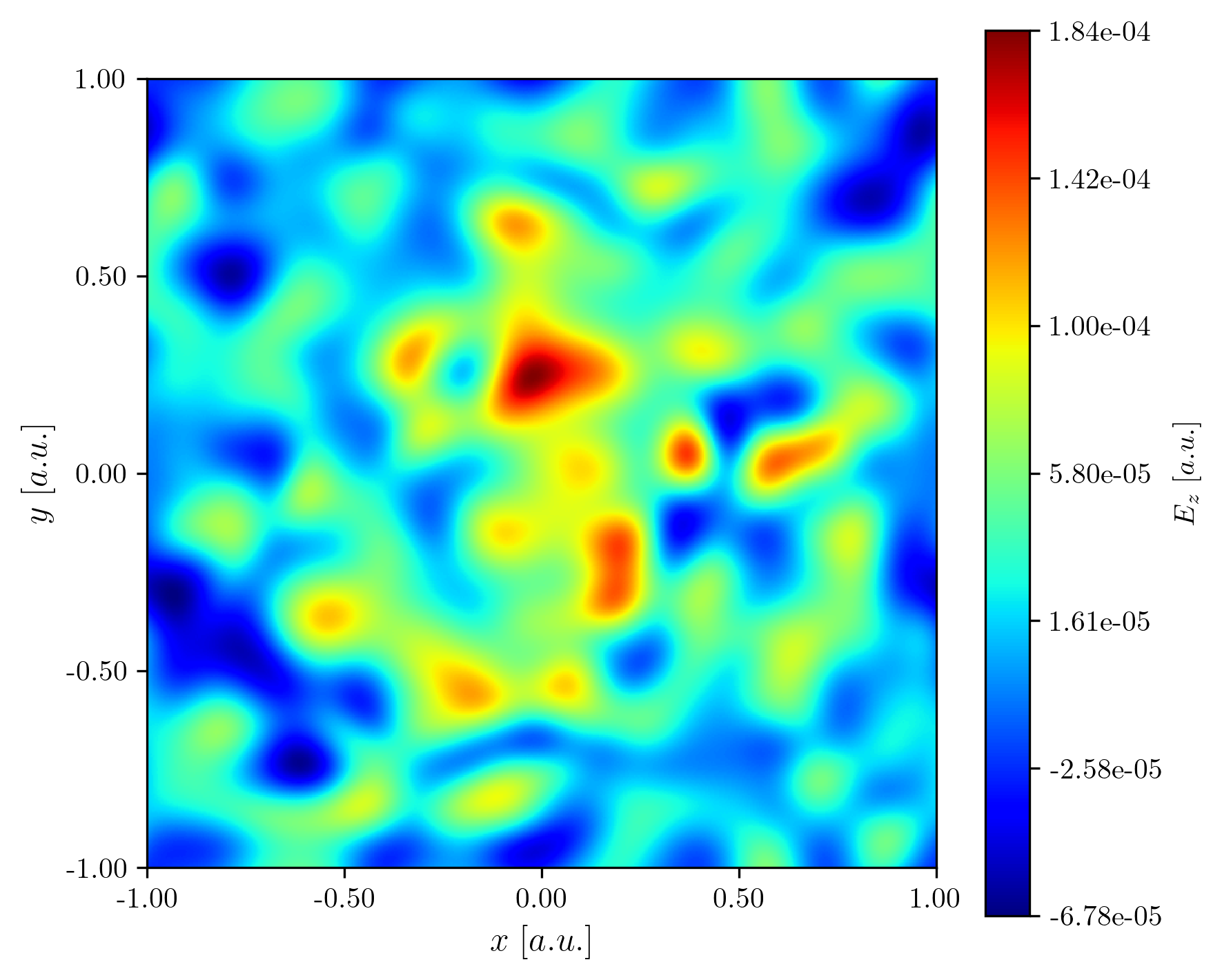} 
    \caption{}
    \label{fig:bh_ez_t03}
  \end{subfigure}\hfill
  \begin{subfigure}[t]{0.33\textwidth}
    \centering
    \includegraphics[width=\linewidth]{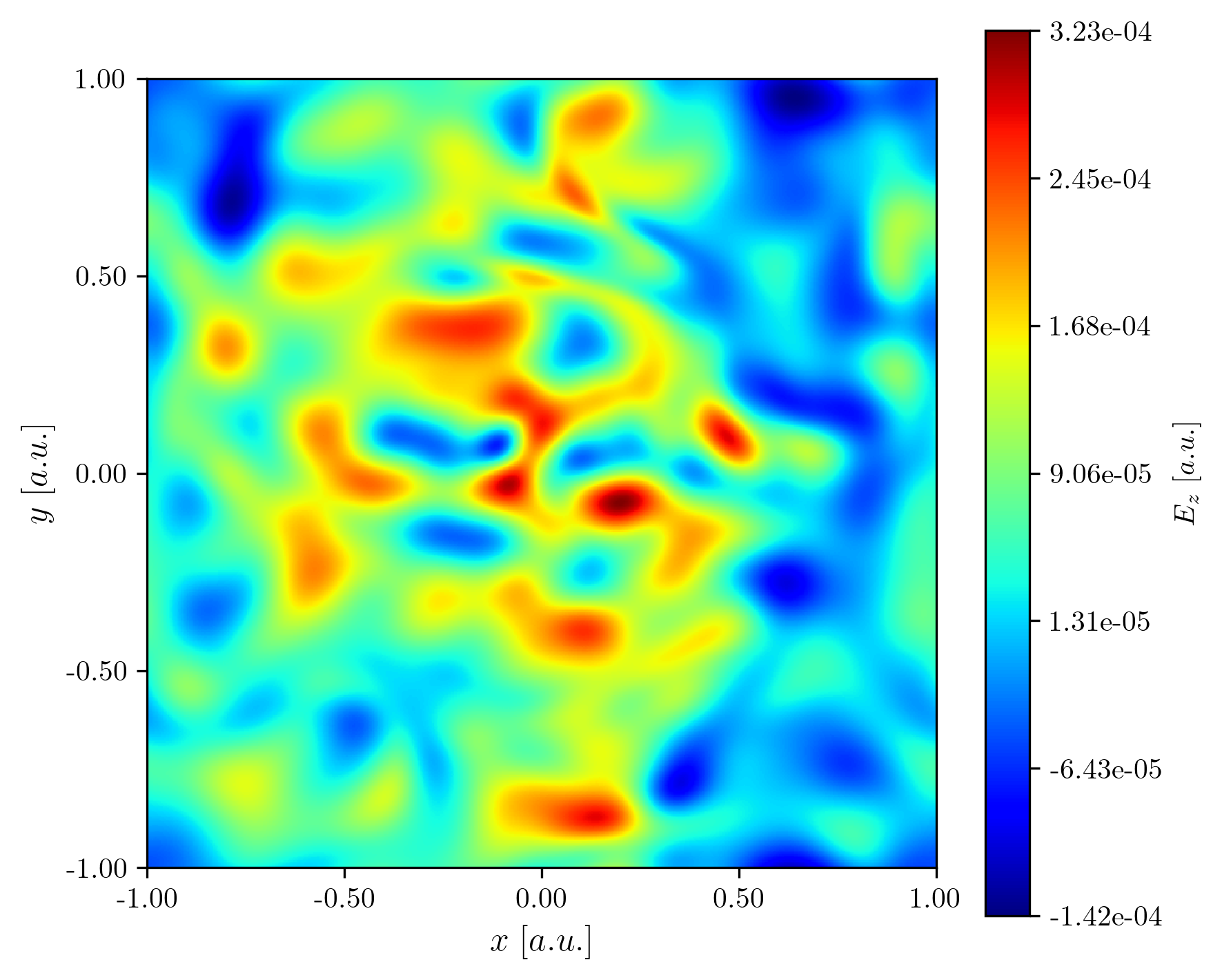} 
    \caption{}
    \label{fig:bh_ez_t15}
  \end{subfigure}
  \caption{$E_z$ at (a) $t=0$, (b) $t=0.3$, and (c) $t=1.5$ (final time slice) in a QPINN run without energy conservation loss. In (b) and (c), note the scale in the colorbar, which is approximately $0$.}
  \label{fig:bh_wo_en_loss_figs}
\end{figure}

In the vacuum case without the energy-conservation loss, we consistently observe a catastrophic training collapse to the trivial solution (TS). This TS is characterized by satisfying only the initial condition, and the field amplitudes become nearly zero for all $t>0$, matching the $t=0$ initial state and yielding a physically meaningless solution (Fig. \ref{fig:bh_wo_en_loss_figs}). When it occurs consistently, we term this phenomenon the “black hole” loss landscape (BH) phenomenon , as the initial conditions don't matter - the loss landscape will eventually pull the solution towards this collapse. This BH effect is distinct from known issues such as barren plateaus (BPs) \cite{mcclean2018barren, Larocca2025review}, weak barren plateaus (WBPs)~\cite{sack2022avoiding}, and laziness training regime \cite{liu2024laziness}. Unlike BPs and WBPs, which are often associated with high-entanglement states or global cost functions, the BH collapse shows no clear correlation with entanglement metrics, apart from that it doesn't occur with the no entanglement ansatz. As presented in Fig. \ref{fig:bh_lin}, the Meyer-Wallach entanglement entropy~\cite{meyer2002global} remains essentially unchanged throughout training and specifically during the collapse, and it is very similar between the runs with and without energy conservation loss. Furthermore, the BH emerges despite the use of local observables, a single Pauli-Z measurement per qubit as the objective. Moreover, whereas traditional BPs or laziness phenomena typically manifest as vanishing gradients from the start of optimization, the BH phenomenon only occurs after an initial period of successful learning. 

To make the notion of BH and the TS precise, an operational criterion based on the total electromagnetic energy is defined. The total energy in time operator $U_\theta(t)$, a sum over the spatial dimensions of the operator $u$, defined in Eq.~\ref{eq:Poynting_value_u}, defined as:
\begin{equation}
\begin{aligned}
    U_\theta(t)&=\int_{\Omega}\tfrac12\big(\varepsilon E_z^2(x,y,t;\theta)+H_x^2(x,y,t;\theta)+H_y^2(x,y,t;\theta)\big)\,dx\,dy \\
    &\approx \sum_{x\in N_x} \sum_{y\in N_y} \tfrac12\big(\varepsilon E_z^2(x,y,t;\theta)+H_x^2(x,y,t;\theta)+H_y^2(x,y,t;\theta)\big)
    ,
\end{aligned}
\end{equation}
and the normalized energy:
\begin{equation}
    \widetilde{U}_\theta(t):=U_\theta(t)/U_\theta(0)    
\end{equation}
We say that \emph{training exhibits collapse to the TS} if there exists a small $\delta>0$ such that $\min_{t\in[\delta,T]}\widetilde{U}_\theta(t)\ll1$ while $\mathcal{L}_{\mathrm{phys}}(\theta)$ remains small.
Equivalently, we report the scalar for cases with a constant energy amount (i.e., there is no dissipation and no additional energy added after initialization):
\begin{equation}
    I_{\mathrm{BH}}\ :=\ 1-\min_{t\in[\delta,T]}\widetilde{U}_\theta(t),    
\end{equation}
so that $I_{\mathrm{BH}}$ close to $1$ indicates collapse to the TS.
A \textit{BH phenomenon} is said to occur if over 95\% of runs with randomly chosen seeds collapse to the TS.

Experimentally, on the order of $10^2$ training epochs of steadily decreasing $L_2$ error, which means the model does learn correctly, the quantum model without $\mathcal{L}_{\mathrm{energy}}$ begins collapsing to a TS. This collapse is evident from the sharp drop in gradient norms and variances (Figs. \ref{fig:bh_grad_norm}, \ref{fig:bh_grad_var}) and the sudden drop in loss once the network falls into the TS (Fig. \ref{fig:bh_average_loss}). Importantly, adding the energy-conservation term completely mitigates this failure mode: with the $\mathcal{L}_{\mathrm{energy}}$ term included, none of the vacuum QPINN runs exhibited BH behavior, and the training instead continued to converge to the correct physical solution, while yielding better accuracy than the PINN run on average.
Including $\mathcal{L}_{\mathrm{energy}}$ restores the skew-adjoint conservative structure, i.e., a global energy invariance in the periodic, source-free case, making uniform ``fade-to-zero'' expensive and eliminating the BH attractor; the result is stable convergence and improved accuracy (Fig.~\ref{fig:avg_loss_and_bar_chart_all_vacuum}).

\subsection{Stabilizing the dielectric test case by the loss function} \label{subsec:stable_diel_loss}
Unlike the QPINN runs in the vacuum test case, in the dielectric case, the BH phenomenon does not occur when the $\mathcal{L}_{\text{phys, diel}}$ as defined in Eq.~\ref{eq:phys_loss_diel} is used, thanks to a non-homogeneous loss that makes the zero solution less favorable. When $res_{1,\;diel}$ and therefore $\mathcal{L}_{\text{phys, diel}}$ are defined more intuitively (as $res_{1,\;\text{intuitive}}$, and $\mathcal{L}_{\text{phys, intuitive}}$) by:
\begin{equation} \label{eq:residual_1_intuitive}
\begin{aligned}
res_{1,\;\text{intuitive}}(x,y,t)=\frac{\partial \tilde{E_z}}{\partial t} - \frac{1}{\varepsilon(x)} \left( \frac{\partial H_y}{\partial x} - \frac{\partial H_x}{\partial y} \right), 
\end{aligned}
\end{equation}
Where $\varepsilon(x)$ has a value of $1$ for collocation points in the vacuum part of the domain and a value of $\varepsilon_r$ in the dielectric part of the domain. The intuitive physics loss is defined by:
\begin{equation}
\begin{aligned}
\mathcal{L}_{\text{phys, intuitive}} = MSE(res_{1,\;\text{intuitive}}) + MSE(res_2) + MSE(res_3),
\label{eq:phys_loss_intuitive}
\end{aligned} 
\end{equation}
where all collocation points weigh the same, the QPINN and PINN runs in the dielectric case experience the same behavior as the QPINN runs in the vacuum case, in which, without adding the energy loss, they don't converge, but with it, they do. Overall, the results for runs used $\mathcal{L}_{\text{phys, intuitive}}$ and energy conservation loss are worse than results for runs with $\mathcal{L}_{\text{phys, diel}}$ (as the ones reported in Fig.~\ref{fig:dielectric_bar_chart_all}). Hence, the loss used was the one defined in Eq.~\ref{eq:phys_loss_diel}.

\subsection{Impact of parameter initialization on ``black holes''} \label{subsec:param_init_bh}
We had several assumptions for why in the vacuum case, the QPINN runs suffer from BH, while the PINN runs and the QPINN with the No Entanglement Ansatz (which would be referred to as classical for brevity for this discussion) don't. One was the tendency of the outputs of PQCs, as used in this work, with random parameter initialization, to cluster around zero due to several effects.
First, for expressive ans\"atze with randomly initialized parameters that approximate unitary 2-designs, the expectation of any traceless observable (e.g., Pauli-Z) over the parameter ensemble is zero, and typical values concentrate near zero in high dimension; formally, the Haar average satisfies: 
\begin{equation}
    \mathbb{E}[\langle O\rangle]=\mathrm{Tr}(O)/d=0
\end{equation}
and the distribution narrows with system size \cite{mcclean2018barren,cerezo2021cost,sim2019expressibility}. 
Second, with a single Pauli-rotation data-encoding layer (e.g., $R_X(\theta)$ per qubit), a fixed trainable block, and a Pauli readout, the model output is a first-harmonic trigonometric polynomial without a constant term; hence its average over $\theta$ (uniform on $[0,2\pi]$) is exactly zero \cite{schuld2021effect}. 
Combined with the bounded range of Pauli expectations ($\langle O\rangle\in[-1,1]$), this bias can create a ``zero-field'' basin that impedes escape at early training. 
This provides intuition for why adding the energy-conservation penalty improves optimization in the vacuum setting by reshaping gradients away from the trivial solution.

Opposed to that, the outputs of a classical linear, fully connected layer with hyperbolic tangent activation function, which is the second-to-last layer in our PINN comparison, "spread" much wider around zero even though those outputs are bounded in $[-1,1]$ as in the QPINN's cases.

\begin{figure}[!htbp]
  \centering
  \begin{subfigure}[t]{0.33\textwidth}
    \centering
    \includegraphics[width=\linewidth]{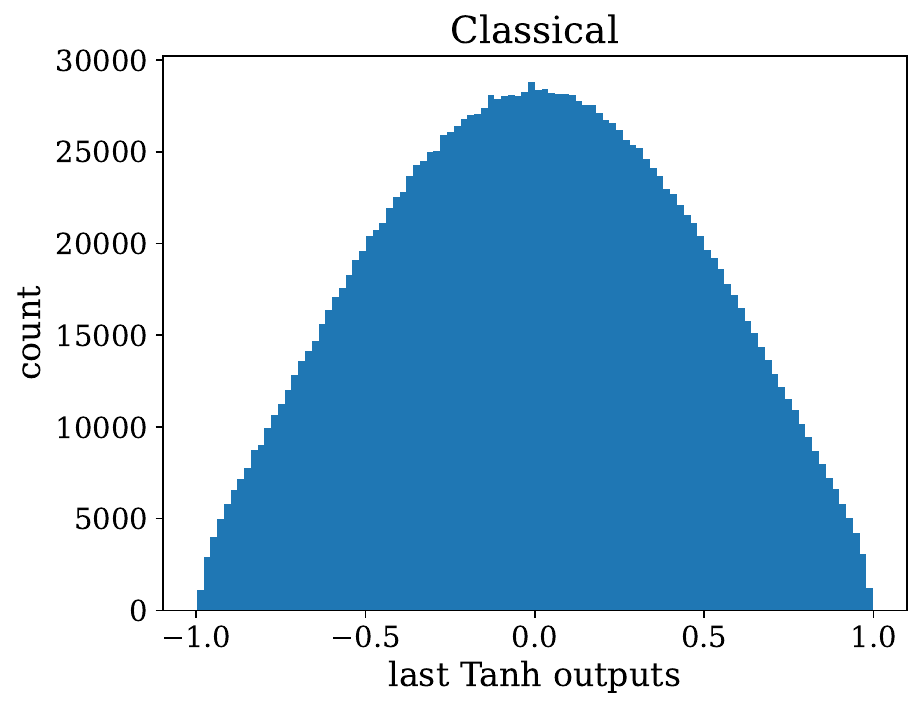}
    \caption{}
    \label{fig:init_params_classical}
  \end{subfigure}\hfill
  \begin{subfigure}[t]{0.33\textwidth}
    \centering
    \includegraphics[width=\linewidth]{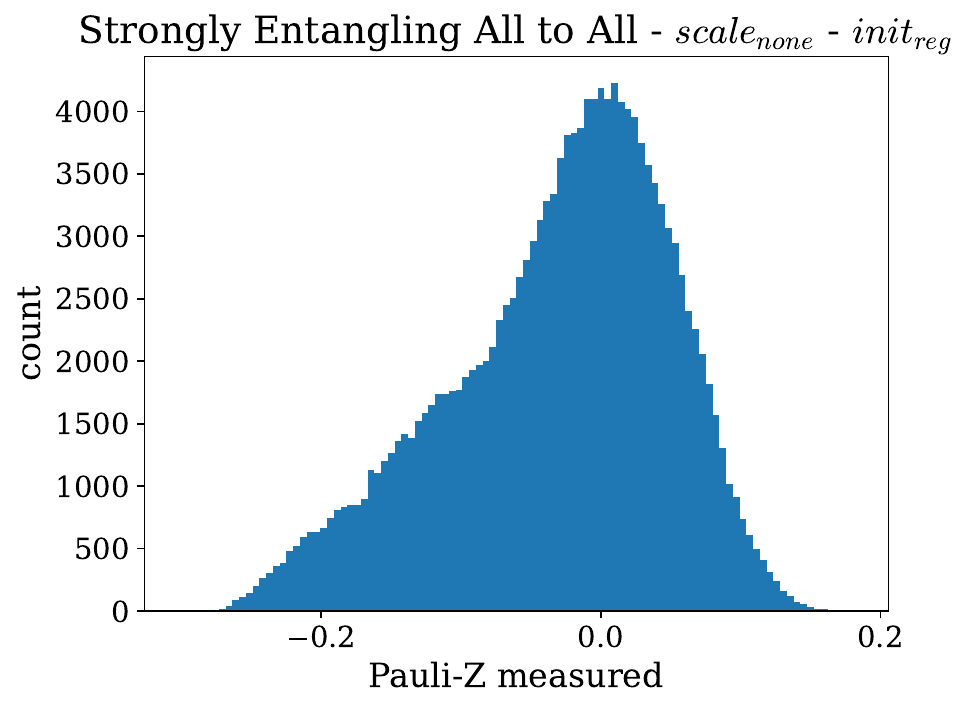}
    \caption{}
    \label{fig:init_params_all_reg}
  \end{subfigure}\hfill
  \begin{subfigure}[t]{0.33\textwidth}
    \centering
    \includegraphics[width=\linewidth]{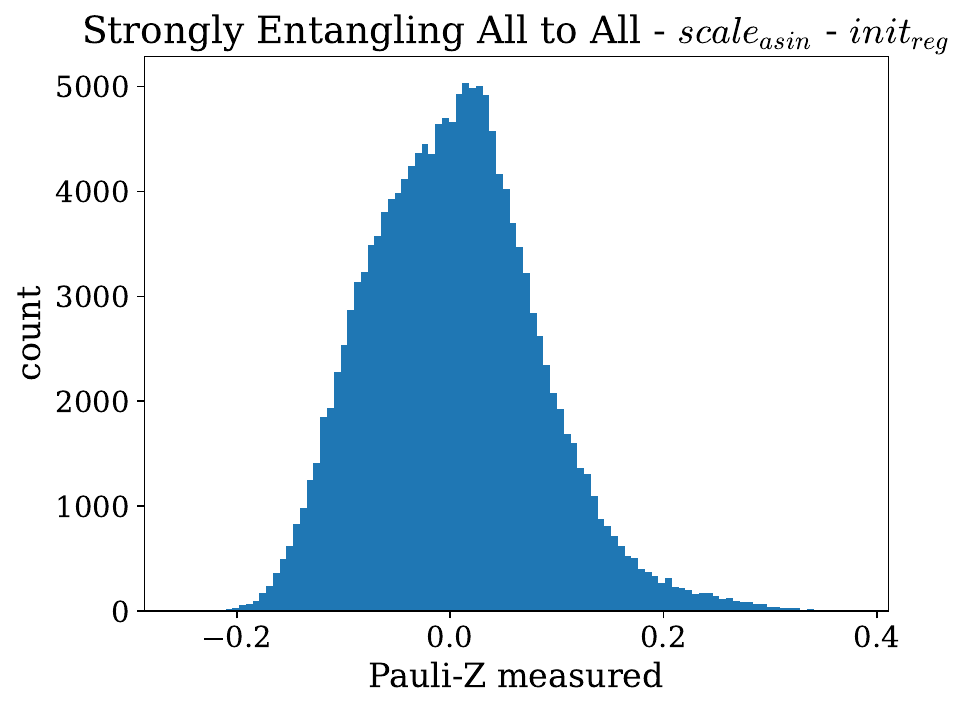}
    \caption{}
    \label{fig:init_params_all_asin}
  \end{subfigure}\hfill

  \medskip

  \begin{subfigure}[t]{0.33\textwidth}
    \centering
    \includegraphics[width=\linewidth]{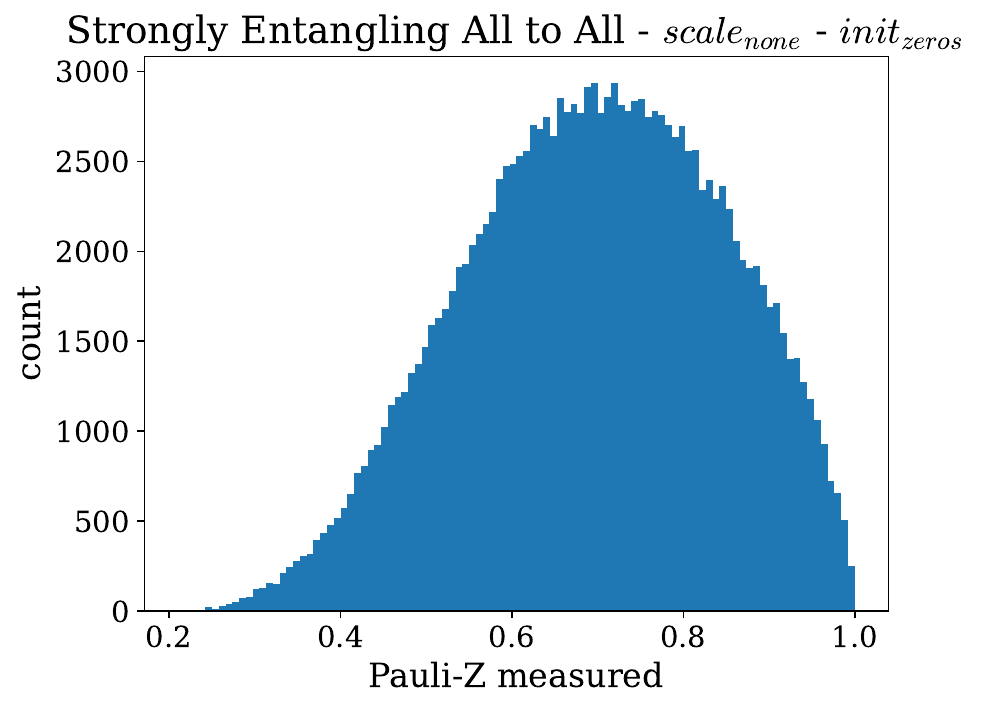}
    \caption{}
    \label{fig:init_params_all_zeros}
  \end{subfigure}\hfill
  \begin{subfigure}[t]{0.33\textwidth}
    \centering
    \includegraphics[width=\linewidth]{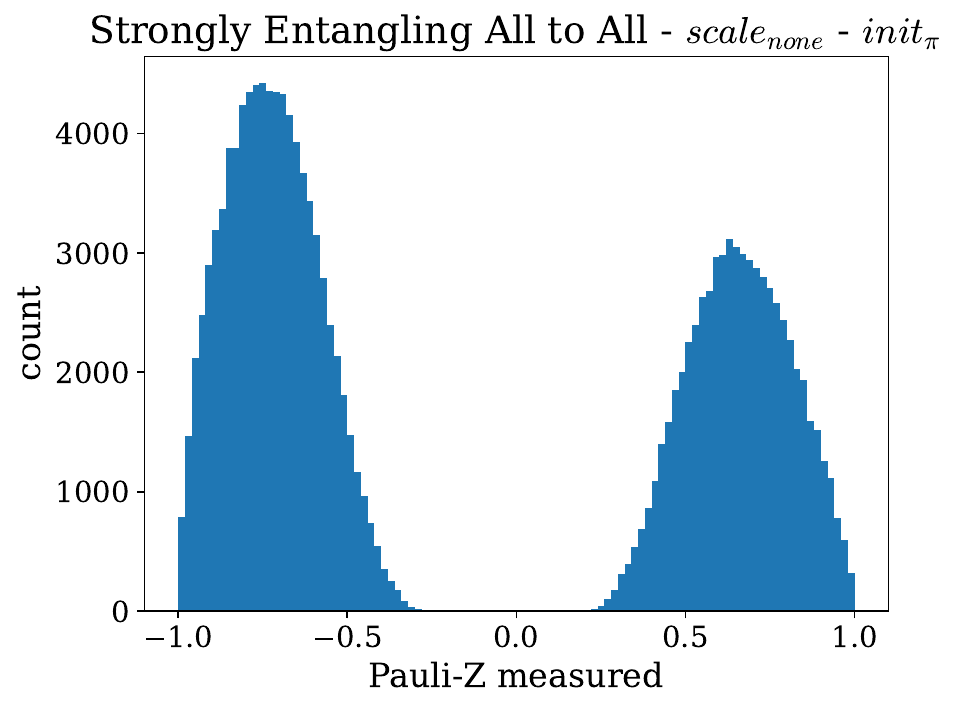}
    \caption{}
    \label{fig:init_params_all_ones}
  \end{subfigure}\hfill
  \begin{subfigure}[t]{0.33\textwidth}
    \centering
    \includegraphics[width=\linewidth]{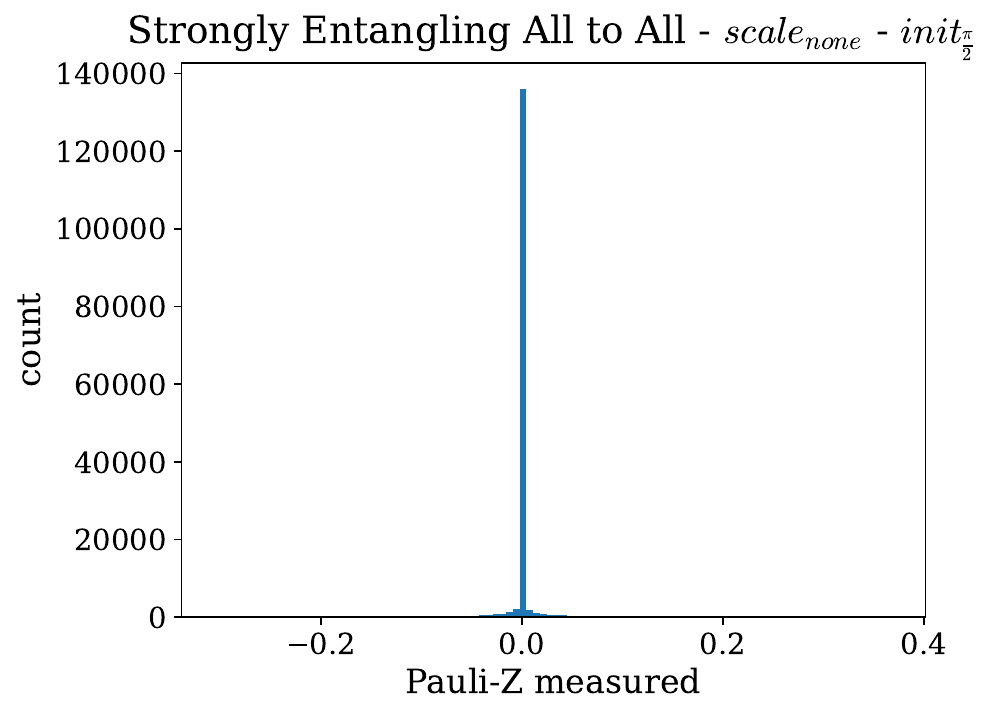}
    \caption{}
    \label{fig:init_params_all_pi_half}
  \end{subfigure}\hfill

  \medskip

  \begin{subfigure}[t]{0.33\textwidth}
    \centering
    \includegraphics[width=\linewidth]{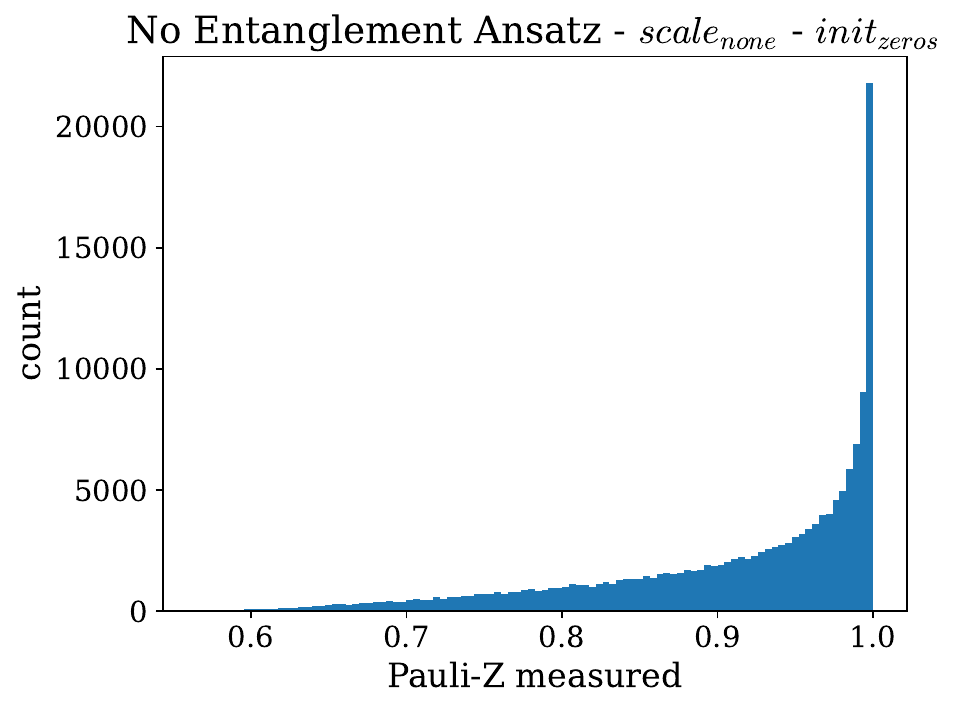}
    \caption{}
    \label{fig:init_params_no_ent_zeros}
  \end{subfigure}\hfill
  \begin{subfigure}[t]{0.33\textwidth}
    \centering
    \includegraphics[width=\linewidth]{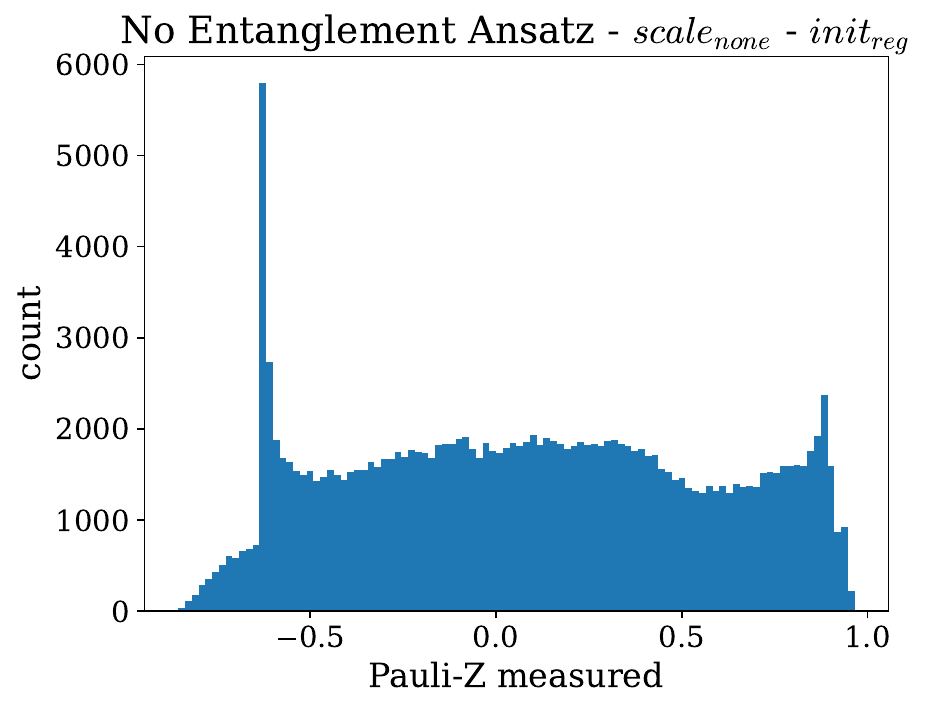}
    \caption{}
    \label{fig:init_params_no_ent_reg}
  \end{subfigure}\hfill
  \begin{subfigure}[t]{0.33\textwidth}
    \centering
    \includegraphics[width=\linewidth]{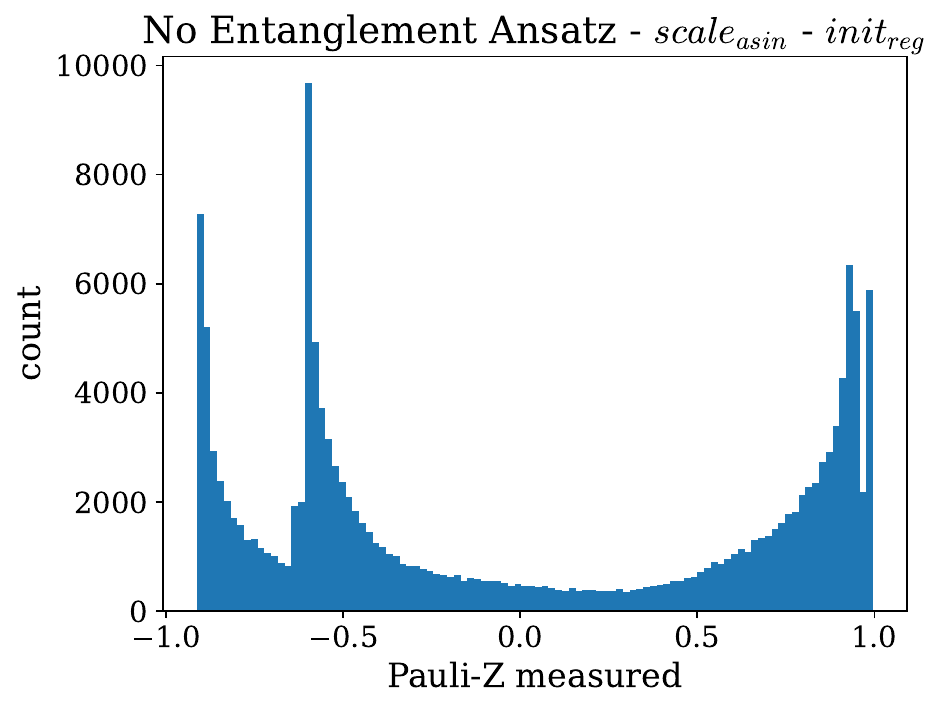}
    \caption{}
    \label{fig:init_params_no_ent_asin}
  \end{subfigure}\hfill

  \caption{Different outputs generated in the second-to-last layer in the network, either the quantum layer or the hyperbolic tangent in the PINN runs. Each run was made with a different combination of ansatz, scaling method, and parameter initialization method. The addition of the No Entanglement Ansatz was to investigate the origin of BH further, as it is a quantum circuit, but it does not experience BH. Please note the x-axis range.}
  \label{fig:all_init_params_figs}
\end{figure}

We assumed that this difference was the root cause for the BH occurrence in the QPINN runs, as most of its outputs are around zero and the TS is geometrically close to the random initialization on the loss landscape. To test this assumption, several parameter initialization strategies were employed, and we tracked the quantum layer's outputs in the initial epoch (Fig.~\ref{fig:all_init_params_figs}).
The initialization strategies tested were:
\begin{itemize}
     \item $init_{reg}$: The method used throughout this work was random parameter initialization between $[0,2\pi]$.
    \item $init_{zeros}$: Initializing all quantum parameters to 0. Without the angle embedding layer, this would result in all output values equal to 1.
    \item $init_{\pi}$: Initializing all quantum parameters to $\pi$.
    \item $init_{\pi/2}$: Initializing all quantum parameters to $\pi/2$.
\end{itemize}

The different initializations did not affect BH at all, and each run kept its original BH behavior as before. According to this and Fig.~\ref{fig:all_init_params_figs}, we can infer that the different output spread at initialization is not the root cause of BH in the QPINN runs, nor is it what mitigates BH in PINN runs; hence, this assumption is refuted. Understanding why the QPINN runs exhibit BH but the classical ones don't remains as a future work direction.

\section{Discussion and Conclusion} \label{sec:discussion_and_conclusion}
\subsection{Discussion} \label{subsec:discussion}
Our results demonstrate that a hybrid quantum-classical PINN can successfully learn solutions for 2D Maxwell's equations, with better accuracy compared to a fully classical PINN, by more than $19\%\;L_2$ error reduction, and even better than much larger networks - the classical extra layer setup. The inclusion of a quantum circuit does not degrade performance; in fact, in some runs, we observe that the QPINN achieves exceptionally low errors, which hints at the potential representational advantages of the quantum ansatz. However, these advantages do not consistently manifest across all trials, pointing to current optimization challenges. Furthermore, much faster learning is observed in most QPINN runs, especially for the vacuum case with an energy loss, that demonstrates even higher potential than before.

One clear finding is the importance of enforcing physical constraints, especially energy conservation, in QPINN training. For the vacuum case, adding a global energy conservation term is essential for a stable training of the quantum model. This suggests that QPINNs can significantly benefit from additional physical ``guidance'' beyond just raw PDE residuals.

Regarding the quantum circuit ansatz, our ablation study yields several insights. In the vacuum case, the best performing are ans\"atze based on mid-depth entangling layers - the Basic and Strongly Entangling Layers. On the other hand, in the dielectric case, those ans\"atze perform worse than the rest. 
We do not claim a monotonic relation between entanglement strength and accuracy. In the vacuum (with the energy-conservation term), mid-depth entangling ans\"atze achieve the best L2 errors, whereas in the dielectric case, less-entangled or differently connected variants perform better. We attribute this to problem- and loss-specific inductive biases rather than entanglement per se.

The effect of different input scales has a significant impact in the vacuum case, where all schemes are substantially better than the PINN run except for the case with $scale_{\pi}$. In case 2 with a dielectric medium, as in the rest of the criteria compared, the input scaling has a more subtle effect, where only $scale_{asin}$ (best) and $scale_{none}$ achieve better results than the PINN run. Following this behavior, we observe that: (a) the ML model can overcome the input and measurement distribution problems as $scale_{none}$ is the second best in both cases. (b) The scaling is problem-specific, and the performance is better when it is coupled with the correct model.

Computationally, we observe a significant reduction in the number of learnable parameters in quantum models, even when they outperform all classical models. Specifically, in the regular model, there is a  $19\%$ reduction in the QPINN's case, and a decrease of $33\%$ compared to the extra layer setup. Moreover, if one can tolerate a lower accuracy, the difference in lower epochs is far greater in favor of the QPINN runs.

Developing solutions for large cases, such as those in this study, required us to optimize the simulator using our own in-house developed quantum simulator, which was more than $50$ times faster than the current industry's most common development packages, while reducing the memory needed by more than 6 times, leveraging the tremendous power of GPU-based simulations.

Another primary consideration is the parameter reduction scaling, although we were unable to run significantly larger simulations due to computational limitations and therefore could not explore this aspect more rigorously. In our 2D unsteady test cases (pulse in a box and dielectric medium), the variational circuits achieved a parameter reduction of approximately $19\%$ (66,932-67,072 in the QPINN's case, and 82,820 in the PINN's case) relative to classical baselines at comparable, and even lower, error levels. As mentioned, due to computational limitations, we were unable to explore significantly deeper or wider circuits; however, we confirmed consistent trends in the shallower configurations studied. 
Moreover, other works, such as Trahan et al.~\cite{trahan2024quantum}, which have worked on significantly smaller cases, support similar findings. In their 1D Burger's equation test case with a similar architecture to our network architecture, they observed approximately $58\%$ (1,456 in the QPINN's case, and 3,441 in the classical case) parameter amount reduction. Still, it should be noted that those are substantially smaller networks compared to the one used in this work.

\subsection{Why might the quantum layer help?} \label{subsec:why_qlayer_help}

Across our ablations, the hybrid model frequently converges faster and attains lower $L_2$ errors than the purely classical PINN while using fewer trainable parameters (Table~\ref{tab:parameters_amount}; Figs.~\ref{fig:avg_loss_and_bar_chart_all_vacuum}-\ref{fig:bar_charts_avg_dielectric}). We outline below several \emph{plausible} mechanisms by which the parametrized quantum circuit (PQC) may contribute to these gains in our setting. These mechanisms are consistent with our observations, but were not directly verified in this study.
\begin{itemize}
    \item \textbf{Harmonic feature expansion aligned with wave physics:}
    With single-qubit $R_x$ data encoding and Pauli-Z readout, PQC outputs are trigonometric polynomials in the encoded angles; different input scalings modulate this effective harmonic basis (see Fig.~\ref{fig:scales} and Subsec.~\ref{subsec:QPINN}). This is theoretically expected: the measurement expectations produced by such encodings generate Fourier-like features whose span grows with depth \cite{schuld2021effect,perez2020data}. Because time-dependent Maxwell solutions contain rich oscillatory content and we enforce periodicity in space/time, placing a periodic, high-frequency basis late in the network can plausibly mitigate spectral bias and lower the depth/width needed to represent wavefields \cite{tancik2020fourier,wang2021eigenvector}.
    \item \textbf{Multiplicative, high-order interactions via entanglement (in moderation):} Entangling layers cause the measured observables to involve multi-qubit Pauli strings, which correspond to products of sines/cosines across the encoded coordinates. This yields multiplicative cross-terms (e.g., in $x$, $y$, and $t$) akin to higher-order features that may compactly capture the coupled curl relations in the TE$_z$ system. Prior analyses relate an ansatz's entangling capability and expressibility to the richness of such interactions \cite{sim2019expressibility,schuld2021effect}. Our results suggest a ``sweet spot'': mid-depth, moderately entangling ans\"atze perform best in vacuum, whereas the no-entanglement variant is competitive in the heterogeneous case (Figs.~\ref{fig:avg_loss_and_bar_chart_all_vacuum}-\ref{fig:bar_charts_avg_dielectric}), indicating that the optimal interaction order is problem- and loss-dependent rather than “more entanglement is always better”.
    \item \textbf{Alignment between learned periodic embeddings and late-stage quantum harmonics:} The classical front-end already enforces strict spatial periodicity and a learned temporal periodicity (Subsec~\ref{subsec:PINN}). The PQC then injects additional periodic (trigonometric) features \emph{near the output}, effectively refining the spectrum after the PDE-informed residuals have shaped lower-frequency structure. This division of labor, where classical layers capture the coarse structure and the quantum layer contributes targeted harmonic refinements, offers a plausible explanation for the faster loss decay we observe when the PQC is present.
\end{itemize}

\emph{Scope and limitations.} We emphasize that these points are \emph{hypotheses} consistent with our ablations and with existing theory on encoding-induced expressivity and entangling capability \cite{schuld2021effect,sim2019expressibility,perez2020data}. We did not directly measure the spectral content, interaction order, or effective model dimension; therefore, we do not claim a definitive causal mechanism. In line with recent guidance on benchmarking quantum ML \cite{bowles2024better}, we view our evidence as suggestive rather than conclusive. Moreover, we do not claim that adding the PQC will be beneficial for other test cases beyond those studied here; further research is needed to generalize these findings.

\emph{Suggested follow-ups.} To validate the points above, future work could (a) quantify the frequency spectra of the learned fields and of the PQC outputs over $(x,y,t)$; (b) compare against a classical control that augments the penultimate layer with an equal-size trigonometric basis; (c) use data-reuploading cycles in varying sizes and measure changes in accuracy/parameter efficiency; (d) probe interaction order by ablating entanglement patterns while keeping parameter counts fixed; and (e) monitor gradient norms/variances versus depth to locate an expressivity-trainability sweet spot \cite{mcclean2018barren,larocca2025barren, chang2025primer}.

\subsection{Conclusions and future work} \label{subsec:conclusion}
We present a QPINN approach for solving two-dimensional Maxwell's equations and analyze the effects of the ansatz design, input scaling, and energy conservation enforcement. The study finds that under appropriate physical constraints and architectural settings, QPINNs can achieve better accuracy with lower parameters compared to classical PINNs. The enforcement of energy conservation is critical for physically consistent solutions for the QPINN runs in the vacuum case, which would experience BH without it, highlighting that domain knowledge can substantially enhance PINN training. 

We demonstrate how quantum machine learning can be leveraged to enhance and accelerate scientific computing for field simulations. While our experiments were performed on a classical simulator, they lay the groundwork for future tests on quantum hardware as it matures. Key directions for future research include scaling up the problem size (both in finer resolution and 3D problems) to test whether QPINNs can retain performance with fewer parameters than classical networks. Additionally, investigating more advanced quantum circuit training techniques, such as quantum natural gradient or circuit-specific regularization, is crucial to improve convergence. It would also be worthwhile to explore other physics-informed tasks, such as inverse problems (e.g., identifying material properties from field observations), to see whether QPINNs offer any advantage there. Another important direction for future work would be to incorporate noise into the quantum circuits and investigate the impact of noise mitigation and error correction methodologies on QPINNs.

The reason the PINN runs don't suffer from the BH phenomenon in the vacuum test case is still not fully clear. Several assumptions were tested, including whether it occurs in the QPINNs but not in the PINNs due to the different spread of outputs from the second-to-last layer (the quantum layer in QPINN runs and tanh activation function in the PINN runs) right from initialization. It turns out that this isn't the root cause of BH, as even when similar spreads were used, the QPINN still suffered from BH, while the classical didn't. Therefore, another future work direction is to understand why this happens in the QPINN runs and not in the PINNs. By understanding this, it may be possible to improve other aspects of both classical and quantum networks.

We present a quantum-enhanced physics-informed neural network (QPINN) that reduces $L_2$ solution error for Maxwell's equations by up to $19\%$ relative to classical PINNs. Our model's energy conservation loss mitigates the ``black hole'' phenomenon, enabling stable training in high-dimensional spatiotemporal domains. This work utilizes an in-house, GPU-accelerated framework implementation that delivers substantial end-to-end training speedups, bringing high-dimensional electromagnetic simulations within practical reach on existing hardware.

\section*{Author Contributions Statement}
Z.C. wrote the main manuscript text, and collected all data, and developed the code for the model and the quantum simulator based on a classical PINN code developed by G.S., H.C. and S.H.. H.C. reviewed the manuscript several times and was a significant part of shaping and correcting it. S.F. and U.P. have jointly supervised the work and reviewed the manuscript.

\section*{Data availability statement}
Data sets and results generated during the current study are available from the corresponding author on reasonable request.

\section*{Code availability}
The in-house developed quantum library, \textit{TorQ - Tensor Operations for Research in Quantum systems}, used for all simulations in this work, is available at: \url{https://github.com/zivchen9993/TorQ.git}. Also, for performance benchmarking against PennyLane's simulators, an expansion is required, and can be found at: \url{https://github.com/zivchen9993/TorQ-bench.git}.\nocite{torq}

\section*{Conflicts of interest}
The authors declare that they have no conflicts of interest.

\section*{Acknowledgements}
We would like to thank Yigal Ilin for providing an early version of the quantum simulator code, and we are grateful to the Israel Innovation Authority for funding this research. 
Z. C. and U. P. were supported by the Israel Science Foundation (ISF), Grants n. 939/23 and 2691/23, German-Israeli Project Cooperation (DIP) n. 2032991, Ollendorff-Minerva Center of the Technion n. 86160946, and the Helen Diller Quantum Center at the Technion. 2033613. U. P. was also supported by Chaya Chair n. 8776026, and Planning and Budgeting Committee of the Council for Higher Education of Israel (VATAT) Program for Quantum Science and Technology through Grant n. 86636903.

\bibliographystyle{unsrt}
\bibliography{references}

\clearpage

\appendix
\section{Asymmetrical pulse in vacuum} \label{app:asymmetrical}  

This asymmetrical test case is identical to the vacuum case, except for the initial condition that enforces an asymmetrical solution. Also, as in the vacuum case, the boundary conditions are periodic. It was a test case added to check if the solution's behavior changes with respect to whether the energy conservation loss (Eq.~\ref{eq:energy_loss}) is added in an asymmetrical test case. Obviously, the symmetry loss (Eq.~\ref{eq:2D_symLoss}) wasn't used here. For this test case, only the best performing QPINN runs in the vacuum case (Fig~\ref{fig:vacuum_bar_chart_all}) - Strongly Entangling Layers ansatz with $scale_{acos}$ scaling method was chosen.

\begin{figure}[!htbp]
  \centering
  \begin{subfigure}[t]{1.0\textwidth}
    \centering
    \includegraphics[width=\linewidth]{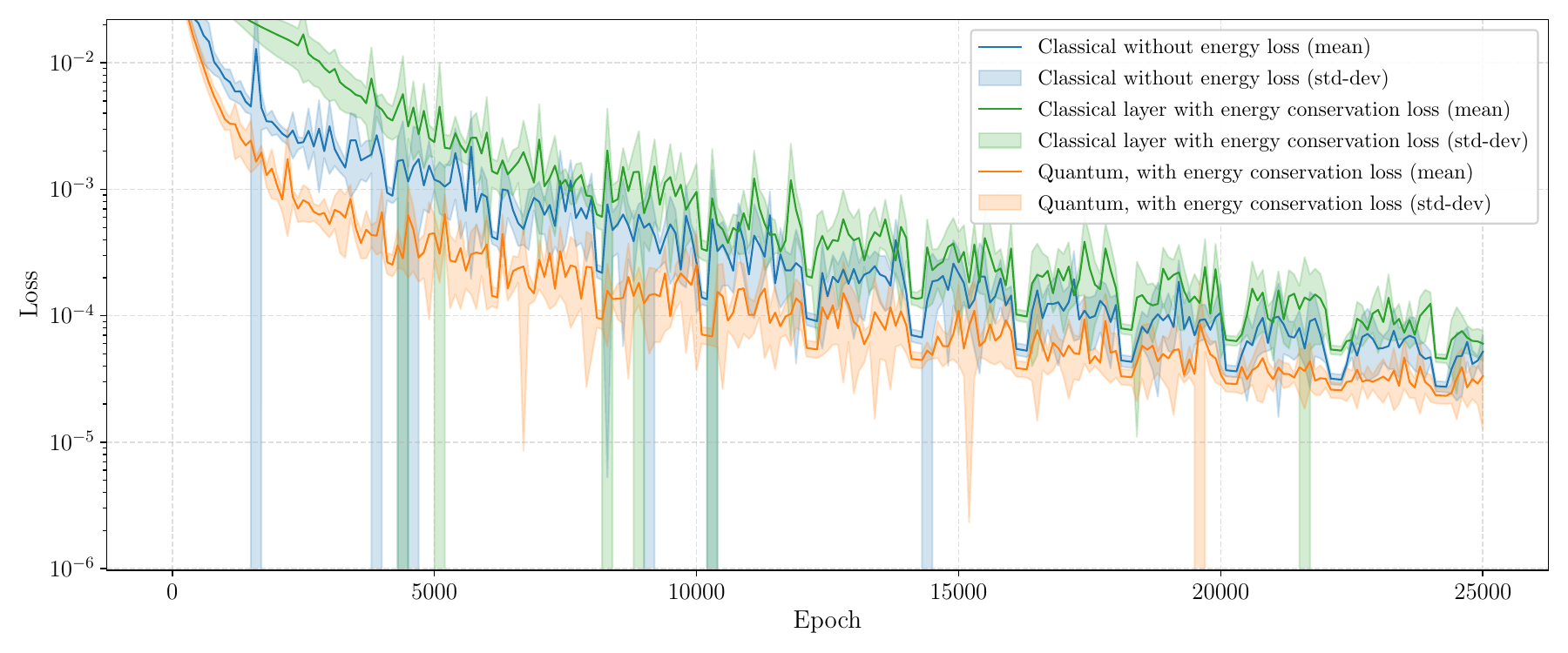}
    \caption{}
    \label{fig:asymmetrical_avg_loss}
  \end{subfigure}\hfill

  \medskip
  
  \begin{subfigure}[t]{0.8\textwidth}
    \centering
    \includegraphics[width=\linewidth]{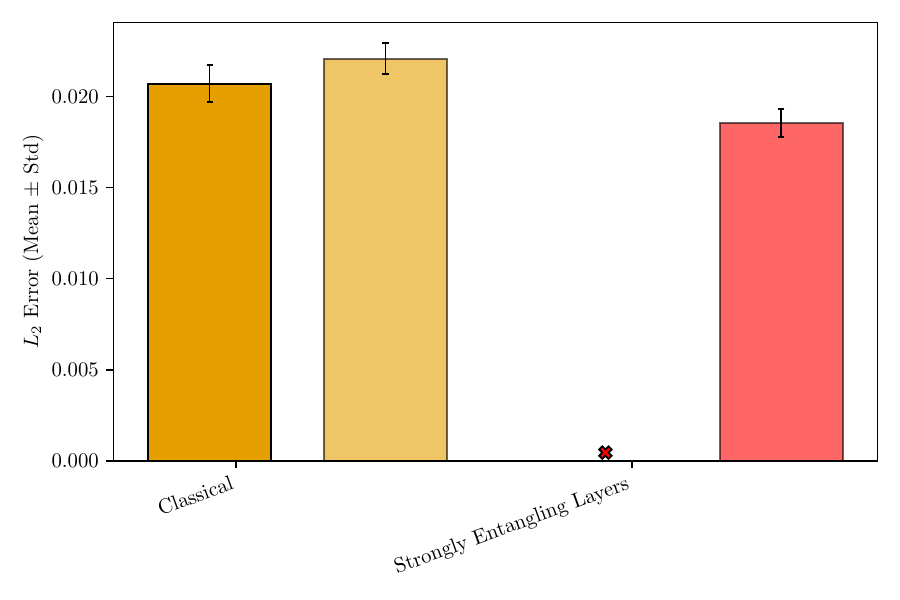} 
    \caption{}
    \label{fig:asymmetrical_bar_chart_all}
  \end{subfigure}\hfill
  
  \caption{(a) Mean loss (log scale) for the runs made in this asymmetric test case, which was the best combination for the vacuum case: Strongly Entangling Layers ansatz with $scale_{acos}$ and energy loss included. The semi-transparent band indicates the standard deviation. (b) $L_2$ errors for the classical regular with energy loss (dark yellow), without energy loss (light yellow), and the Strongly Entangling Layers ans\"atze, where only the runs with energy conservation converged and the ones without are marked with ``X''. The dashed yellow line shows the classical result for the regular-depth network. The dash-dot line indicates the best run's error (highlighted by a gold star). An ``X'' denotes configurations where none of the 5 runs converged.}
  \label{fig:avg_loss_and_bar_chart_all_asymmetrical}
\end{figure}

According to the results in Fig.~\ref{fig:avg_loss_and_bar_chart_all_asymmetrical}, the same behavior as in the vacuum case is observed: the QPINN runs without energy conservation loss experienced BH, but with it, the QPINN model outperformed both PINN runs, with and without it. Also, the PINN run without the energy conservation loss achieved better results than the one with.

\begin{figure}[!htbp]
  \centering
  \begin{subfigure}[t]{0.49\textwidth}
    \centering
    \includegraphics[width=\linewidth]{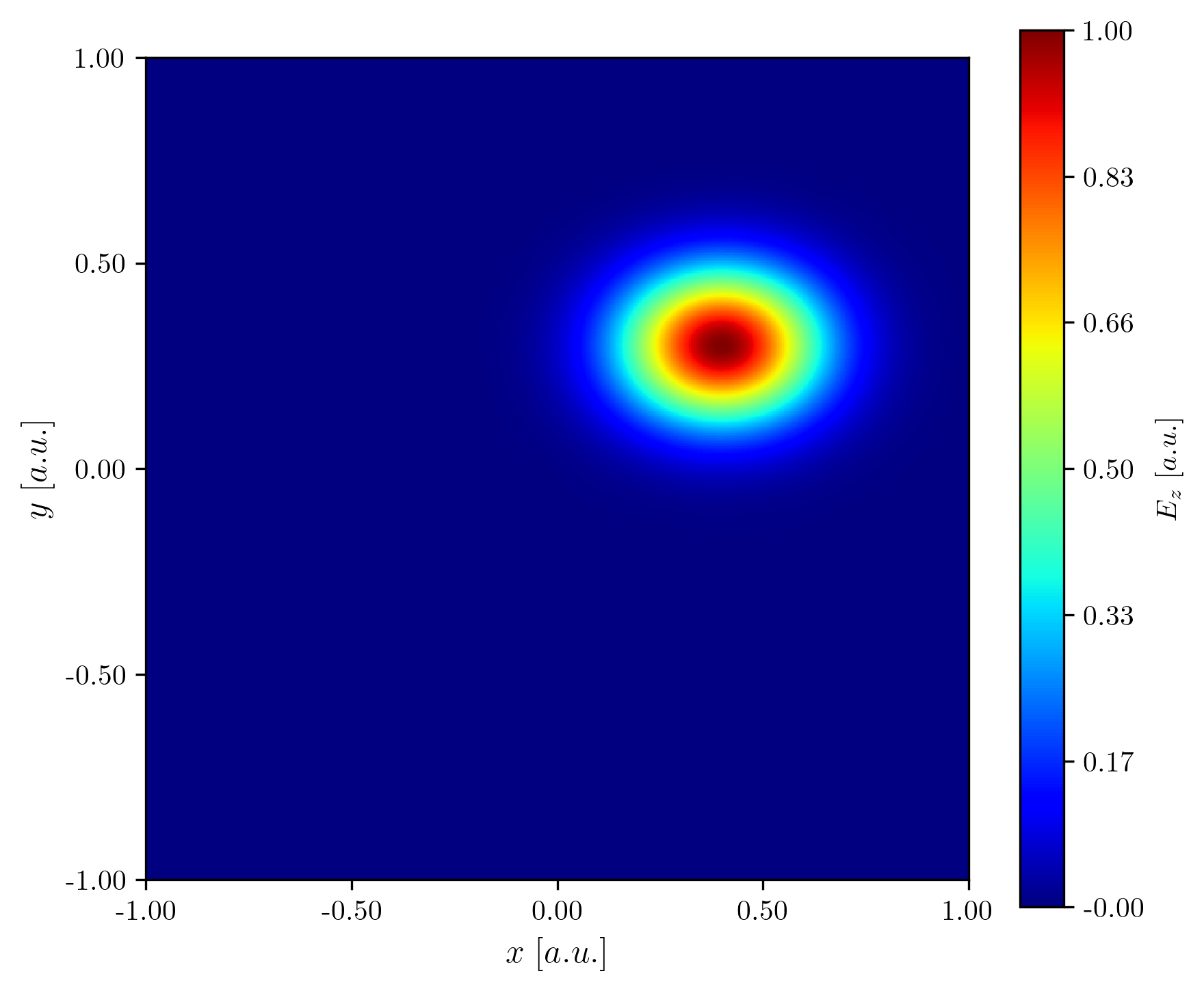}
    \caption{}
    \label{fig:asymmetrical_res_init}
  \end{subfigure}
  \begin{subfigure}[t]{0.49\textwidth}
    \centering
    \includegraphics[width=\linewidth]{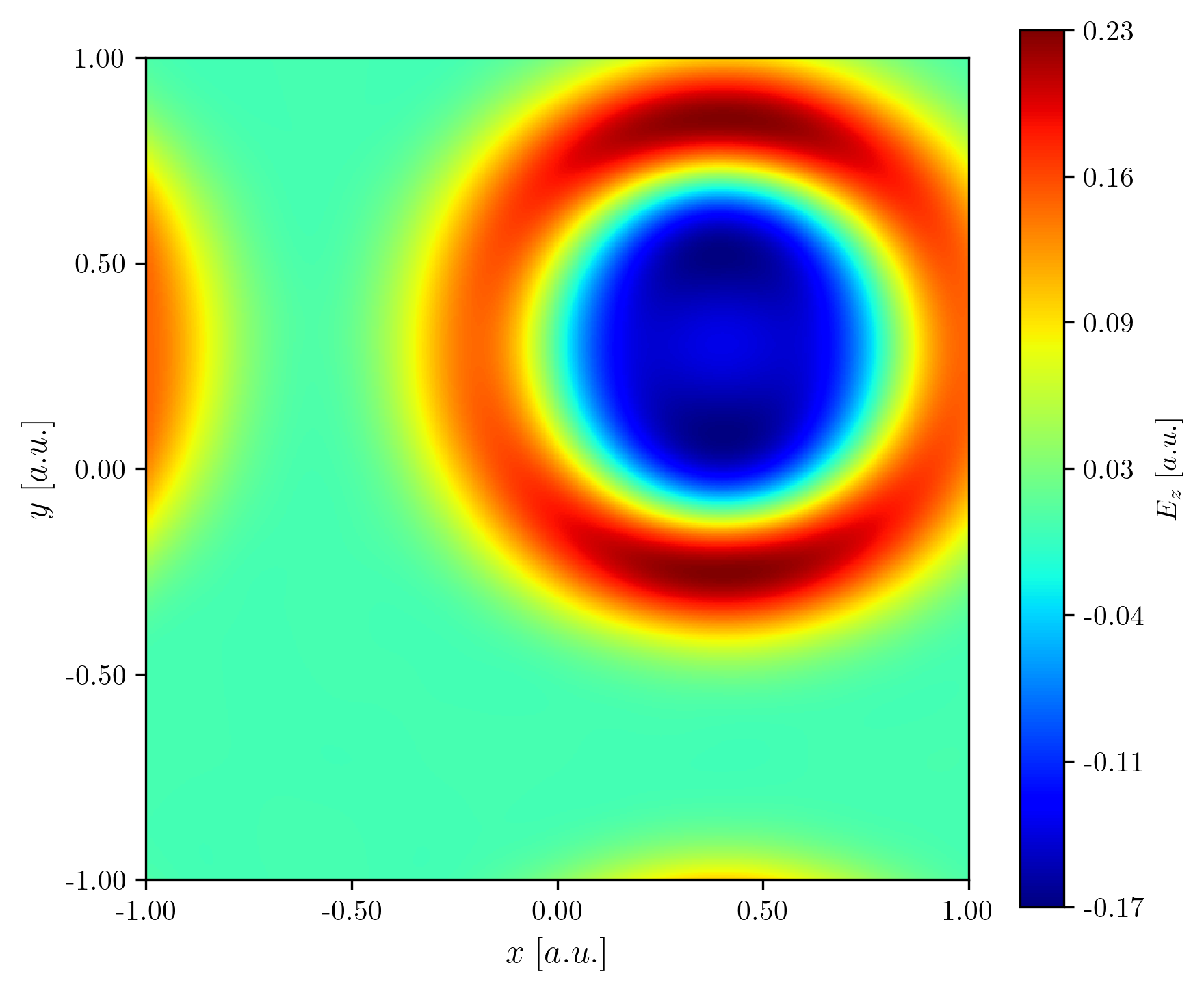}
    \caption{}
    \label{fig:asymmetrical_res_05}
  \end{subfigure}\hfill

    \medskip

  \begin{subfigure}[t]{0.49\textwidth}
    \centering
    \includegraphics[width=\linewidth]{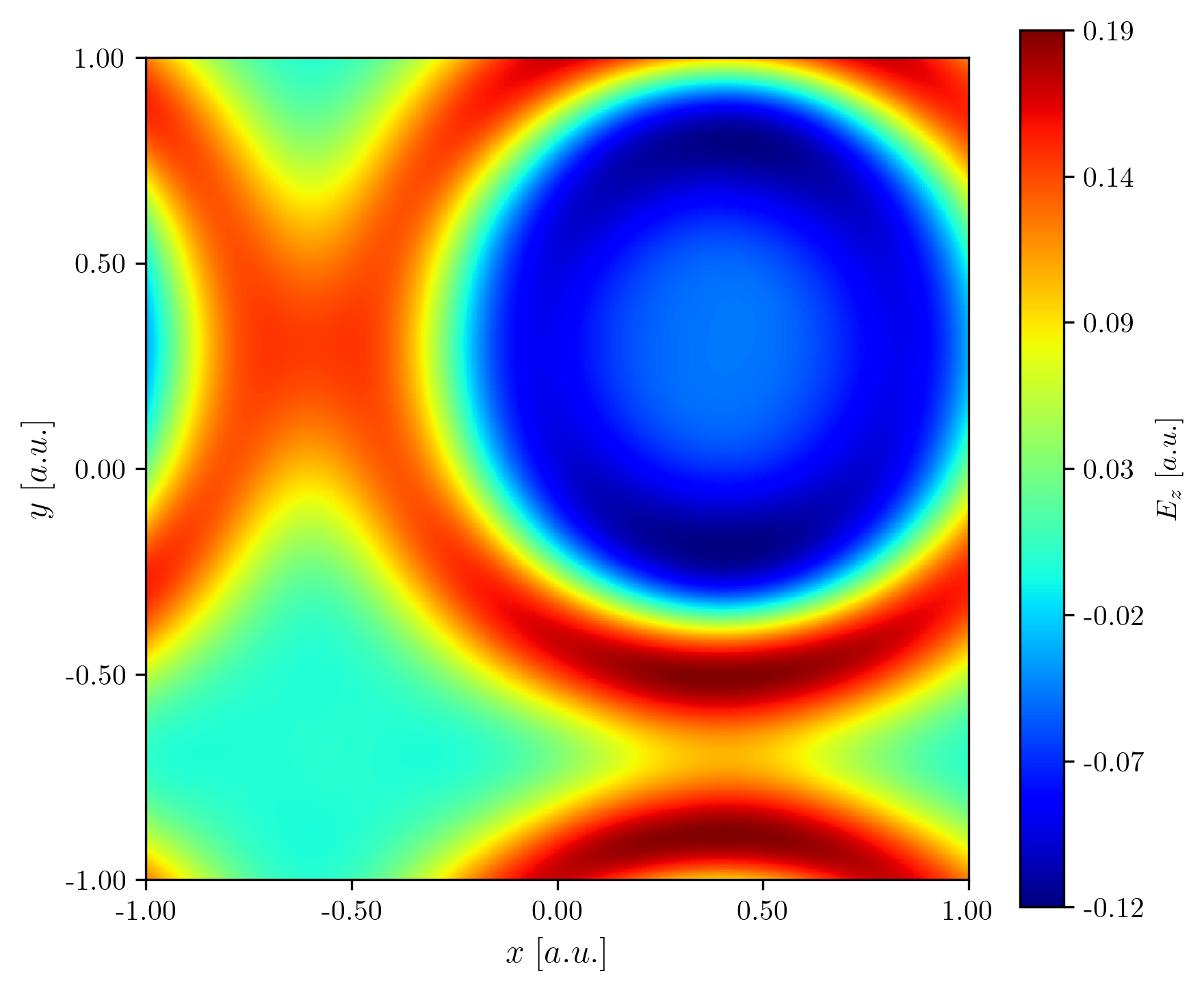}
    \caption{}
    \label{fig:asymmetrical_res_08}
  \end{subfigure}\hfill
  \begin{subfigure}[t]{0.49\textwidth}
    \centering
    \includegraphics[width=\linewidth]{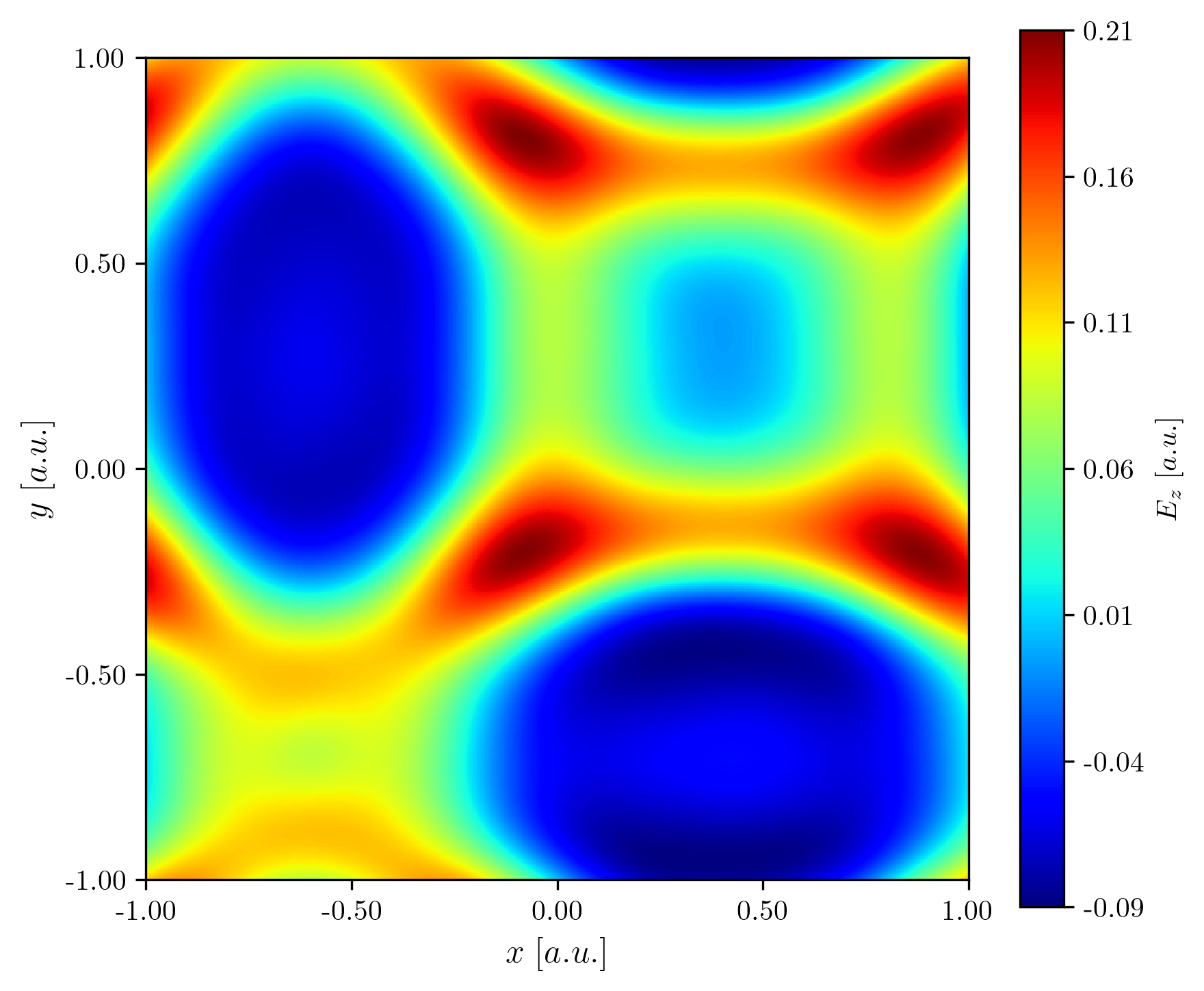}
    \caption{}
    \label{fig:asymmetrical_res_15}
  \end{subfigure}  

  \caption{(a) Initial conditions (i.e. $t=0$) for the asymmetrical pulse test case. The Gaussian pulse is initialized at $(x, y)=(0.4, 0.3)$ and stretched by $(\sigma_x, \sigma_y)=(0.85,0.65)$. (b) Shows the contours of the electric field $E_z$ at the moment the wave's propagation reaches the boundaries of the domain, where $t=0.5$. (c) Captures the propagation at time $t=0.8$, in which waves from nearby domains reach the wave in the domain simulated (the boundaries are periodic, one can view it as infinite tiles of the same domain tiled next to each other). (d) Shows the final simulation time at $t=1.5$.}
  \label{fig:asymmetrical_results}
\end{figure}

\end{document}